\documentclass[11pt, prd, aps, showpacs, superscriptaddress, floatfix, nofootinbib]{revtex4-1}
\usepackage{amsmath,amssymb}
\usepackage{geometry}
\usepackage{graphicx}
\usepackage{epstopdf}
\usepackage{bbold}
\usepackage{braket}

\newcommand{\RomatreINFN}{Istituto Nazionale di Fisica Nucleare, Sezione di Roma Tre,\\ Via della Vasca Navale 84, I-00146 Rome, Italy}
\newcommand{\LaSapienza}{Physics Department and INFN Sezione di Roma La Sapienza,\\ Piazzale Aldo Moro 5, 00185 Roma, Italy}
\newcommand{\SNS}{Scuola Normale Superiore,\\ Piazza dei Cavalieri 7, 56126 Pisa, Italy}
\newcommand{\INFNPisa}{Istituto Nazionale di Fisica Nucleare, Sezione di Pisa,\\ Largo Bruno Pontecorvo 3, I-56127 Pisa, Italy}

\newcommand{\be}{\begin{equation}}
\newcommand{\ee}{\end{equation}}
\newcommand{\bea}{\begin{eqnarray}}
\newcommand{\eea}{\end{eqnarray}}

\begin{document}

\title{$\vert V_{cb} \vert$ and $R(D^{(*)})$ using lattice QCD and unitarity}

\author{G.\,Martinelli}\affiliation{\LaSapienza}
\author{S.\,Simula}\affiliation{\RomatreINFN}
\author{L.\,Vittorio}\affiliation{\SNS}\affiliation{\INFNPisa}

\begin{abstract}
The Cabibbo-Kobayashi-Maskawa (CKM) matrix element $\vert V_{cb}\vert$ is extracted from exclusive semileptonic $B \to D^{(*)}$ decays adopting a novel unitarity-based approach which allows to determine in a full non-perturbative way the relevant hadronic form factors (FFs) in the whole kinematical range. By using existing lattice computations of the $B \to D^{(*)}$ FFs at small recoil from FNAL/MILC and JLQCD Collaborations, we show that it is possible to extrapolate their behavior also at large recoil without assuming any specific momentum dependence and without constraining their shape using experimental data. Thus, we address the extraction of $\vert V_{cb}\vert$ from the experimental data on the semileptonic $B \to D^{(*)} \ell \nu_\ell$ decays, obtaining $\vert V_{cb}\vert = (41.0 \pm 1.2 ) \cdot 10^{-3}$ from $B \to D$ using as input the final FNAL/MILC lattice data for the FFs and $\vert V_{cb}\vert = (40.4 \pm 1.8 ) \cdot 10^{-3}$ from $B \to D^*$ using the preliminary JLQCD lattice data. Our result from $B \to D$ is consistent within $\sim 1$ standard deviation with the most recent inclusive determination $\vert V_{cb} \vert_{incl} = (42.00 \pm 0.65) \cdot 10^{-3}$. The resulting uncertainty is comparable with those obtained in literature using experimental data to constrain the shape of the FFs. Our result from $B \to D^*$, though consistent with $\vert V_{cb} \vert_{incl} $, is still based on preliminary lattice data for the FFs and its uncertainty is greater than the ones obtained in literature using experimental data to constrain the shape of the FFs. We investigate also the issue of Lepton Flavor Universality thanks to new theoretical estimates of the ratios $R(D^{(*)})$, namely $R(D) = 0.296(8)$ using final FNAL/MILC lattice results, and $R(D^{*}) = 0.261(20)$ using preliminary JLQCD and FNAL/MILC lattice data. Our findings differ by $\sim 1.4\sigma$ from the latest experimental determinations.
\end{abstract}

\maketitle

\section{Introduction}

Exclusive semileptonic $B \to D^{(*)} \ell \nu$ decays are among the most important and challenging processes in the phenomenology of flavor physics. There are two reasons that justify their importance. On the one hand, we have the so-called $\vert V_{cb} \vert$ \emph{puzzle}, $i.e.$ the tension between the inclusive \cite{Gambino:2013rza,Alberti:2014yda, Gambino:2016jkc} and the exclusive determinations of the CKM matrix element $|V_{cb}|$ \cite{Aubert:2007qs,Aubert:2007rs,Aubert:2008yv,Aubert:2009ac,Dungel:2010uk,Glattauer:2015teq,Abdesselam:2017kjf,Abdesselam:2018nnh}. On the other hand, a discrepancy exists between the theoretical expectation value and the measurements of $R(D^{(*)})$  \cite{HFLAV}, defined as the ratios of  the branching fractions $B \to D^{(*)} \tau \nu$ over $B \to D^{(*)} \ell \nu$, $\ell = e, \mu$, performed by Belle, BaBar and LHCb \cite{Lees:2012xj,Lees:2013uzd,Aaij:2015yra,Huschle:2015rga,Sato:2016svk,Hirose:2016wfn,Aaij:2017uff,Hirose:2017dxl,Aaij:2017deq}. 
 
From the theoretical point of view, the extraction of $\vert V_{cb} \vert$ from exclusive $B \to D^{(*)} \ell \nu$ decays and the theoretical estimates of $R(D^{(*)})$ depend on the value and the shape of the Form Factors (FFs) entering the matrix elements and amplitudes. These functions encode the non-perturbative strong interactions between the quarks and can be determined through lattice QCD (LQCD) computations. However, for the $B \to D^{(*)} \ell \nu$ decays the kinematical regions accessible to LQCD calculations are still restricted to small values of the recoil\footnote{Only recently\,\cite{McLean:2019qcx,Harrison:2021tol} in the case of the $B_s \to D_s^{(*)}\ell \nu$ decays LQCD simulations have been extended to the full kinematical range for a series of heavy-quark masses adopting truncated z-expansions to parametrize the momentum dependence.}. In this work we make use of lattice computations for $B \to D \ell \nu$ decays in final form~\cite{Bailey_2015} and for the $B\to D^* \ell \nu$ decays in preliminary unblinded  ~\cite{Kaneko:2019vkx} and blinded~\cite{Aviles-Casco:2019zop} forms available at non-zero, but small recoil.

The fundamental question is whether we can describe the FFs in the full kinematical range. To this aim, in the past some parametrisation of the form factors has always been chosen. The two most famous ones are Boyd-Grinstein-Lebed (BGL)~\cite{ Boyd:1995cf,Boyd:1995sq,Boyd:1997kz} and Caprini-Lellouch-Neubert (CLN)~\cite{Caprini:1995wq,Caprini:1997mu}. For example, the authors of Refs.~\cite{Bigi:2016mdz, Grinstein:2017nlq,Bigi:2017njr, Gambino:2019sif, Bordone:2019vic, Jaiswal:2020wer, Iguro:2020cpg} have used these parametrizations (in some cases with some modifications) in order to obtain their theoretical estimates of $\vert V_{cb} \vert$ from $B \to D^{(*)} \ell \nu$ decays. The bottom line of these analyses is that the determination of $\vert V_{cb} \vert$ from $B\to D \ell \nu$ transitions is compatible with the inclusive one, while the estimate from $B\to D^* \ell \nu$ is in strong tension with it.

In this paper, we re-analyse the lattice and the experimental data about $B \to D^{(*)} \ell \nu$ decays, by adopting the model-independent method introduced for lattice calculations in Ref.\,\cite{Lellouch_1996} and recently reappraised in Ref.\,\cite{DiCarlo:2021dzg}, and the new non-perturbative determination of the susceptibilities presented in Ref.\,\cite{Martinelli:2021frl}. Our fundamental assumption relies on a different treatment of the lattice results and of the experimental measurements. To be more specific, our approach is to use lattice calculations alone, combined  with unitarity,  to determine the dependence of  the form factors on the momentum transfer. In other words  the FFs are known theoretically in the whole kinematical region used to determine $\vert V_{cb} \vert$. We will develop a specific treatment of experimental correlations in the $B \to D^*$ case, since we have encountered some problems with the comparison of some set of data with our results for the FFs. As discussed in the following, our understanding is that the problem arises from the correlation matrix of the data of Ref.\,\cite{Abdesselam:2017kjf}.

The main outputs of the present work are both methodological and phenomenological ones. We apply a new approach to extract $\vert V_{cb} \vert$ from exclusive experiments and to determine the ratios $R(D^{(*)})$ from the theory alone. We get values of $\vert V_{cb} \vert$ lower than the inclusive one, $\vert V_{cb} \vert_{incl} = (42.00 \pm 0.65) \cdot 10^{-3}$\,\cite{Gambino:2016jkc, Aoki:2019cca}, but still compatible with it. Indeed, our results are $\vert V_{cb} \vert = (40.4 \pm 1.8 ) \cdot 10^{-3}$ from $B \to D^*$ and $\vert V_{cb} \vert = (41.0 \pm 1.2) \cdot 10^{-3}$ from $B \to D$. The latter one is based on the final FNAL/MILC\,\cite{Bailey_2015} lattice data for the FFs, while for the former one we still make use of the preliminary JLQCD\,\cite{Kaneko:2019vkx}  lattice results for the FFs. In the case of $B \to D$ the uncertainty of our result for  $\vert V_{cb} \vert$ is comparable with those obtained in literature using experimental data to constrain the shape of the FFs (see Refs.\,\cite{Bigi_2016,Jaiswal_2017,Aoki:2019cca}), while for $B \to D^*$ it is greater, but nevertheless still remarkably good (see Refs.\,\cite{Gambino:2019sif,Jaiswal:2020wer,Aoki:2019cca}). Furthermore, our method for the description of the FFs leads to important novelties also in the prediction of the ratios $R(D^{(*)})$, that are now completely independent from the experimental measurements. We obtain $R(D) = 0.296(8)$ using final FNAL/MILC\,\cite{Bailey_2015} lattice results, and $R(D^{*}) = 0.261(20)$ using preliminary JLQCD\,\cite{Kaneko:2019vkx} and FNAL/MILC\,\cite{Aviles-Casco:2019zop} lattice data. We stress that the tension between theoretical and experimental determinations of such quantities\,\cite{HFLAV} is reduced.

The paper is organized as follows. In Section II we review the main properties of the matrix method for the description of the FFs, as described in \cite{DiCarlo:2021dzg}. In Section III we show the results of the application of this method to the FFs entering $B \to D^* \ell \nu$ decays. We also describe a new way to extract an estimate of $\vert V_{cb} \vert$ from the experimental data. New theoretical expectation values for $R(D^*)$ and the polarization observables $P_{\tau}$ and $F_L$ are also presented. In Section IV the same machinery is applied to $B \to D \ell \nu$ decays. Our conclusions can be found in Section V, where we highlight the main results of our analysis of semileptonic $B \to D^{(*)} \ell \nu$ decays and sketch possible future developments and improvement of the accuracy of the theoretical predictions. 

\section{A recap of the matrix approach to the Form Factors}

In this Section, we briefly summarize the main properties of the non-perturbative dispersive matrix (DM) method for the FFs. For more details, see the original paper \cite{Lellouch_1996} and Ref.\,\cite{DiCarlo:2021dzg}. We will also describe a \emph{sceptical} approach \cite{DAgostini:2020vsk, DAgostini:2020pim} for the treatment of the LQCD data, which is particularly relevant in the calculation of the semileptonic $B \to D^{(*)}$ transitions. 

\subsection{The main ingredients}

Let us introduce an inner product defined as \cite{BOURRELY1981157, Lellouch_1996}
\begin{equation}
\langle g\vert h\rangle =\frac{1}{2\pi i } \oint_{\vert z\vert=1 } \frac{dz}{z}   \bar {g}(z) h(z)\, ,  \label{eq:inpro}
\end{equation}
where $\bar{g}(z)$ is the  complex conjugate of the function $g(z)$. Then, the dispersion relation for a generic spin-parity quantum channel can be written as 
\begin{equation}
\label{eq:JQ2z}
\frac{1}{2\pi i } \int_{\vert z\vert =1} \frac{dz}{z} \vert\phi(z,q^2) f(z)\vert^2 \leq \chi(q^2)\, , 
\end{equation}
where $f(z)$ is the generic FF, the kinematical functions $\phi(z,q^2)$ for the different FFs entering $B \to D^{(*)}$ decays are given below and $\chi(q^2)$ is related to the derivative with respect to $q^2$ of  the Fourier transform of suitable Green functions of bilinear quark operators~\cite{Boyd:1997kz, Caprini:1997mu}. From the physical point of view, $\phi(z,q^2)$ depends on the phase space and on the spin-parity quantum numbers of the channel we are looking at. The expression (\ref{eq:JQ2z}) can be equivalently written as
\begin{equation}
\label{eq:JQinpro}
0\leq \langle\phi f\vert \phi f\rangle\leq \chi(q^2)\, .
\end{equation}
Hereafter, we will refer to $\chi(q^2)$ as \emph{susceptibility}. In this paper we fix $q^2=0$, however in principle our analysis can be repeated for whatever value of $q^2$ one has in mind.

Following refs.\,\cite{BOURRELY1981157,Lellouch_1996}, we introduce a function $g_t(z)$ as 
\begin{equation*}
g_t(z) \equiv \frac{1}{1-\bar{z}(t) z}\, , 
\end{equation*}
where $z$ is the integration variable of Eq.\,(\ref {eq:inpro}) and $\bar{z}(t)$ is the complex conjugate of the variable $z(t)$, defined as\footnote{More generally~\cite{Boyd:1997kz} the conformal variable $z$ is related to the momentum transfer $t$ by the relation $z = (\sqrt{t_+ - t} - \sqrt{t_+ - t_0}) / (\sqrt{t_+ - t} + \sqrt{t_+ - t_0})$, where $t_0 < t_+$ is an arbitrary value. In this work we adopt $t_0 = t_-$.}
\begin{equation}
\label{conf}
z(t) = \frac{\sqrt{t_+ -t} - \sqrt{t_+ - t_-}}{\sqrt{t_+ -t} + \sqrt{t_+ - t_-}},
\end{equation}
where we have defined
\begin{equation*}
t_{\pm}=(m_B \pm m_{D^{(*)}})^2 \, .
\end{equation*}
Equivalently, it can be also expressed in terms of the recoil $w$ as
\begin{equation*}
z=\frac{\sqrt{w+1}-\sqrt{2}}{\sqrt{w+1}+\sqrt{2}},
\end{equation*}
since the momentum transfer and the recoil are related through the expression
\begin{equation*}
t=m_B^2 + m_{D^{(*)}}^2 - 2 m_B m_{D^{(*)}} w.
\end{equation*}
Then 
\begin{equation}
\label{CI}
\langle g_t|\phi f \rangle  = \phi(z(t),q^2)\, f\left(z(t)\right)\, , \qquad   \langle g_{t_m} | g_{t_l} \rangle  = \frac{1}{1- \bar{z}(t_l) z(t_m)}.
\end{equation}
At this point, we introduce the matrix 
\be
\mathbf{M} \equiv   \left (
\begin{array}{ccccc}
\langle\phi f | \phi f \rangle  & \langle\phi f | g_t \rangle  & \langle\phi f | g_{t_1} \rangle  &\cdots & \langle\phi f | g_{t_N}\rangle  \\
\langle g_t | \phi f \rangle  & \langle g_t |  g_t \rangle  & \langle  g_t | g_{t_1} \rangle  &\cdots & \langle g_t | g_{t_N}\rangle  \\
\langle g_{t_1} | \phi f \rangle  & \langle g_{t_1} | g_t \rangle  & \langle g_{t_1} | g_{t_1} \rangle  &\cdots & \langle g_{t_1} | g_{t_N}\rangle  \\
\vdots & \vdots & \vdots & \vdots & \vdots \\ 
\langle g_{t_N} | \phi f \rangle  & \langle g_{t_N} | g_t \rangle  & \langle g_{t_N} | g_{t_1} \rangle  &\cdots & \langle g_{t_N} | g_{t_N} \rangle  \\
\end{array}\right )  ~ . ~
\label{eq:Delta}
\ee
Since the variable $z$ can assume only real values in the allowed kinematical region, $\mathbf{M}$ can be expressed in a simpler way through the Eqs.\,(\ref{eq:JQinpro}) and (\ref{CI}) as
\be
\mathbf{M} = \left( 
\begin{tabular}{cccccc}
   $\chi$ & $\phi f$                            & $\phi_1 f_1$                             & $\phi_2 f_2$                           & $...$ & $\phi_N f_N$ \\[2mm] 
   $\phi f$     & $\frac{1}{1 - z^2}$     & $\frac{1}{1 - z z_1}$      & $\frac{1}{1 - z z_2}$     & $...$ & $\frac{1}{1 - z z_N}$ \\[2mm]
   $\phi_1 f_1$ & $\frac{1}{1 - z_1 z}$  & $\frac{1}{1 - z_1^2}$     & $\frac{1}{1 - z_1 z_2}$ & $...$ & $\frac{1}{1 - z_1 z_N}$ \\[2mm]
   $\phi_2 f_2$ & $\frac{1}{1 - z_2 z}$  & $\frac{1}{1 - z_2 z_1}$  & $\frac{1}{1 - z_2^2}$    & $...$ & $\frac{1}{1 - z_2 z_N}$ \\[2mm]
   $... $  & $...$                           & $...$                              & $...$                              & $...$ & $...$ \\[2mm]
   $\phi_N f_N$ & $\frac{1}{1 - z_N z}$ & $\frac{1}{1 - z_N z_1}$ & $\frac{1}{1 - z_N z_2}$ & $...$ & $\frac{1}{1 - z_N^2}$
\end{tabular}
\right) ~ . ~
\label{eq:Delta2}
\ee
In this expression, $\phi_i f_i \equiv \phi(z_i) f(z_i)$ (with $i = 1, 2, ... N$) represent the known values of the quantity $\phi(z) f(z)$ corresponding to the values $z_i$ of the kinematical variable $z$. In order to use a compact notation let us indicate $z$ and the corresponding unknown values of $\phi f$ as $z_0$ and $\phi_0 f_0 \equiv \phi(z_0) f(z_0)$, respectively, so that the index $i$ now runs from $0$ to $N$. 

The positivity of the determinant of this matrix allows to compute the lower and the upper bounds for the FF of interest. We rephrase the condition $\det \mathbf{M} \geq 0$ into an inequality that interests the quantities in the r.h.s. of the Eq.\,(\ref{eq:Delta2}). For the details of the computation, see the Appendix A of \cite{DiCarlo:2021dzg}. One finds that
\be
  \beta - \sqrt{\gamma} \leq f_0 \leq \beta + \sqrt{\gamma} ~ , ~
    \label{eq:bounds}
\ee 
where, by introducing the quantities 
\bea
   \label{eq:d0}
    d_0 & = & \prod_{m = 1}^N \frac{1 - z_0 z_m}{z_0 - z_m}  ~ , ~ \\[2mm]
    \label{eq:di}
    d_j & = & \prod_{m \neq j = 1}^N \frac{1 - z_j z_m}{z_j - z_m}  ~   , 
\eea
we have that
\bea
      \label{eq:beta_final}
      \beta & = & \frac{1}{\phi_0 d_0} \sum_{j = 1}^N \phi_j f_j d_j \frac{1 - z_j^2}{z_0 - z_j} ~ , ~ \\[2mm]
      \label{eq:gamma_final}
      \gamma & = &  \frac{1}{1 - z_0^2} \frac{1}{\phi_0^2 d_0^2} \left( \chi - \chi_0 \right) ~ , ~ \\[2mm]
      \label{eq:chi0_final}
      \chi_0 & = & \sum_{i, j = 1}^N \phi_i f_i \phi_j  f_j d_i d_j \frac{(1 - z_i^2) (1 - z_j^2)}{1 - z_i z_j} ~ . ~
\eea
Unitarity is satisfied only when $\gamma \geq 0$, which implies $\chi \geq \chi_0$.
Since $\chi_0$ does not depend on $z_0$, the above condition is either never verified or always verified for any value of $z_0$.

We remind an important feature of the DM approach (see Ref.~\cite{DiCarlo:2021dzg}).
When $z_0$ coincides with one of the data points, i.e.~$z_0 \to z_j$, one has $\beta \to f_j$ and $\gamma \to 0$.
In other words the DM method reproduces exactly the given set of data points.
This is at variance with what may happen using the (truncated) BGL or the CLN parametrisations, since there is no guarantee that such parametrizations can reproduce exactly the set of input data.
Thus, it is worthwhile to stress the following important feature of the DM approach: the DM band given by Eqs.~(\ref{eq:bounds}), (\ref{eq:beta_final}) and (\ref{eq:gamma_final}) is equivalent to the results of all possible BGL fits which satisfy unitarity and at the same time reproduce exactly the input data.

\subsection{Implementation of the kinematical constraints}

Some of the FFs entering semileptonic $B \to D^{(*)}$ decays are related to each other. These relations are called kinematical constraints (KCs) and add an important piece of information to be included in the DM method. In fact, they usually relate the relevant FFs at zero momentum transfer, namely in the region not accessible by LQCD computations.

In what follows, we focus on the production of a pseudoscalar meson, in which case the two FFs $f_{+,0}(t)$ are constrained by the relation 
\begin{equation*}
f_0(0) = f_+(0).
\end{equation*}
Following the Reference \cite{Lellouch_1996}, we define
\begin{eqnarray*}
f_{lo}^*(0)&=&\max[f_{+,lo}(0),f_{0,lo}(0)],\\
f_{up}^*(0)&=&\min[f_{+,up}(0),f_{0,up}(0)],
\end{eqnarray*}
so that
\begin{equation}
\label{rangeFFKC}
f_{lo}^*(0) \leq f(0) \leq f_{up}^*(0),
\end{equation}
where $f(0) \equiv f_0(0) = f_+(0)$.
We now consider the FFs at zero momentum transfer to be uniformly distributed in the range given by Eq.\,(\ref{rangeFFKC}) and we take it as a new input at $t_{N+1} = 0$. For each of the two FFs, we then consider a modified matrix, $\mathbf{M}_C$, that has one more row and one more column with respect to $\mathbf{M}$ in Eq.\,(\ref{eq:Delta}) and contains the common value $f(t_{N+1} = 0)$. To be more specific, $\mathbf{M}_C$ has the form 
\be
\mathbf{M}_C = \left( 
\begin{tabular}{ccccccc}
   $\chi$ & $\phi_0 f_0$                            & $\phi_1 f_1$                             & $\phi_2 f_2$                           & $...$ & $\phi_N f_N$ & $\phi_{N+1} f_{N+1}$ \\[2mm] 
   $\phi_0 f_0$     & $\frac{1}{1 - z_0^2}$     & $\frac{1}{1 - z_0 z_1}$      & $\frac{1}{1 - z_0 z_2}$     & $...$ & $\frac{1}{1 - z_0 z_N}$ & $\frac{1}{1 - z_0 z_{N+1}}$ \\[2mm]
   $\phi_1 f_1$ & $\frac{1}{1 - z_1 z_0}$  & $\frac{1}{1 - z_1^2}$     & $\frac{1}{1 - z_1 z_2}$ & $...$ & $\frac{1}{1 - z_1 z_N}$ & $\frac{1}{1 - z_1 z_{N+1}}$ \\[2mm]
   $\phi_2 f_2$ & $\frac{1}{1 - z_2 z_0}$  & $\frac{1}{1 - z_2 z_1}$  & $\frac{1}{1 - z_2^2}$    & $...$ & $\frac{1}{1 - z_2 z_N}$ & $\frac{1}{1 - z_2 z_{N+1}}$ \\[2mm]
   $... $  & $...$                           & $...$                              & $...$                              & $...$ & $...$ \\[2mm]
   $\phi_N f_N$ & $\frac{1}{1 - z_N z_0}$ & $\frac{1}{1 - z_N z_1}$ & $\frac{1}{1 - z_N z_2}$ & $...$ & $\frac{1}{1 - z_N^2}$ & $\frac{1}{1 - z_N z_{N+1}}$ \\[2mm]
$\phi_{N+1} f_{N+1}$ & $\frac{1}{1 - z_{N+1} z_0}$ & $\frac{1}{1 - z_{N+1} z_1}$ & $\frac{1}{1 - z_{N+1} z_2}$ & $...$ & $\frac{1}{1 - z_{N+1} z_N}$ & $\frac{1}{1 - z_{N+1}^2}$
\end{tabular}
\right) ~ . ~
\label{eq:Delta2KC}
\ee
For any point $t$ at which we want to predict the dispersive bands of $f_{+,0}(t)$, we compute the matrix $\mathbf{M}_C$ and using Eq.(\ref{eq:bounds}) we get the corresponding lower and upper bounds. Note that the extension of the above procedure to the $B \to D^*$ case is straightforward. Furthermore, for a general treatment of the statistical and the systematic errors of LQCD computations of the FFs, see the details in Section V of \cite{DiCarlo:2021dzg}.
 
\subsection{The sceptical approach to the DM method}

The machinery described in Sections IIA-IIB allows us to compute the lower/upper bounds of the FFs once we have chosen our set of input data, i.e. the susceptibility and the LQCD computations of the same FFs.  In order to propagate the uncertainties related to these quantities to the evaluation of the  FFs, we propose the following method.  First we build up a multivariate Gaussian distribution whose mean value and covariance matrix are $\mu = \{ f(t_1),\cdots,f(t_n)\}$ and $ \Sigma_{ij} = \rho_{ij} \sigma_i \sigma_j$,  where the  (average) values of the  $f(t_i)$ are the form factors extracted on the lattice, the $\sigma_i$s the corresponding uncertainties, and the $\rho_{ij}$ their correlation matrix. Thus, we generate bootstrap events according to this probability distribution. At the same time we will also generate the same number of values of the susceptibilities through normal distributions defined by their mean values and standard deviations. For each of the bootstrap events, we verify if unitarity is satisfied. If this is not the case, then the event is eliminated from the sample. From the physical point of view, this step can be read as a consistency check between all the input data, namely the susceptibilities and the FFs on that particular bootstrap. 

For what concerns semileptonic $B \to D^{(*)}$ decays, a problem may occur in the application of the DM method to the FFs. This happens when only a small percentage of the generated bootstraps verify both the unitarity filter and the kinematical constraints. We can then ask ourselves whether the final bands of the FFs obtained with our method can be considered reliable or not. In these cases the unitarity constraint has a crucial impact on the covariance matrix of the input data. This may be due to the fact that the output of a lattice calculation with its uncertainties and correlations does not contain necessarily all the effects of unitarity and/or to the possibility that systematic effects (eventually lattice artefacts), that have not been properly corrected for, come into play and may jeopardise the unitary relations.

In order to recover a sufficiently large percentage of bootstraps passing the unitarity (and/or kinematical) constraint, we introduce the \emph{sceptical} approach\,\cite{DAgostini:2020vsk, DAgostini:2020pim} to the analysis of the lattice data for the FFs. The idea is to modify the standard deviations $\sigma_i$ of the LQCD points, by assuming {\it new} values $\sigma_i^t$ which are related to the original ones by a factor $r_i$, one for each of the measured points, so that $\sigma_i^t = r_i \sigma_i$, whereas the average values of the LQCD computations are kept the same. In this way we generate a larger set of bootstrap events, among which we search for those bootstraps passing the unitarity (and/or kinematical) constraint.  
 In the present work we started from the simplest choice of a unique $r$ for all the LQCD values of the FFs. A posteriori such a choice turned out to be very successful in recovering a sufficiently large percentage of bootstrap events passing the given unitarity and/or kinematical filters\footnote{When the data points of different FFs obey independent unitarity constraints and they are not connected by kinematical constraints, one can easily use different values of r for different FFs, as later in Section~\ref{sceptical} it will be the case for the FF $g$ with respect to the FFs $f$, $\mathcal{F}_1$ and $P_1$.   We have checked that in this work, when different FFs obey kinematical constraints, the use of different values of r for different FFs does not lead to any significant improvement of our procedure and it does not change our final results.}. This positive-definite variable $r$ has a Gamma probability distribution, $i.e.$
\begin{equation*}
P(r) \propto e^{-r/\beta} r^{\alpha-1}.
\end{equation*}
The parameters $\alpha$ and $\beta$ are fixed by imposing that this distribution has a unitary mean value and a unitary variance. A simple calculation shows that this request corresponds to the choice $\alpha=\beta=1$. Then we build up a multivariate Gaussian distribution, whose covariance matrix now is
\begin{equation*}
 \Sigma_{ij} = \rho_{ij} \sigma_{i} \sigma_{j} \times r^2,
\end{equation*}
where $\sigma_{i}$ ($i=1,\ldots,N$) are the $N$ LQCD points uncertainties and $\rho_{ij}$ is the correlation matrix. We adopt a similar prescription for the susceptibilities, namely we modify their uncertainties as
\begin{equation*}
\hat{\sigma}_{\chi} = \sigma_{\chi} \times r.
\end{equation*}
Hence, we extract $N_r$ values of $r$ and, for each of them, $N_b$ bootstrap events for both the FFs values and the susceptibilities. 

To decide whether a single bootstrap event is accepted or rejected, let us fix the bootstrap event, $i.e.$ the $i$-th event $N_{i}$, and we consider $N_i^r$ values of $r$. We check the unitarity constraint for all the FFs for each of the $N_i^r$ events, then, we compute the lower and the upper bounds for the survived $\hat{N}_r^i \leq N_r^i$ events and check whether the KCs are verified or not. This second step will leave us with $\tilde{N}_r^i \leq \hat{N}_r^i$ bootstraps. Our prescription is thus the following: \emph{the event $N_{i}$ is considered as accepted if $\tilde{N}_r^i \neq 0$, namely if there exists at least one value of $r$ which passes both the unitarity and the KC filters}. Adopting this ansatz, we see that a much larger fraction of the generated events is accepted. For example, in the $B \to D$ case we pass from a $\sim 15\%$ of accepted bootstraps without the sceptical approach to $\sim 100\%$ with the sceptical approach.

At this point, we combine the values of the accepted $r$ in a unique value. We proceed in three steps. For each bootstrap we compute the mean value of $r$ over the $\tilde{N}_i^r$ extractions. Then, we find the $r$ (among the $\tilde{N}_i^r$ extracted) closest to that mean value. Finally, we save the event corresponding to that $r$ as representative of the bootstrap that we have fixed. Note that in this way we are guaranteed that the new bootstrap events will pass both the unitarity and the KC filters. 

\section{Semileptonic $B \to D^*$ decays}

%%%%%%%%%%%%%%% BD* %%%%%%%%%%%%%%%%%%%

Let us now apply the non-perturbative DM method to semileptonic $B \to D^* \ell \nu$ decays. We first describe how to characterize the differential decay width of $B \to D^* \ell \nu$ decays through the FFs. Then we apply the DM method to describe their behaviours as functions of $q^2$. At present, the inputs for our matrices are the preliminary unblinded JLQCD \cite{Kaneko:2019vkx} and blinded FNAL/MILC \cite{Aviles-Casco:2019zop} lattice data, available also at non-zero recoil. For what concerns the susceptibilities, we will use the results of our non-perturbative computation on the lattice \cite{Martinelli:2021frl}. We present new theoretical estimates of $\vert V_{cb} \vert$ and of the ratio $R(D^*)$. We also compute new predictions for two polarization observables, $i.e.$ the $\tau$-polarization $P_{\tau}$ and the $D^*$ longitudinal polarization $F_L$. 

\subsection{Theoretical expression of the differential decay width}

In the $B \to D^*$ case, the vector current $V^{\mu}\equiv \bar{b}\gamma^{\mu}c$ and the axial current $A^{\mu}\equiv \bar{b}\gamma^{\mu}\gamma^5c$ give the following contributions to the amplitude
\begin{eqnarray} 
\label{eq:matrix_el_Dstar}
\langle D^{*} (p,\epsilon)| \bar{c} \gamma^{\mu} \left(1 \mp \gamma_{5}\right) b |\bar{B}(p_{B})\rangle 
 & = & - \frac{2 }{m_{B}+m_{D^{*}}}  \varepsilon^{\mu}_{\alpha \beta \gamma} \epsilon^{*\alpha}p^{\beta} q^{\gamma} V(q^{2}) \nonumber \\
& \, & \pm \, i\, \frac{2 m_{D^{*}}}{q^2} (\epsilon^{*} \cdot q ) q^{\mu}  A_{0}(q^{2}) \\
& \, & \mp \frac{i}{m_{B}-m_{D^{*}}} \left[(m_{B}-m_{D^{*}}) \epsilon^{*\mu}-(\epsilon^{*} \cdot q) (p+p_{B})^{\mu}\right] A_{1} (q^{2}) \nonumber \\
& \, & \mp \, i \, \frac{ 2 m_{D^{*}}}{q^{2}} (\epsilon^{*} \cdot q) \left[\frac{q^{2}}{m_{B}^{2}-m_{D^{*}}^{2}} (p+p_{B})^{\mu}-q^{\mu}\right] A_{3}(q^{2}) \ ,
\nonumber
\end{eqnarray}
where we can also re-express $A_3(q^2)$ as
\begin{equation}
\label{A1A2A3}
2\,m_{D^{*}}A_{3}(q^{2}) = [(m_{B}+m_{D^{*}}) A_{1}(q^{2})-(m_{B}-m_{D^{*}})A_{2}(q^{2})].
\end{equation}
As we want to use a BGL-like nomenclature \cite{Boyd:1995sq, Boyd:1995cf, Boyd:1997kz}, we express the FFs in Eqs.(\ref{eq:matrix_el_Dstar})-(\ref{A1A2A3}) as 
\begin{eqnarray} 
\label{eq:ff_BtoDstar}
V(w) & = & \frac{m_{B}+m_{D^{*}}}{2} g(w)  \ , \\
\label{eq:ff_BtoDstar2}
A_{1}(w) & = &  \frac{f(w)}{m_{B} + m_{D^{*}}}  \ ,\\ 
\label{eq:ff_BtoDstar3}
A_{2}(w) & = & \frac{1}{2} \frac{m_{B}+m_{D^{*}}}{(w^{2}-1)m_{B} m_{D^{*}}}\left[  
\left(w-\frac{m_{D^{*}}}{m_{B}}\right) f(w) - \frac{\mathcal{F}_{1}(w)}{m_{B}} \right] \ ,\\  
\label{eq:ff_BtoDstar4}
A_{0}(w) & = & \frac{1}{2} \frac{m_{B}+m_{D^{*}}}{\sqrt{m_{B} m_{D^{*}}}} P_1(w) \ . 
\end{eqnarray}
There is a precise relation between the BGL-like FFs and the CLN ones (described in the Appendix A of \cite{Caprini:1997mu}), namely
\begin{eqnarray*}
f(w) &=& \sqrt{m_B m_{D^*}} (1+w) h_{A_1}(w),\\
g(w) &=& h_V(w) / \sqrt{m_B m_{D^*}},\\
\mathcal{F}_1(w) &=& m_B^2 (1+w) \sqrt{r} \left[(w-r) h_{A_1}(w) - (w-1) ( rh_{A_2}(w+h_{A_3}(w)) \right],\\
P_1(w) &=& \left[(w+1) h_{A_1}(w) - (1-wr) h_{A_2}(w) - (w-r)  h_{A_3}(w) \right]/(1+r),
\end{eqnarray*}
where $r \equiv m_{D^*}/m_B$. These relations are necessary for our analysis since the most recent lattice computations \cite{Aviles-Casco:2019zop, Kaneko:2019vkx} give the values of the FFs $h_V, h_{A_1}, h_{A_2}, h_{A_3}$ at non-zero recoil.

The FFs are characterized by the following \emph{kinematical constraints}. The first one applies at zero recoil, where we have at our disposal the results of the LQCD computations
\begin{equation}
\label{KC1}
\mathcal{F}_1(1) = (m_B-m_{D^*})f(1).
\end{equation}
Instead, the second one applies in the opposite regime, namely at maximum recoil
\begin{equation}
\label{KC2}
P_1 (w_{max}) = \frac{\mathcal{F}_1(w_{max})}{(1+w_{max})(m_B-m_{D^*}) \sqrt{m_B m_{D^*}}},
\end{equation}
where
\begin{equation*}
w_{max} = \frac{m_B^2 + m_{D^*}^2}{2 m_B m_{D^*}},
\end{equation*}
under the assumption that the mass of the final state lepton is negligible.

\begin{figure}[htb!]
\centering{\includegraphics[width=0.90\textwidth]{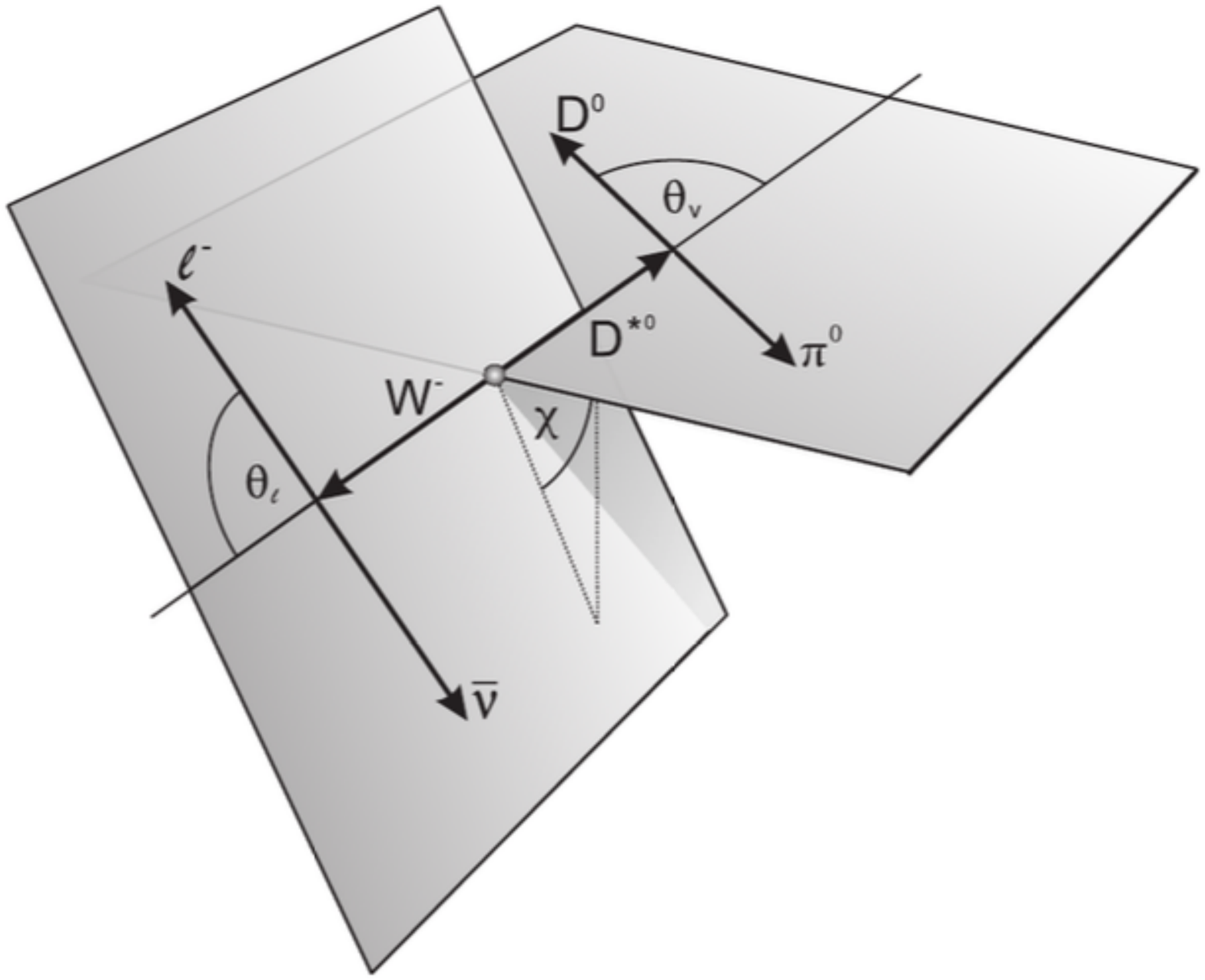}}
\vspace{-1cm}
\caption{\textit{Representation of the semileptonic $B \rightarrow D^{*} \ell \nu$ decay. $\theta_{\ell}$ is the angle between the lepton and the direction opposite the $B$ meson in the virtual $W$ boson rest frame. $\theta_v$ is the angle between the $D$ meson and the direction opposite the $B$ meson in the $D^{*}$ rest frame. $\chi$ is the angle between the two decay planes spanned by the $W - \ell$ and $D^{*} - D$ systems in the $B$ meson rest frame.}
\hspace*{\fill} \small}
\label{bellefig}
\end{figure}

To conclude this Section, from the matrix element (\ref{eq:matrix_el_Dstar}) we are able to compute the differential decay width
\begin{equation}
\begin{aligned}
\label{finaldiff333BDst}
&\frac{d\Gamma(B \rightarrow D^{*}(\rightarrow D\pi) \ell \nu)}{dw d\cos \theta_{\ell} d\cos \theta_v d\chi} = \frac{G_F^2 \vert V_{cb} \vert^ 2 \eta_{EW}^2}{4(4\pi)^4} 3 m_B m_{D^*}^2 \sqrt{w^2-1} \left( 1 - 2 r w + r^2 \right)\\
&\hskip 3.98truecm  \cdot B(D^{*} \rightarrow D\pi) \Big\{ (1-\cos \theta_{\ell} )^2 \sin^2 \theta_v \vert H_{+} \vert^2\\
&\hskip 3.98truecm + (1+\cos \theta_{\ell} )^2\sin^2 \theta_v \vert H_{-} \vert^2+ 4 \sin^2 \theta_{\ell}\cos^2 \theta_v\vert H_{0} \vert^2\\
&\hskip 3.98truecm - 2 \sin^2 \theta_{\ell}\sin^2 \theta_v \cos 2\chi  H_{+}  H_{-} \\
&\hskip 3.98truecm - 4 \sin \theta_{\ell} (1-\cos \theta_{\ell} ) \sin\theta_v\cos\theta_v\cos\chi H_{+}  H_{0}\\
&\hskip 3.98truecm +   4 \sin \theta_{\ell} (1+\cos \theta_{\ell} ) \sin\theta_v\cos\theta_v\cos\chi H_{-}  H_{0} \Big\},
\end{aligned}
\end{equation}
where we have neglected the mass of the lepton and introduced the helicity amplitudes 
\begin{equation}
\label{helampl}
H_0(w) = \frac{\mathcal{F}_1(w)}{\sqrt{m_B^2+m_D^2-2m_Bm_Dw}},\,\,\,\,\,H_{\pm}(w) = f(w) \mp m_B m_{D^*} \sqrt{w^2-1}\,g(w).
\end{equation}
The various helicity angles $\theta_l, \theta_v, \chi$ are defined in Fig.\,\ref{bellefig}. In conclusion, we can obtain the final forms of the four differential decay widths $d\Gamma/dx$ (where $x=w,  \cos \theta_l, \cos \theta_v, \chi$) simply by integrating on three of them in the expression (\ref{finaldiff333BDst}).

\subsection{Review of the available lattice data}

In this work, we use two preliminary sets of lattice results, the unblinded one by the JLQCD Collaboration~\cite{Kaneko:2019vkx} and the blinded one by the FNAL/MILC Collaboration~\cite{Aviles-Casco:2019zop}. We have extracted three data points for each of the FFs $h_V, h_{A_1}, h_{A_2}, h_{A_3}$ off the plots of Refs.~\cite{Aviles-Casco:2019zop,Kaneko:2019vkx}. The number of LQCD inputs has been chosen in analogy with the $B \to D$ case, where, as we will explain later, the available lattice computations of the FFs give three values for each FF. The recoils at which the LQCD computations are considered are $\{w_1,w_2,w_3\}=\{1.00,1.07,1.14\}$ for the FNAL/MILC case and $\{w_1,w_2,w_3\}=\{1.00,1.06,1.12\}$ for the JLQCD one, as shown in Tables \ref{tab:LQCDMILCJLQCD} and \ref{tab:LQCDMILCJLQCD2}, respectively.

\begin{table}[htb!]
\renewcommand{\arraystretch}{1.1}
\begin{center}
{\small
\begin{tabular}{ c|c|c|c|c}
%& FNAL/MILC results\\
$w$ & $h_V$ & $h_{A_1}$ & $h_{A_2}$ & $h_{A_3}$\\ 
\hline
1.00 & 1.274(37) &  0.936(7)  &  -0.562(69) & 1.241(57) \\
1.07& 1.129(37)  &  0.850(10) & -0.508(66) & 1.130(57) \\
1.14& 1.002(41)   & 0.772(16) & -0.434(75) & 1.037(68) \\
\end{tabular}
}
\caption{\textit{Values of the preliminary blinded FNAL/MILC computations of the FFs $h_{V, A_1, A_2, A_3}(w)$ extracted from the plots of Ref.~\cite{Aviles-Casco:2019zop}.}
\hspace*{\fill} \small}\label{tab:LQCDMILCJLQCD}
\end{center}
\end{table}

\begin{table}[htb!]
\renewcommand{\arraystretch}{1.1}
\begin{center}
{\small
\begin{tabular}{ c|c|c|c|c}
%& JLQCD results\\
$w$ & $h_V$ & $h_{A_1}$ & $h_{A_2}$ & $h_{A_3}$\\ 
\hline
1.00& 1.216(33) & 0.882(11)    & -0.176(142)  & 0.949(141) \\
1.06& 1.118(33)  &0.824(11)     &-0.215(138)  & 0.808(137) \\
1.12& 1.009(33)  &0.770(11)     &-0.207(146)  & 0.742(141) \\
\end{tabular}
}
\caption{\textit{Values of the preliminary unblinded JLQCD computations of the FFs $h_{V, A_1, A_2, A_3}(w)$ extracted from the plots of Ref.~\cite{Kaneko:2019vkx}.}
\hspace*{\fill} \small}
\label{tab:LQCDMILCJLQCD2}
\end{center}
\end{table}

Unfortunately the proceedings only contain preliminary results without reference to the correlations between different data. For this reason, we adopt a the  following reasonable assumption: we consider a high correlation between the values of the same FF computed at the three different recoils, while we will assume zero correlation between the values of the different FFs. See Appendix B for the complete form of the correlation matrix. This structure has been inspired by the correlation matrix presented by FNAL/MILC for semileptonic $B \to D$ decays in \cite{Bailey_2015}  (reported in Table \ref{tab:LQCDMILC}), the one by HPQCD for $B \to D$ in \cite{Na_2015} and the preliminary JLQCD one for semileptonic $B \to D^*$ decays in \cite{Ferlewicz:2020lxm} (where only the subset of FFs $h_V,h_{A_1}$ was considered). From these cases, it is evident that, while the diagonal block elements are always $\gtrsim 0.8$, the off-diagonal block ones are subject to large fluctuations since they relate different FFs. Obviously, once further results will be available, the following study will be repeated with the true correlations among the LQCD data. We stress that we have developed our study also assuming that the off-diagonal block elements are equal to 0.5 and that this different assumption does not change the results that we will describe in what follows. 

\subsection{Description of the FFs with the DM method in the $B \to D^*$ case}

In this Subsection, we give the ingredients necessary to implement the matrix description of the FFs. The kinematical functions to be used for each FF matrix in Eq.\,(\ref{eq:Delta}) read

\begin{eqnarray}
\label{eq:fac_outer}
\phi_{f}(z) & = & 4 \, \frac{r}{m_B^2} \sqrt{\frac{2}{3 \pi }} \,
                           \frac{(1+z)(1-z)^{3/2}}{\left[(1+r)(1-z)+2\sqrt{r}(1+z)\right]^4} \ , \nonumber \\ 
\phi_{g}(z) & = & 16 \, r^2 \sqrt{\frac{2}{3 \pi}} \,
                           \frac{(1+z)^{2}}{\sqrt{1-z}\left[(1+r)(1-z)+2\sqrt{r}(1+z)\right]^4}\ , \nonumber \\
\phi_{\mathcal{F}_1}(z) & = & 4 \, \frac{r}{m_B^3} \sqrt{\frac{1}{3 \pi}} \,
                                               \frac{(1+z)(1-z)^{5/2}}{\left[(1+r)(1-z)+2\sqrt{r}(1+z)\right]^5}\ ,   \\
\phi_{P_1}(z) & = &  16 \, (1 + r) r^{3/2} \sqrt{\frac{1}{\pi }} \,
                                 \frac{(1+z)^{2}}{\sqrt{1-z}\left[(1+r)(1-z)+2\sqrt{r}(1+z)\right]^4}\,  \nonumber 
\end{eqnarray}
with $r \equiv m_D^* / m_B$.
As usual, if the assumption of analyticity does not hold, $i.e.$ when each FF has for instance $N$ poles at $t=t_{P1},t_{P2},\cdots...,t_{PN}$, it is sufficient to modify each kinematical function $\phi$ with the transformation \cite{Lellouch_1996}
\begin{equation}
\label{chinesePOLES}
\phi(z) \to \phi_p(z) \equiv \phi(z) \times \frac{z-z(t_{P1})}{1-\bar{z}(t_{P1})z}  \times \cdots  \times  \frac{z-z(t_{PN})}{1-\bar{z}(t_{PN})z}.
\end{equation}
For the masses of the poles corresponding to $B_c^{(*)}$ mesons with different quantum numbers which enter in the FF, we refer to Table III of \cite{Bigi_2017}.

\subsubsection{A specific variant to the sceptical approach for the $B \to D^*$ case}
\label{sceptical}

For the $B \to D^*$ transition, we have implemented a variant of the sceptical approach explained in Section II B. In this case we have four FFs (\ref{eq:ff_BtoDstar})-(\ref{eq:ff_BtoDstar4}). Three of them ($f,\mathcal{F}_1,P_1$) are related to each other for two reasons. On the one hand, $f$ and $\mathcal{F}_1$ share the same spin-parity quantum number and contribute to the same susceptibility. They are also related by the first KC (\ref{KC1}). On the other hand, $\mathcal{F}_1$ and $P_1$ are linked by the second KC (\ref{KC2}). The behaviour of $g$ instead is completely unrelated to that of the other three FFs. 

\begin{figure}[htb!]
\begin{center}
\includegraphics[width=0.85\textwidth]{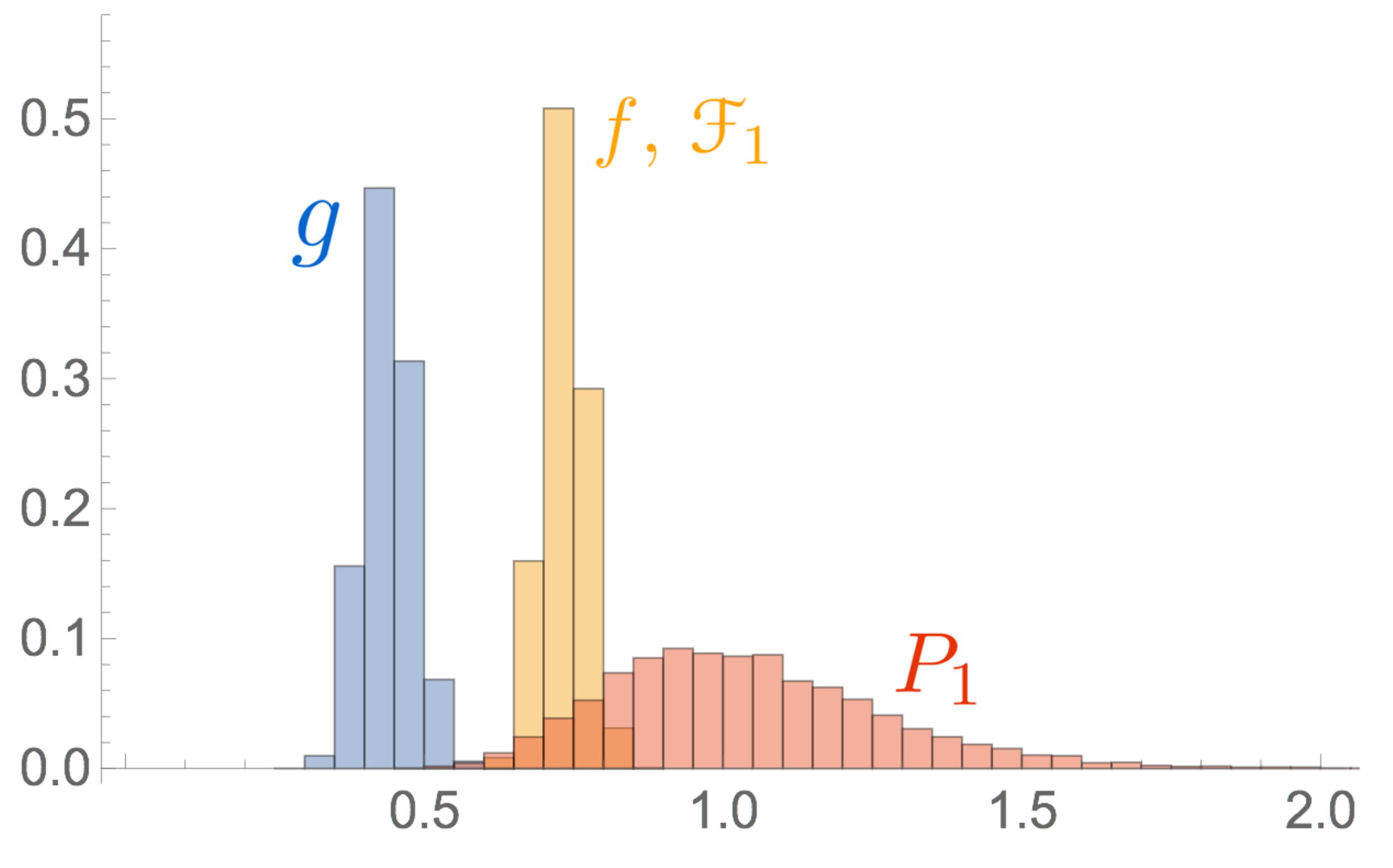}
\caption{\textit{Probability distribution of the mean values of the $r$ survived after the unitarity constraint for each bootstrap for the four FFs. The colour legend is yellow for $f$ and $\mathcal{F}_1$, blue for $g$ and red for $P_1$. The input is FNAL/MILC \cite{Aviles-Casco:2019zop}.}
\hspace*{\fill} \small}
\label{rdistr2}
\end{center}
\end{figure}

\begin{figure}[htb!]
\begin{center}
\includegraphics[width=0.85\textwidth]{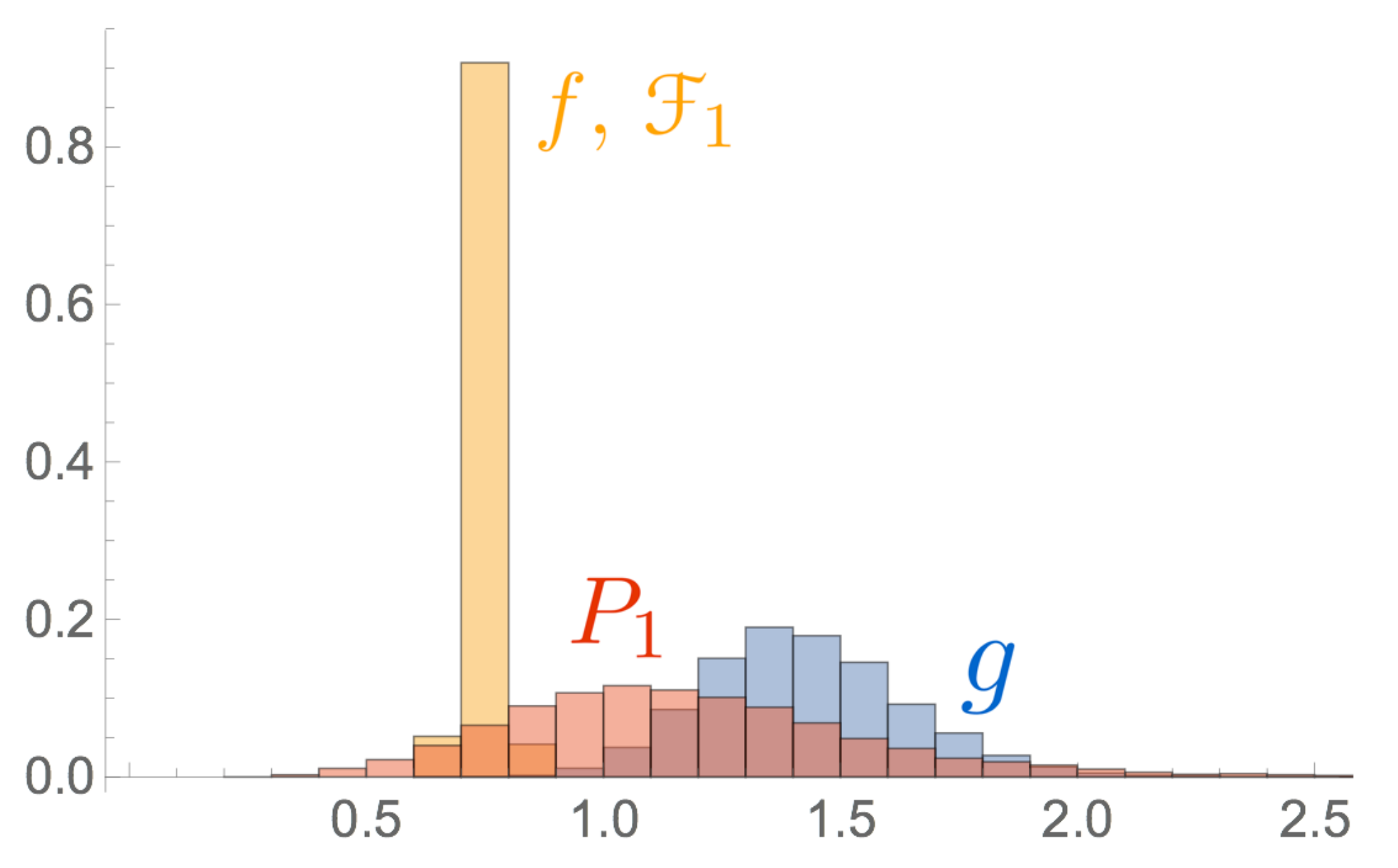}
\caption{\textit{Probability distribution of the mean values of the $r$ survived after the unitarity constraint for each bootstrap for the four FFs. The colour legend is the same of the Fig.\,\ref{rdistr2}. The input is JLQCD \cite{Kaneko:2019vkx}.}
\hspace*{\fill} \small}
\label{rdistr3}
\end{center}
\end{figure}

Figs.\,\ref{rdistr2}-\ref{rdistr3} give a graphical representation of the situation if we use the FNAL/MILC and the JLQCD inputs, respectively. We have plotted the distribution of the $r$ values that allow to pass the unitarity constraint for each FF with a \emph{unique} sceptical parameter common to all the four FFs. The (colour) legend is shown in the caption of the figures. For both cases the FF $g$ prefers values of the $r$ parameter different from those of the other three FFs. Since the unitarity constraint for the FF g is independent from those of the other FFs, we have decided to implement one $r$ variable for $f,\mathcal{F}_1,P_1$ and a \emph{different} $r_g$ parameter, specifically for $g$.

After having extracted 300 values of $r$ and after having implemented the unitarity constraints and the KC (\ref{KC2}), the survived events are around the 100\% (70\%) of the generated bootstraps for $f$, the 100\% (70\%) for $\mathcal{F}_1,P_1$ and the 100\% (100\%) for $g$, when we use the FNAL/MILC (JLQCD) input data. Our understanding is that the JLQCD inputs suffer a more severe filter by the constraints with respect to the FNAL/MILC ones.

We stress that the sceptical approach has been introduced to account for systematic effects  that have not been corrected for, which may manifest as  an  apparent violation of the unitary relations. This is particularly relevant in the study of $B\to D^{(*)}$ decays,  but also for  $B\to \pi$  or $B_s\to K$ decays,   where  discretisation effects  are expected to be rather large. The sceptical  procedure allows to filter only those bootstrap events which satisfy the unitarity bounds, without losing a  huge percentage of the generated  events.   

\subsubsection{Final bands of the FFs entering the $B \to D^*$ decay}

\begin{figure}[htb!]
\begin{center}
{\includegraphics[scale=0.55]{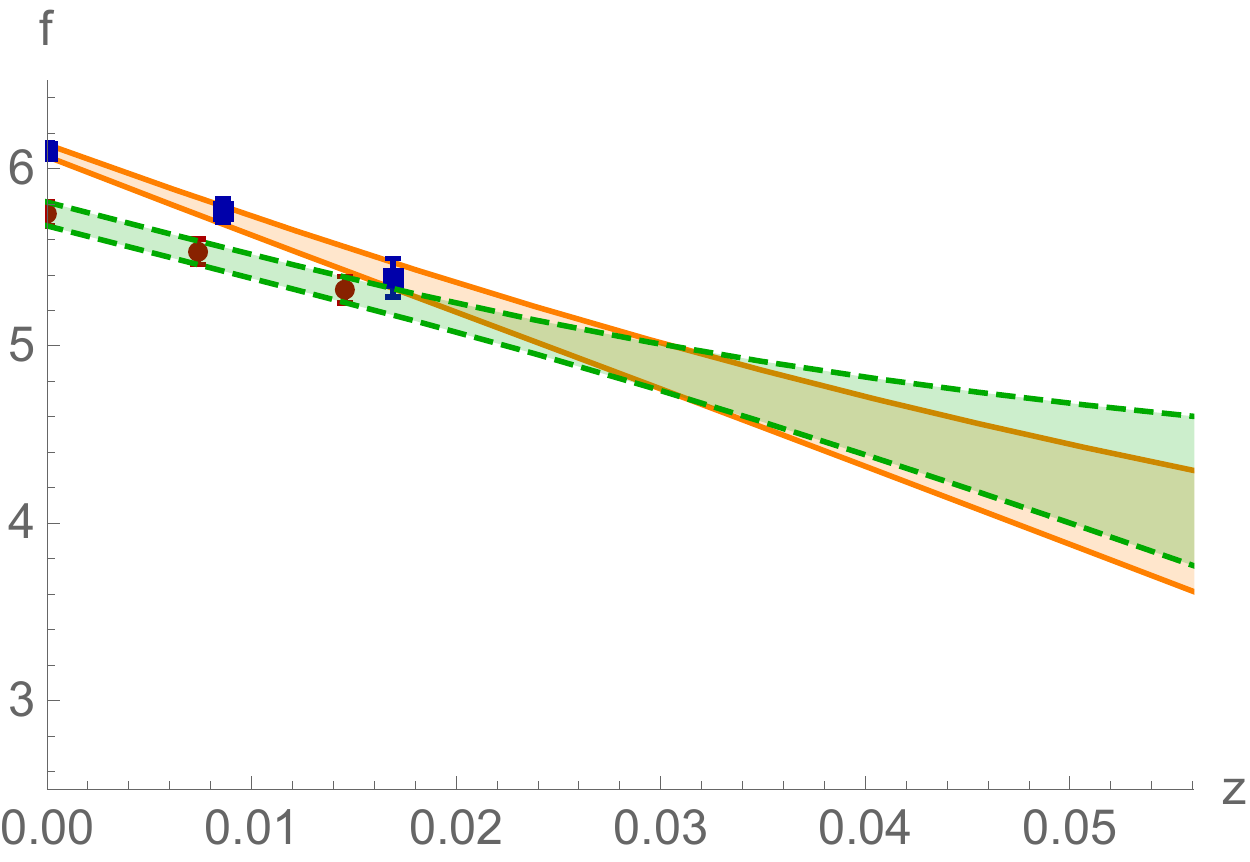}
\label{fig:FFMM_A}}
{\includegraphics[scale=0.55]{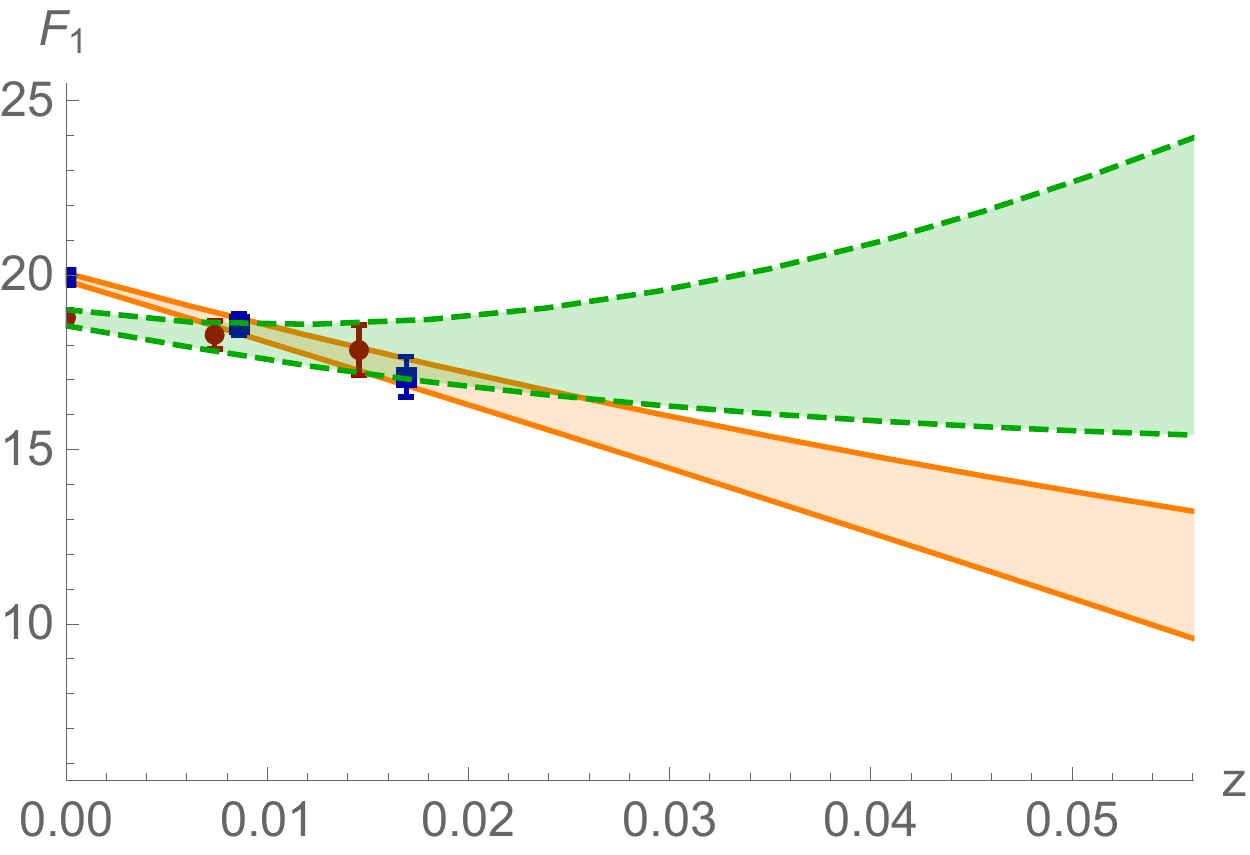}
\label{fig:FFMM_B}}\\ 
{\includegraphics[scale=0.55]{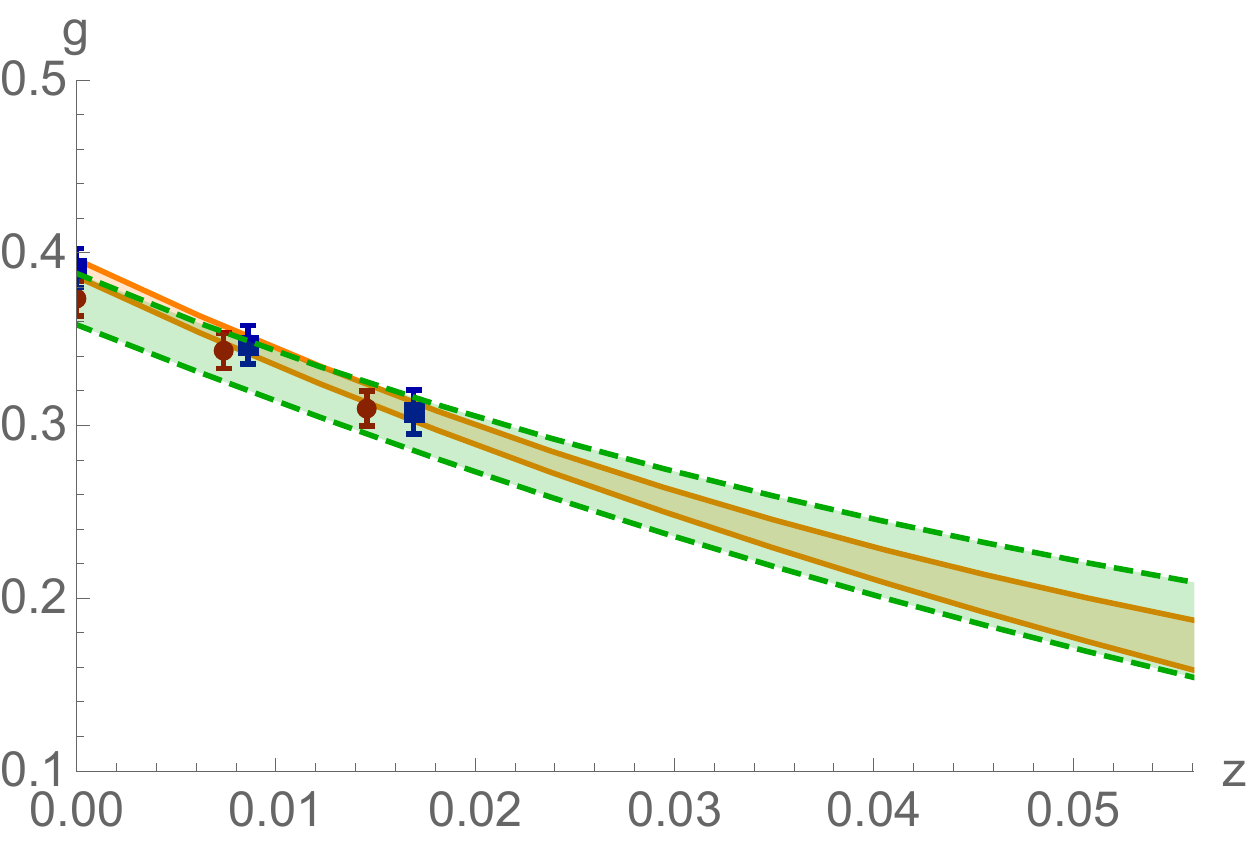}
\label{fig:FFMM_C}}
{\includegraphics[scale=0.55]{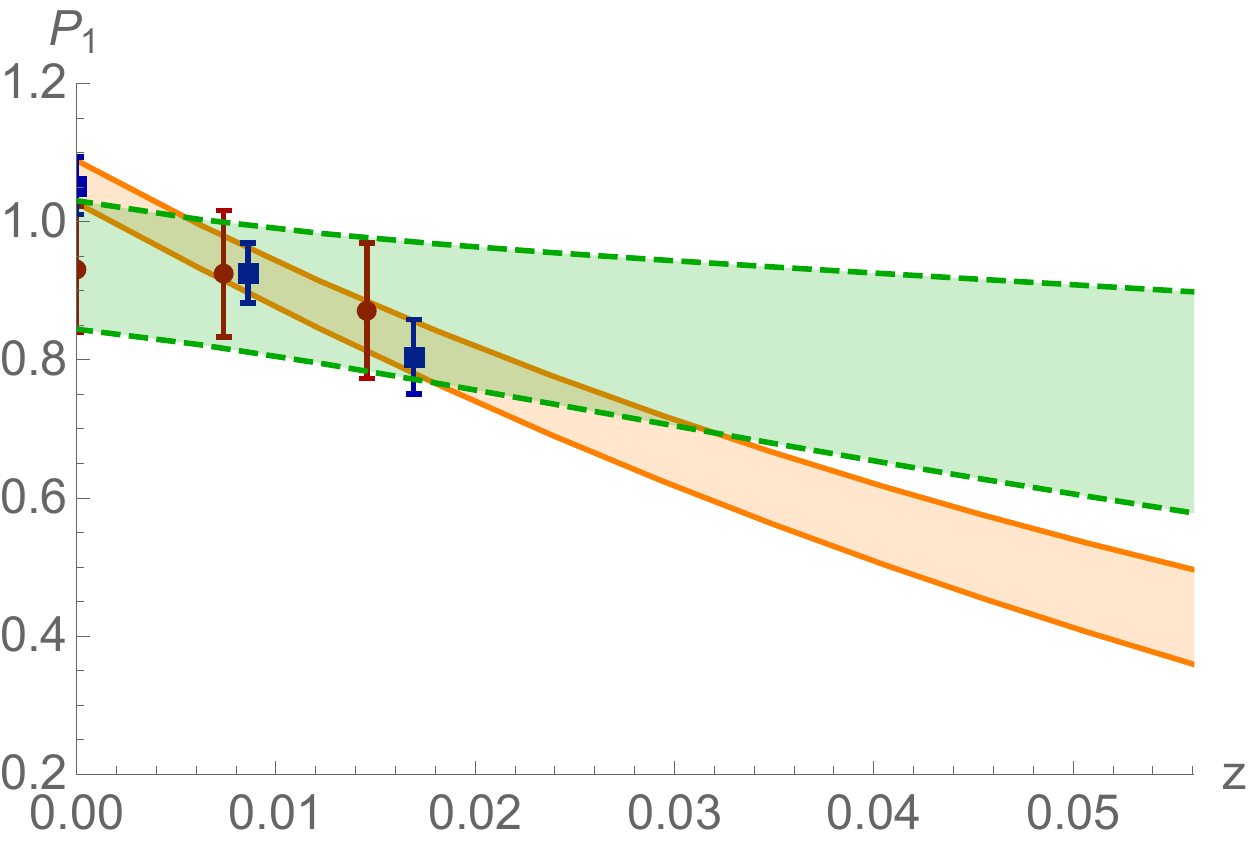}
\label{fig:FFMM_D}}
\caption{\it\small The bands of the four FFs entering $B \to D^* \ell \nu$ decays, $i.e.$ $f(z),\,g(z),\,\mathcal{F}_1(z),\,P_1(z)$, computed through the DM method. The solid orange (dashed green) bands are the results of the study by adopting the FNAL/MILC (JLQCD) results. Moreover, the blue squares (red points) are the FNAL/MILC (JLQCD) input lattice data. We used the non-perturbative values of the susceptibilities in the last column of Table\,\ref{tab:chi} in order to obtain the bands in these figures.\hspace*{\fill}}
\label{FFMM}
\end{center}
\end{figure}

We have first implemented the DM method for semileptonic $B \to D^*$ decays using the FNAL/MILC and the JLQCD inputs, given in Tables \ref{tab:LQCDMILCJLQCD}-\ref{tab:LQCDMILCJLQCD2}, separately. In Fig.\,\ref{FFMM} we compare the resulting bands of the four FFs having definite spin-parity, namely $f$, $g$, $\mathcal{F}_1$ and $P_1$. Throughout the paper it is understood that $f$, $g$ and $\mathcal{F}_1$ are given in units of GeV, GeV$^{-1}$ and GeV$^2$, respectively, while $P_1$ is dimensionless. We have also used the non-perturbative values of the susceptibilities shown in the last column on Table \ref{tab:chi}, where the perturbative estimates present in literature are also shown. The extrapolations at $z_{max}$, which are important for the phenomenological applications discussed below, read 
\begin{eqnarray*}
f(z_{max})\vert_{\rm{FNAL/MILC}}&=& 3.93 \pm 0.34,\\
g(z_{max})\vert_{\rm{FNAL/MILC}}&=& 0.173 \pm 0.014,\\
\mathcal{F}_1(z_{max})\vert_{\rm{FNAL/MILC}}&=& 11.3 \pm 1.8,\\
P_1(z_{max})\vert_{\rm{FNAL/MILC}}&=& 0.42 \pm 0.07,
\end{eqnarray*}
and 
\begin{eqnarray*}
f(z_{max})\vert_{\rm{JLQCD}}&=& 4.18 \pm 0.44,\\
g(z_{max})\vert_{\rm{JLQCD}}&=& 0.182 \pm 0.027,\\
\mathcal{F}_1(z_{max})\vert_{\rm{JLQCD}}&=& 19.6 \pm 4.6,\\
P_1(z_{max})\vert_{\rm{JLQCD}}&=& 0.74 \pm 0.17,
\end{eqnarray*}
for FNAL/MILC and JLQCD lattice data inputs, respectively. First of all, we note that in some case the two data sets are not compatible to each other (see for instance $f(z=0)$ and $\mathcal{F}_1(z=0)$). Furthermore, while the extrapolations of $f$ and $g$ at $z_{max}$ are substantially identical in the two cases, the ones of $\mathcal{F}_1$ and $P_1$ result compatible only at more than $1\sigma$ level.

\begin{table}[htb!]
\renewcommand{\arraystretch}{1.1}
\begin{center}
{\small
\begin{tabular}{ |c||c|c||c|c| }
\hline
& Perturbative & With subtraction & Non-perturbative & With subtraction\\
\hline
\hline
$\chi_{V_L} [10^{-3}]$ & $6.204$& $-$& $7.52 \pm 0.63$ & $7.58 \pm 0.59$\\
$\chi_{A_L} [10^{-3}]$ & $24.1$ & $19.4$& $25.9 \pm 1.8$& $21.9 \pm 1.9$\\
$\chi_{V_T} [10^{-4}$ GeV$^{-2}]$ & $6.486$& $5.131$ & $6.76 \pm 0.40$&  $5.84 \pm 0.44$\\
$\chi_{A_T} [10^{-4}$ GeV$^{-2}]$ & $3.89$& $-$&  $4.68 \pm 0.30$ & $4.69 \pm 0.30$\\
\hline
\end{tabular}
}
\caption{\it\small{Values of the susceptibilities adopted for the DM method. We compare the perturbative estimates \cite{Bigi_2016, Bigi_2017} with the non-perturbative ones. In the former case, the uncertainties are completely negligible. Moreover, for the perturbative values we also show the result of the subtraction of the relevant $B_c^{(*)}$ poles for the $A_L$ and the $V_T$ channels, obtained through the values of the masses and the decay constants present in Table III of \cite{Bigi_2017}. On the contrary, for the non-perturbative values we present the results ($pre-$ and $post-$subtraction of the ground-state contribution) as computed on the lattice \cite{Martinelli:2021frl}}.\hspace*{\fill}}
\label{tab:chi}
\end{center}
\end{table}

Our results can be compared with the blue (FNAL/MILC) and the red (JLQCD) bands of Fig.\,7 of \cite{Jaiswal:2020wer}. There, the authors show the shape of the FFs resulting from a BGL-like analysis. They have taken as inputs, in addition to the lattice data, the result coming from light-cone sum rule (LCSR) \cite{Gubernari:2018wyi} at $q^2 = 0$. In \cite{Gubernari:2018wyi}, the theoretical predictions are given in terms of the FFs $V, A_1, A_2$, appearing in the matrix element (\ref{eq:matrix_el_Dstar}). Thanks to Eqs.\,(\ref{eq:ff_BtoDstar})-(\ref{eq:ff_BtoDstar2})-(\ref{eq:ff_BtoDstar3}), their results can be rephrased as
\begin{eqnarray*}
f(z_{max})\vert_{\rm{LCSR}}&=& 4.37 \pm 0.66,\\
g(z_{max})\vert_{\rm{LCSR}}&=& 0.19 \pm 0.04,\\
\mathcal{F}_1(z_{max})\vert_{\rm{LCSR}}&=& 16.0 \pm 2.1.
\end{eqnarray*}
The main difference between the LCSR results and our extrapolated values of the FFs at $z_{max}$ seem to be the mean values and the uncertainties associated to $\mathcal{F}_1(z_{max})$.

\subsubsection{The FNAL/MILC+JLQCD case}

\begin{figure}[htb!]
\begin{center}
{\includegraphics[scale=0.55]{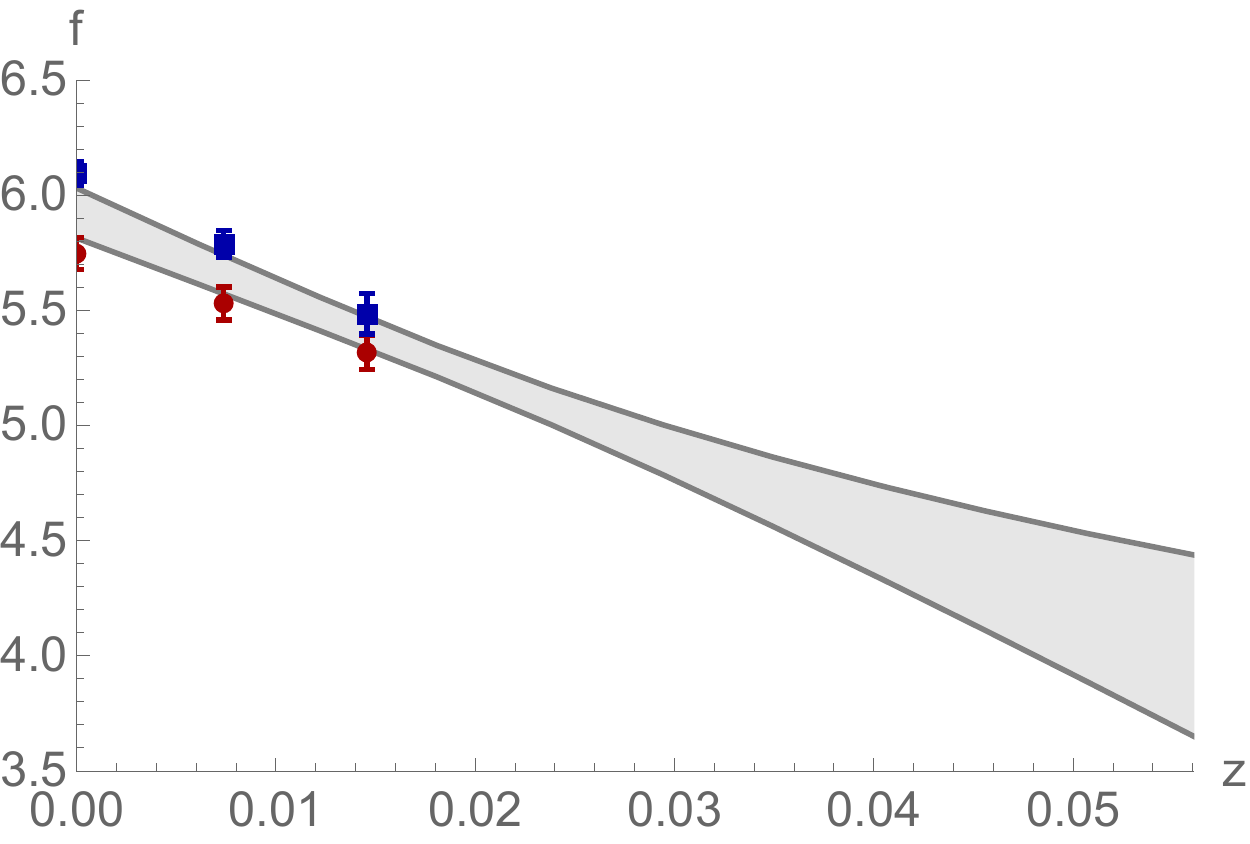}
\label{fig:FFMM222_A}}
{\includegraphics[scale=0.55]{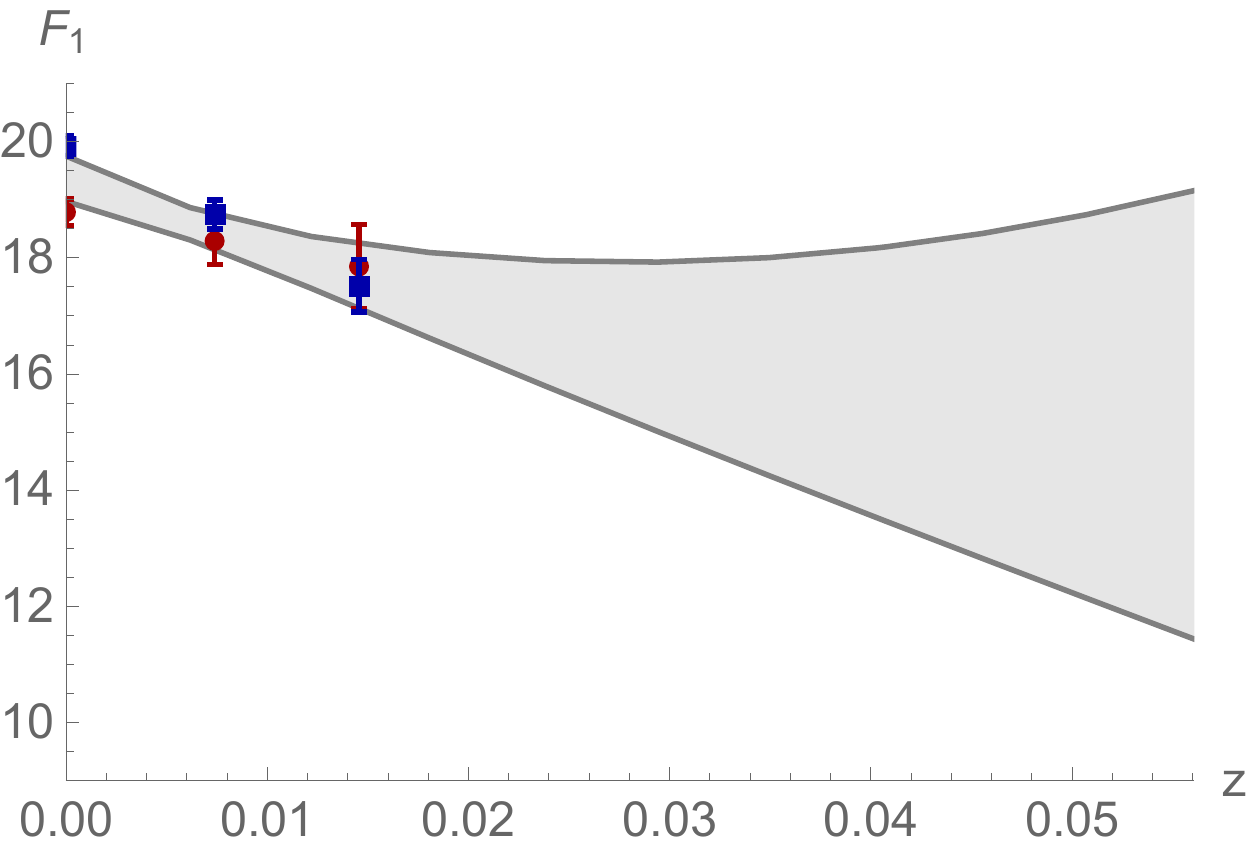}
\label{fig:FFMM222_B}}\\ 
{\includegraphics[scale=0.55]{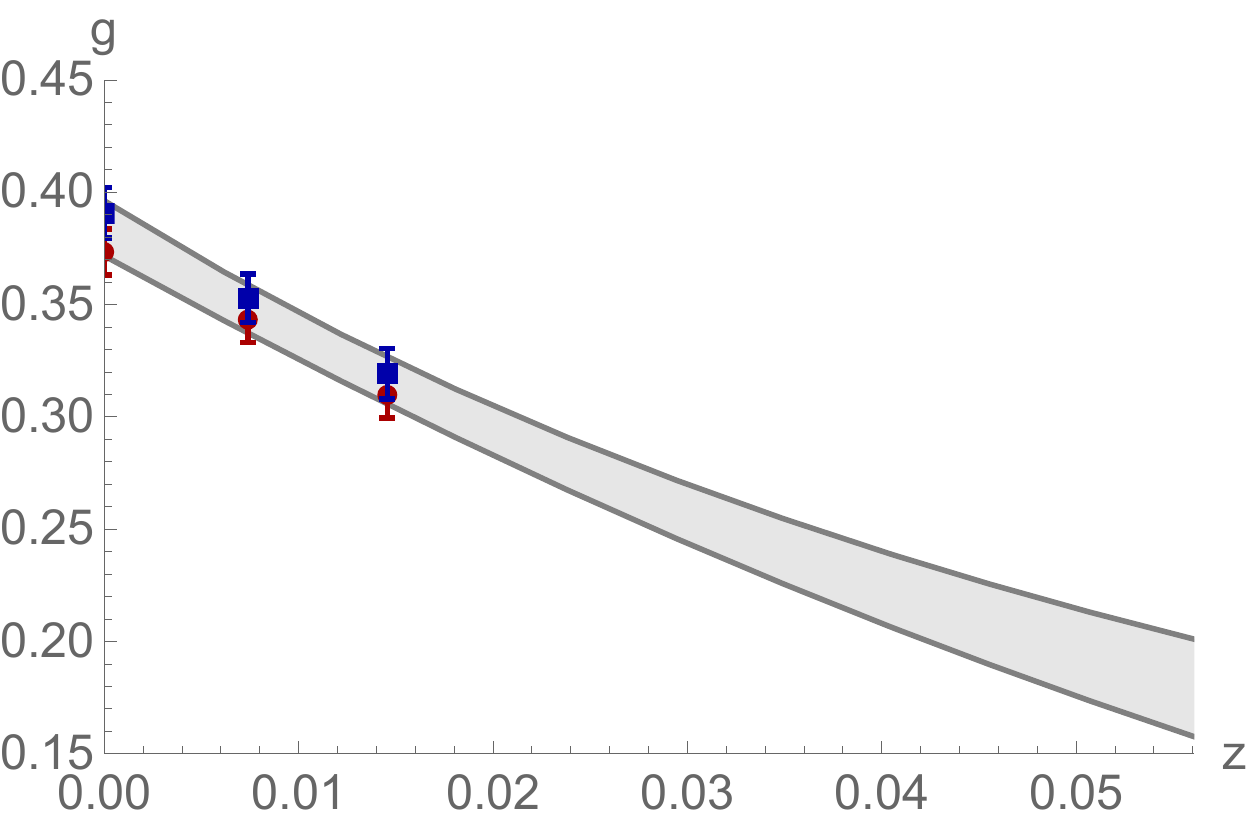}
\label{fig:FFMM222_C}}
{\includegraphics[scale=0.55]{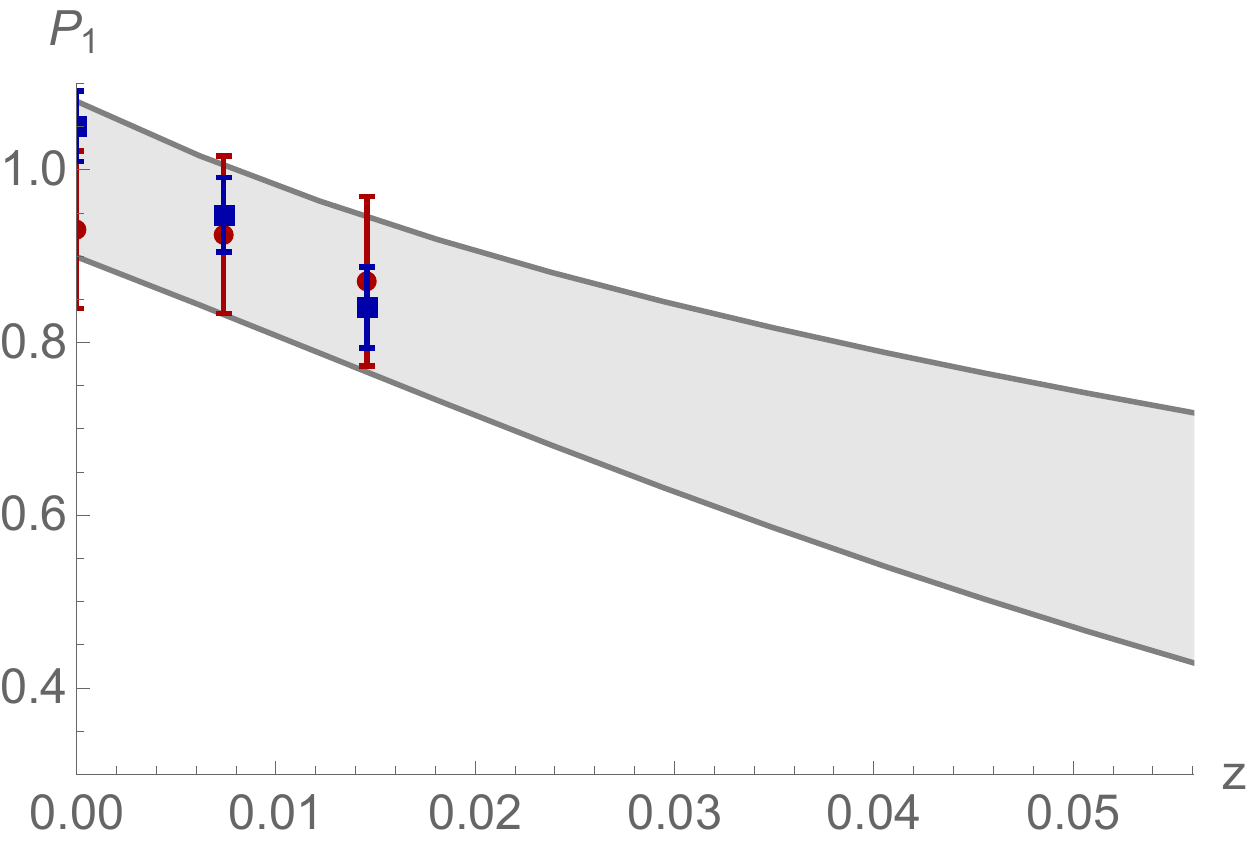}
\label{fig:FFMM222_D}}
\caption{\textit{The bands of the FFs, computed through the DM method, after the combination of the FNAL/MILC and the JLQCD data. As in Fig.\,\ref{FFMM}, the blue squares (red dots) are the FNAL/MILC (JLQCD) input lattice data, before being combined together.}
\hspace*{\fill} \small}
\label{FFMM222}
\end{center}
\end{figure}

As a further case for investigating the sensitivity of our DM method and the stability of the results with respect to changes of the the input lattice data, we have combined together the lattice data coming from the FNAL/MILC and the JLQCD collaborations computations at the same recoil. Thus, we have taken the FNAL/MILC values at the recoils $w=\{1.00,1.06,1.12\}$ from \cite{Aviles-Casco:2019zop} and then we have combined them with the JLQCD ones by adopting Eqs.\,(\ref{eq28Carr})-(\ref{eq28Carrb}), described in the appendix. By repeating the same sceptical analysis described in the previous cases, we get
\begin{eqnarray*}
f(z_{max})\vert_{\rm{FNAL/MILC+JLQCD}}&=& 4.03 \pm 0.39,\\
g(z_{max})\vert_{\rm{FNAL/MILC+JLQCD}}&=& 0.179 \pm 0.021,\\
\mathcal{F}_1(z_{max})\vert_{\rm{FNAL/MILC+JLQCD}}&=& 15.1 \pm 3.9,\\
P_1(z_{max})\vert_{\rm{FNAL/MILC+JLQCD}}&=& 0.57 \pm 0.14.
\end{eqnarray*}
In Fig.\,\ref{FFMM222} we show the bands of the FFs as functions of $z$ in the FNAL/MILC+JLQCD case. Note that the blue squares (red dots) are the single FNAL/MILC (JLQCD) data before being combined together.

\subsubsection{Summary of this analysis}

We summarize the main features of our procedure based on the DM method. First of all, the FFs are described in a \emph{parametrization-independent} way thanks to the DM method. Secondly, we choose to get their shapes (shown in Figs.\,\ref{FFMM}-\ref{FFMM222}) from the theory only, i.e.~\emph{independently} of the experimental data. This fact determines an important difference with respect to other analyses of $B \to D^*$ decays, which add experimental points to constrain the shape of the FFs. Moreover, the knowledge of the pseudoscalar FF $P_1(z)$ is necessary also for $\vert V_{cb} \vert$, since the KC condition (\ref{KC2}) induces a large decrease of the width of the band of $\mathcal{F}_1(z)$ at large $z$. Instead, in other studies present in the literature $P_1(z)$ is neglected for $\vert V_{cb} \vert$, since the experimental data constrain very precisely the shape of $\mathcal{F}_1(z)$ also at large $z$.

\subsection{New estimate of $\vert V_{cb} \vert$}

Let us focus our attention on the experimental decay widths measured in semileptonic $B \to D^*$ decays. For what concerns the experimental state-of-the-art, at present we have at our disposal two different measurements of the differential decay widths, both performed by the Belle Collaboration \cite{Abdesselam:2017kjf, Waheed:2018djm}. The authors report the results of the measurements of the differential decay widths $d\Gamma/dx$, where $x$ is one of the four kinematical variables of interest  ($x= w, \cos \theta_l, \cos \theta_v, \chi$), by dividing the available region for each variable into 10 bins. Hence, we have globally 40 points for each of the two different measurements \cite{Abdesselam:2017kjf, Waheed:2018djm}.  The correlation matrices of the errors are also presented for both the measurements.

First of all, we compute the theoretical  $d\Gamma/dx$ from the expression (\ref{finaldiff333BDst}), using the value of the FFs derived in Section III C. We generate $N_{boot}$ bootstrap values of the FFs $f,g,\mathcal{F}_1, P_1$ for each of the experimental bins through a multivariate Gaussian distribution. In this case the mean values and covariance matrix come directly from the implementation of the DM method. We also generate an independent set of $N_{boot}$ bootstrap values of the experimental differential decay widths for all the bins. For each of them, we fit the histogram of the resulting $N_{boot}$ estimates of $\vert V_{cb} \vert$ with a normal distribution and save the corresponding mean value and uncertainty. Thus, we find 10 values of the CKM matrix element for each of the four kinematical variables ($w, \cos \theta_l, \cos \theta_v, \chi$) and for each of the two experiments \cite{Abdesselam:2017kjf, Waheed:2018djm}.

\subsubsection{FNAL/MILC input}

\begin{figure}[htb!]
\begin{center}
{\includegraphics[scale=0.55]{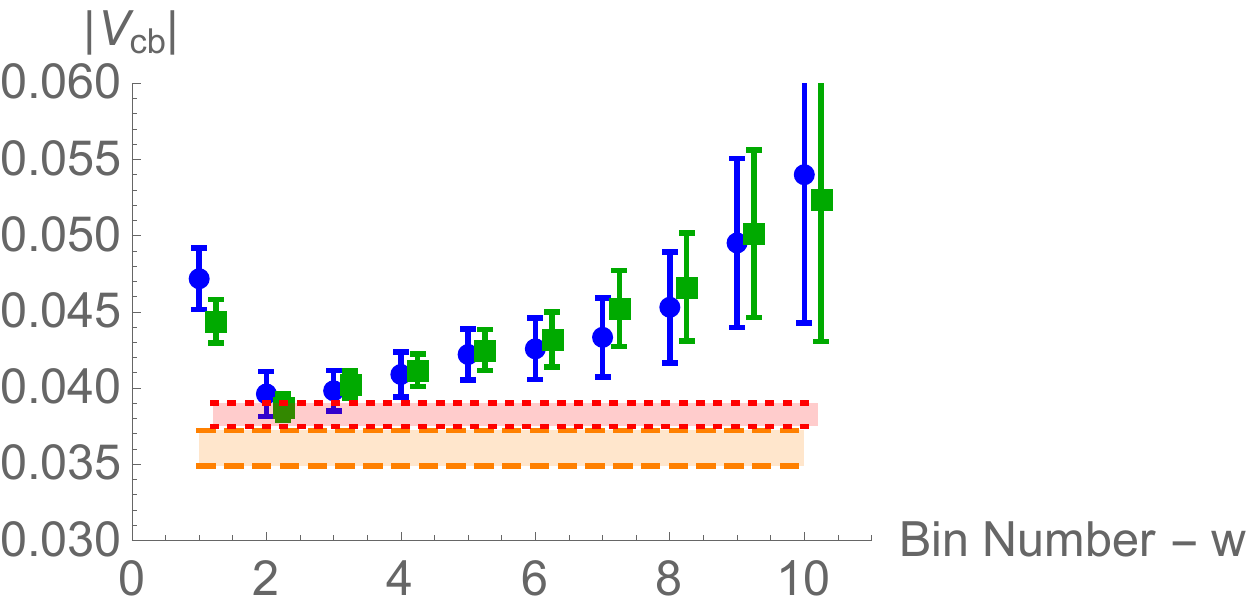}
\label{fig:VcbfinalvalueDstar_A}}
{\includegraphics[scale=0.55]{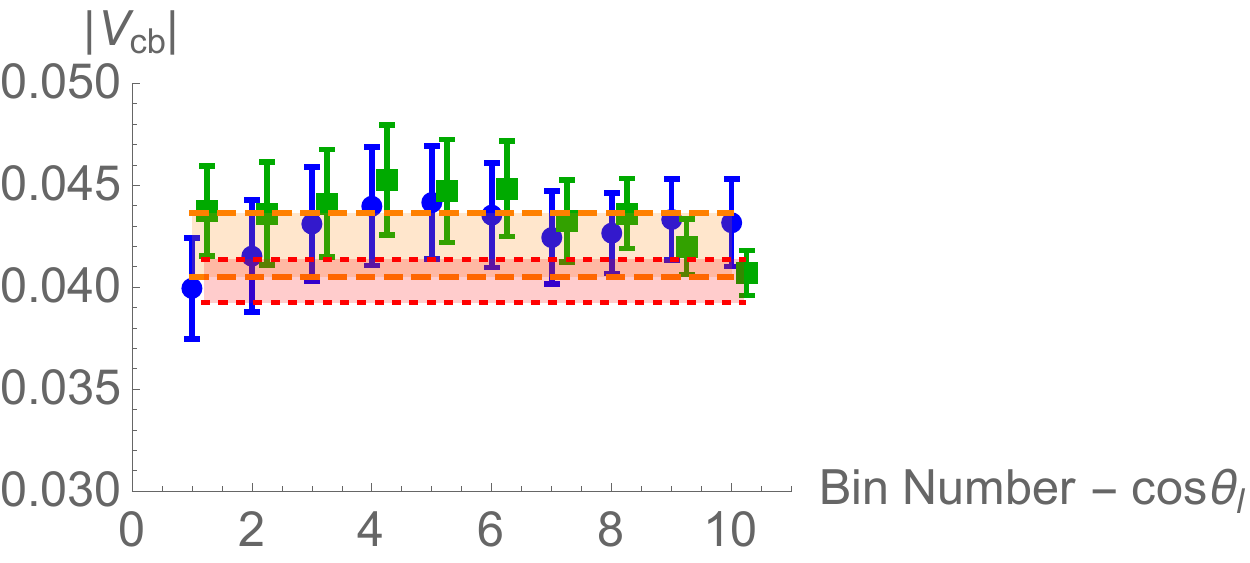}
\label{fig:VcbfinalvalueDstar_B}}\\ 
{\includegraphics[scale=0.55]{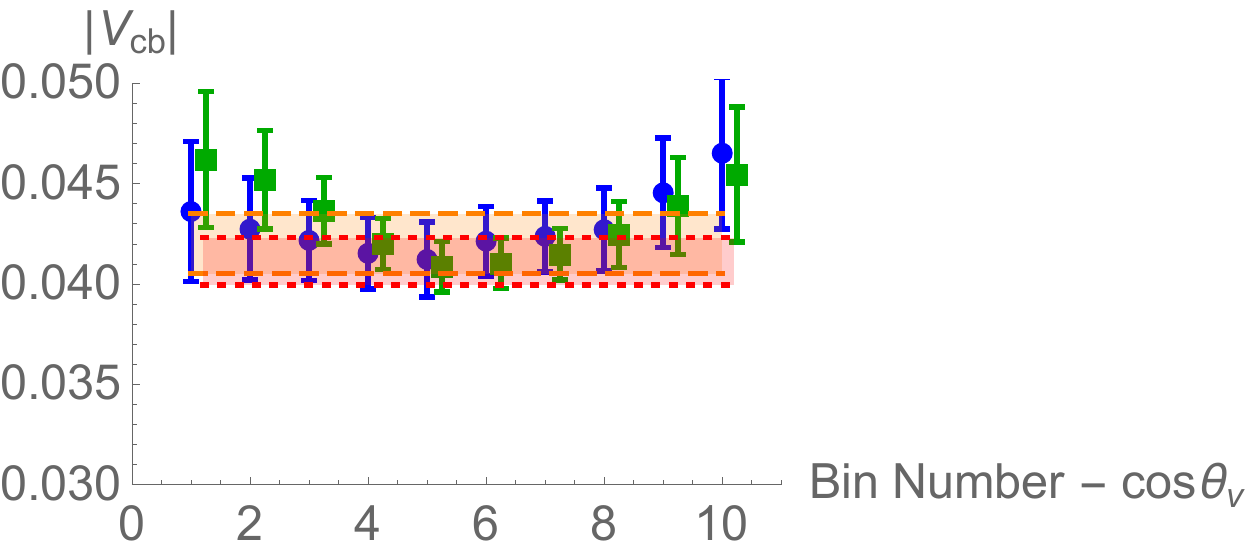}
\label{fig:VcbfinalvalueDstar_C}}
{\includegraphics[scale=0.55]{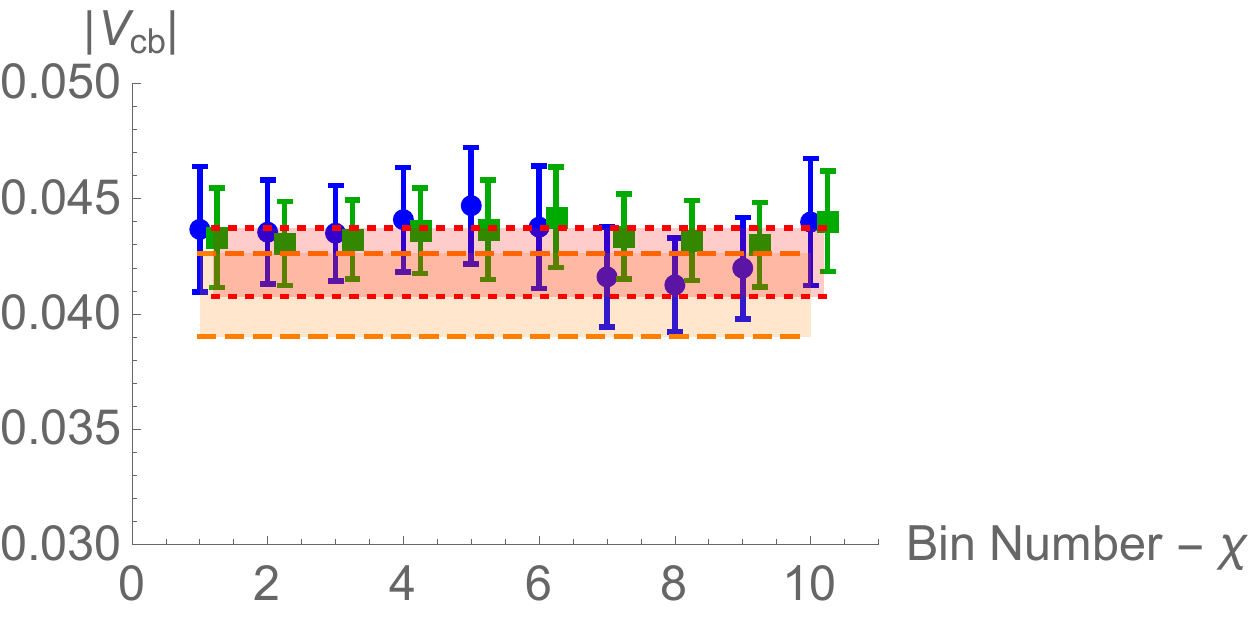}
\label{fig:VcbfinalvalueDstar_D}}
\caption{\textit{The final estimates of $\vert V_{cb} \vert$ for all the experimental bins in $\{w,  \cos \theta_l, \cos \theta_v, \chi\}$, which would result from the use of the blinded FNAL/MILC input~\cite{Aviles-Casco:2019zop}. We show both the blue points, that correspond to the first Belle measurements~\cite{Abdesselam:2017kjf}, and the green squares, which refer to the second set of experimental data~\cite{Waheed:2018djm}.  Finally, the dashed orange (dotted red) bands are the results of the application of Eq.\,(\ref{muVcbfinal}) for each variable, taking into consideration the blue points (green squares).}
\hspace*{\fill} \small}
\label{VcbfinalvalueDstar}
\end{center}
\end{figure}

In Fig.\,\ref{VcbfinalvalueDstar} we show the estimates of $\vert V_{cb} \vert$ for each bin, resulting from our matrix using the FNAL/MILC data as LQCD inputs. For each kinematical variable and for each of the two sets of experimental measurements, we compute a weighted mean of the 10 $\vert V_{cb} \vert$ taking into consideration the correlations. To achieve this goal, calling $\mathbf{C}$ the covariance matrix and $\vert V_{cb} \vert_i$ ($i=1,\cdots,10$) the values of the CKM matrix element for each bin and for each of the two sets of experimental measurements, it is sufficient to compute \cite{Schmelling:1994pz} 
\begin{equation}
\label{muVcbfinal}
\vert V_{cb} \vert = \frac{\sum_{i,j=1}^{10} (\mathbf{C}^{-1})_{ij} \vert V_{cb} \vert_j}{\sum_{i,j=1}^{10} (\mathbf{C}^{-1})_{ij}},\,\,\,\,\,\,\,\,\,\,\,\,\sigma^2_{\vert V_{cb} \vert} = \frac{1}{\sum_{i,j=1}^{10} (\mathbf{C}^{-1})_{ij}}.
\end{equation}
We consider separately the 10x10 diagonal blocks corresponding to each kinematical variable and compute thus four separate mean values for $\vert V_{cb} \vert$. They are combined through Eqs.\,(\ref{eq28Carr})-(\ref{eq28Carrb}), as explained in the appendix. We can use these expressions also to combine the values of $\vert V_{cb} \vert$ coming from the two different experiments  \cite{Abdesselam:2017kjf, Waheed:2018djm}.

In the case of the $w$-distribution, however, the final result for $\vert V_{cb} \vert$ lies below the experimental data as shown by the orange and the red bands in Fig.\,\ref{VcbfinalvalueDstar}a. This problem is well-known in literature \cite{DAgostini:1993arp} and is usually related to some systematic effects in the correlation matrix. It results to be more pronounced for the "blue" experimental measurements \cite{Abdesselam:2017kjf} rather than for the "green" ones \cite{Waheed:2018djm}. Let us focus for a moment only on the latter ones. In this case Eqs.\,(\ref{eq28Carr})-(\ref{eq28Carrb}) give the result
\begin{equation*}
\vert V_{cb} \vert \times 10^{3} = 40.5 \pm 1.9.
\end{equation*}
Instead, in the former case from the expressions (\ref{eq28Carr})-(\ref{eq28Carrb}) we find
\begin{equation*}
\label{VcbestLASTBD*milcold}
\vert V_{cb} \vert \times 10^{3} = 40.2 \pm 2.9.
\end{equation*}
The uncertainty is increased, in fact we are combining very different values of $\vert V_{cb} \vert$ for each kinematical variable. The main effect in this sense is given by the orange band in Fig.\,\ref{VcbfinalvalueDstar}a. If we then consider together both the experiments \cite{Abdesselam:2017kjf, Waheed:2018djm}, we find
\begin{equation}
\label{VcbestLASTBD*milcold_full}
\vert V_{cb} \vert \times 10^{3} = 40.3 \pm 2.5.
\end{equation}

We now illustrate a procedure that helps to overcome the underestimation of $\vert V_{cb} \vert$ observed in this subsection. Let us consider the relative differential decay width given by the ratio $(d\Gamma/dx)/\Gamma$ (where $x=w,  \cos \theta_l, \cos \theta_v, \chi$) for each bin. The advantage of this procedure is that, if there is a calibration error in the data \cite{DAgostini:1993arp}, computing the ratio $(d\Gamma/dx)/\Gamma$ will help to reduce it since all the points enter in the evaluation of $\Gamma$. 

\begin{figure}[htb!]
\begin{center}
{\includegraphics[scale=0.55]{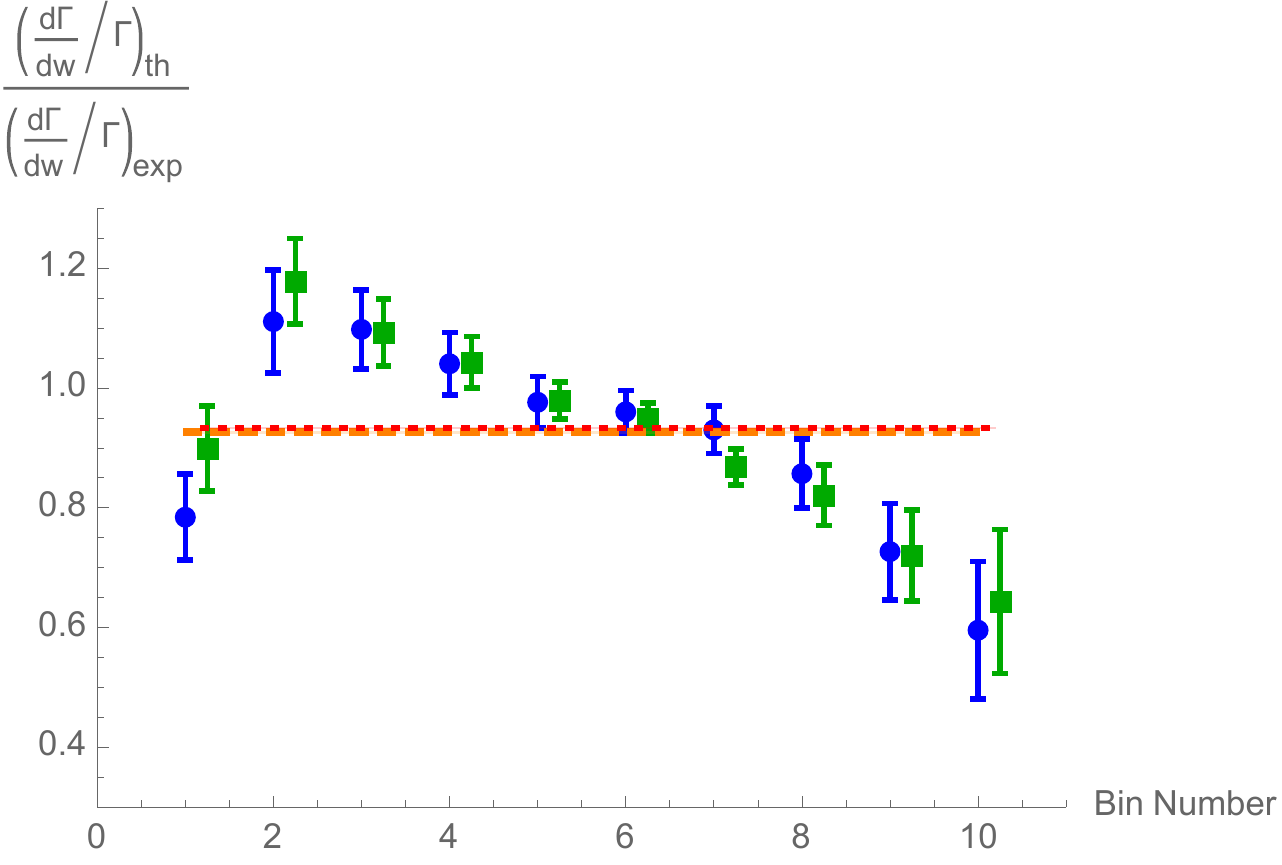}
\label{fig:dGammaratiofinalvalue_A}}
{\includegraphics[scale=0.55]{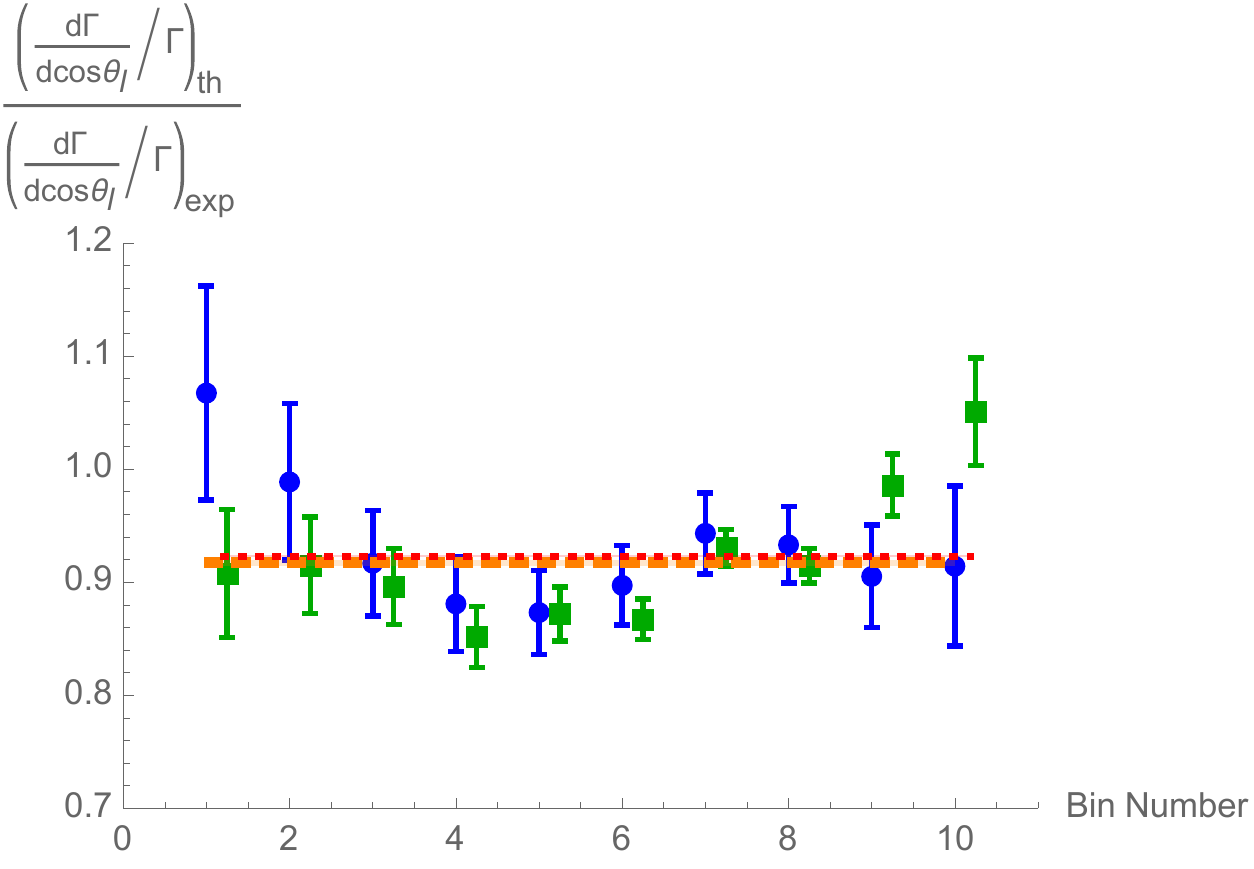}
\label{fig:dGammaratiofinalvalue_B}}\\ 
{\includegraphics[scale=0.55]{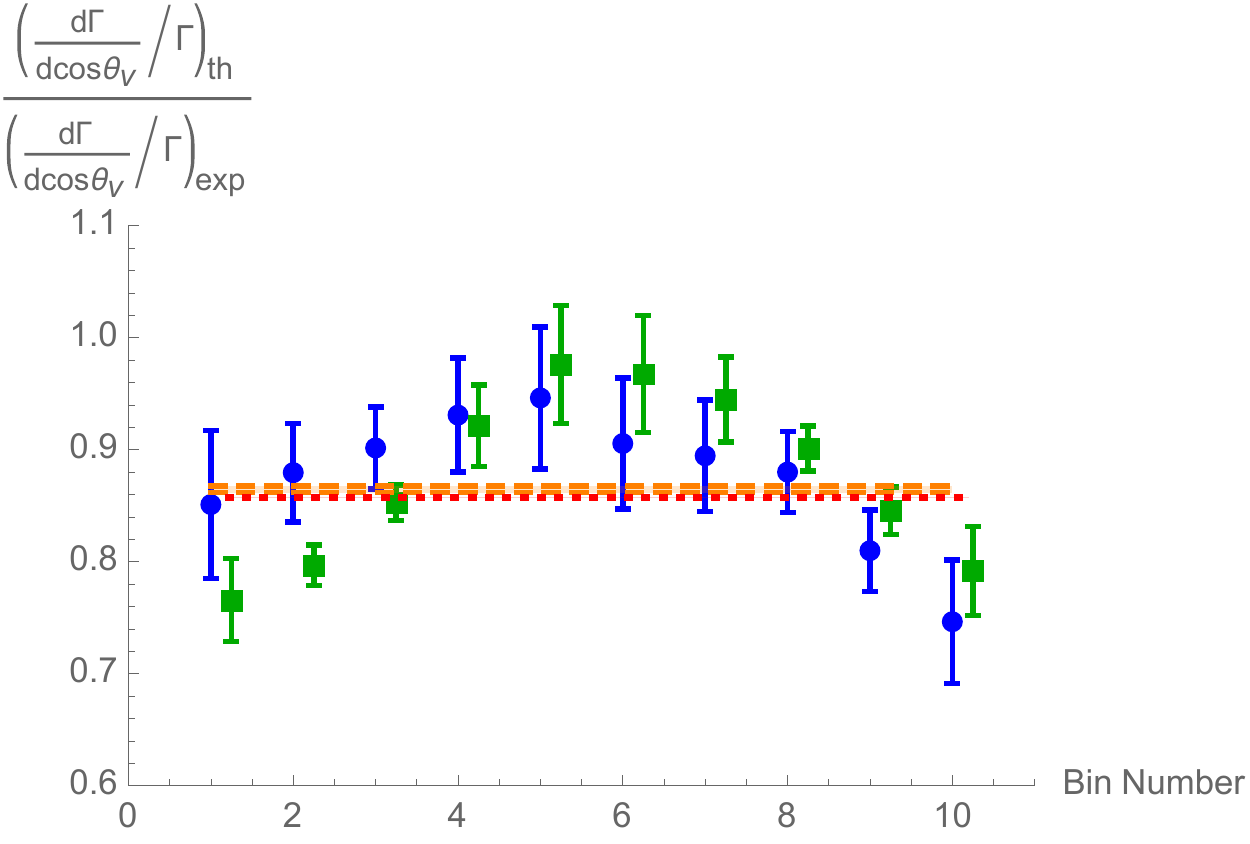}
\label{fig:dGammaratiofinalvalue_C}}
{\includegraphics[scale=0.55]{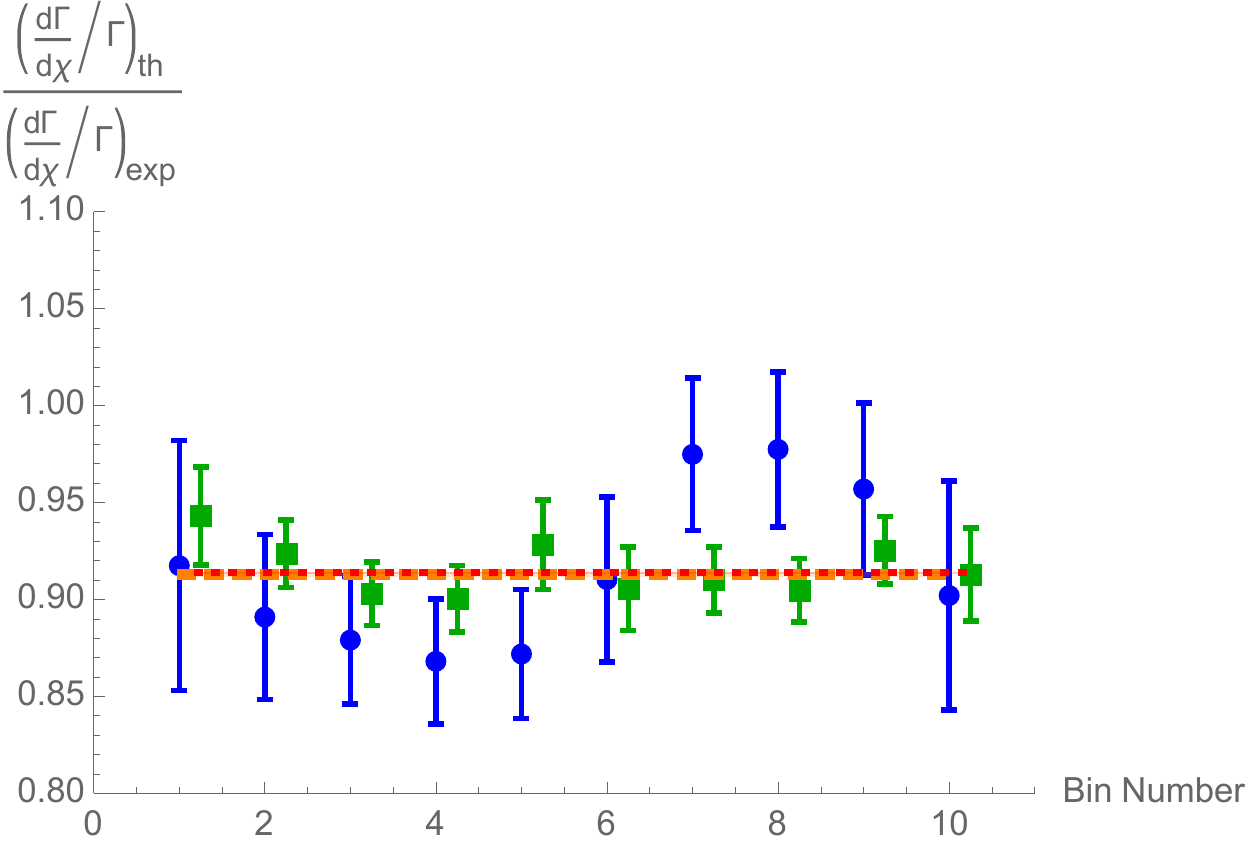}
\label{fig:dGammaratiofinalvalue_D}}
\caption{\textit{The final estimates of the quantities (\ref{dGammadxGammaEXPR}) for all the experimental bins and for each kinematical variable, which would result from the use of the blinded FNAL/MILC input~\cite{Aviles-Casco:2019zop}. As in Fig.\,\ref{VcbfinalvalueDstar}, the blue points correspond to~\cite{Abdesselam:2017kjf} and the green squares to~\cite{Waheed:2018djm}.}
\hspace*{\fill} \small}
\label{dGammaratiofinalvalue}
\end{center}
\end{figure}

Let us compute this ratio both with the extrapolated values of the FFs, $\left[ (d\Gamma/dx) / \Gamma \right]_{th}$, and with the measured data, $\left[ (d\Gamma/dx) / \Gamma \right]_{exp}$. We then compute the double ratio
\begin{equation}
\label{dGammadxGammaEXPR}
\left[ \frac{1}{\Gamma} \frac{d\Gamma}{dx} \right]_{th} ~ / ~ \left[ \frac{1}{\Gamma} \frac{d\Gamma}{dx} \right]_{exp} ~ . ~
\end{equation}
The double ratio should be equal to unity if there is no tension between theory and experiments. Hence, we estimate the mean values and uncertainties for each bin through the extractions of the experimental measurements and of the FFs and, if the calibrations errors have been reduced, we expect then that the final mean values, computed from Eqs.\,(\ref{eq28Carr})-(\ref{eq28Carrb}), will not be systematically underestimated in the $w$ bins. This is the case and the results of this test are illustrated in Fig.\,\ref{dGammaratiofinalvalue}.

\begin{figure}[htb!]
\begin{center}
{\includegraphics[scale=0.55]{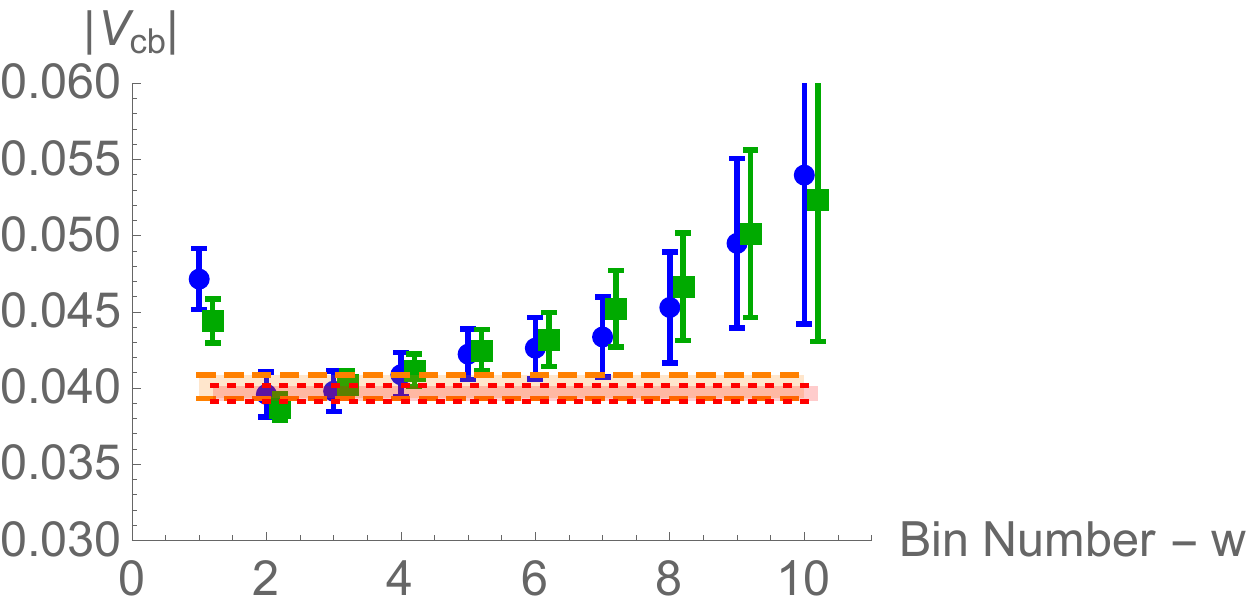}
\label{fig:VcbfinalvalueBIS_A}}
{\includegraphics[scale=0.55]{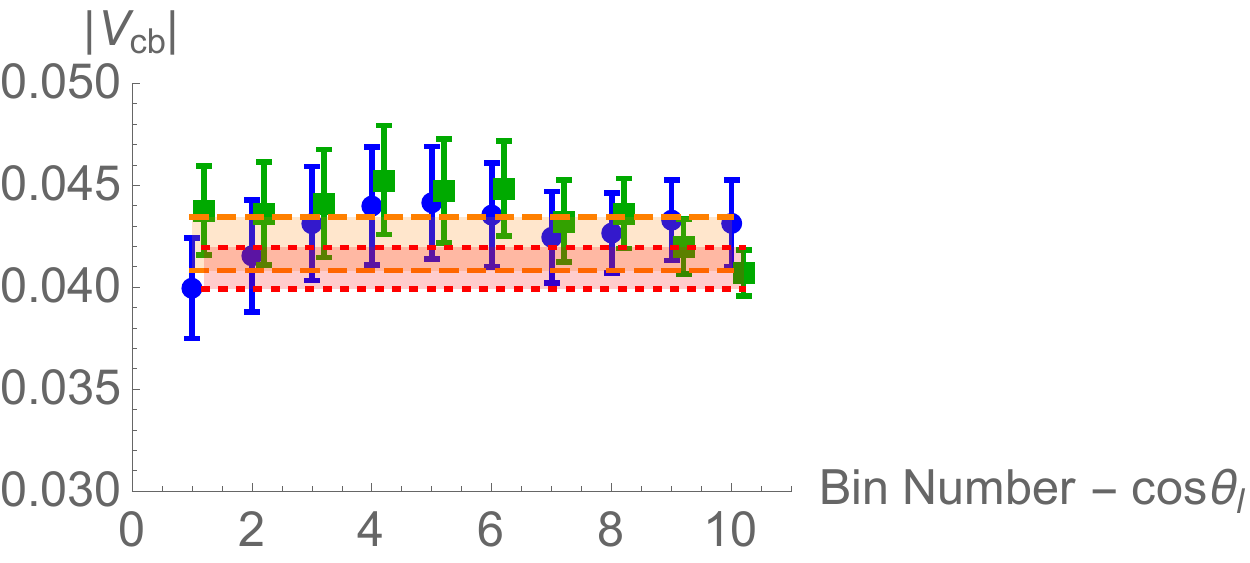}
\label{fig:VcbfinalvalueBIS_B}}\\ 
{\includegraphics[scale=0.55]{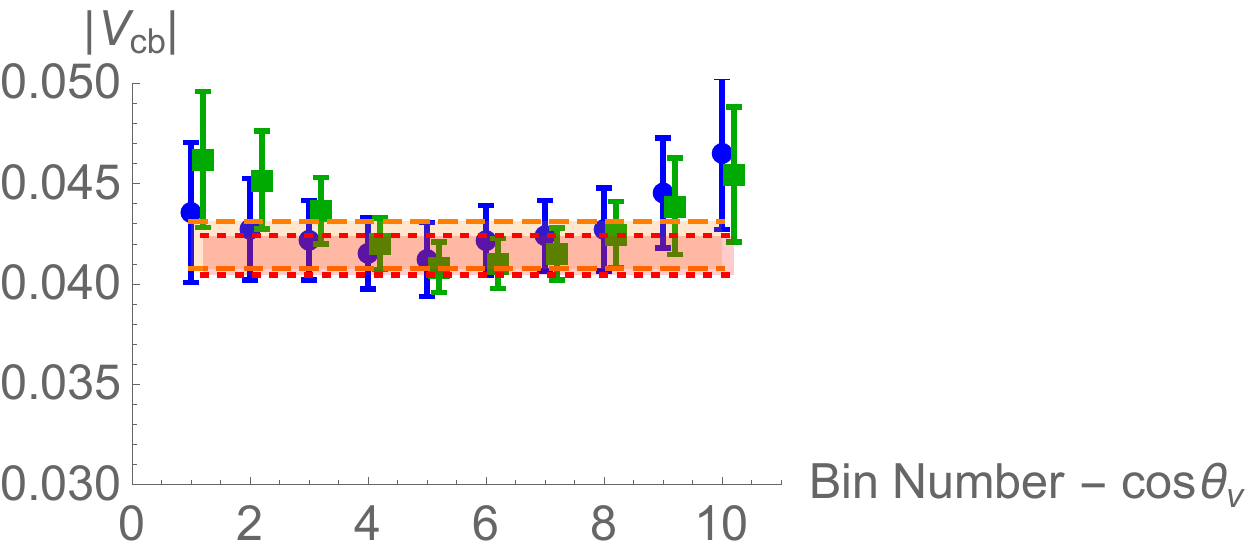}
\label{fig:VcbfinalvalueBIS_C}}
{\includegraphics[scale=0.55]{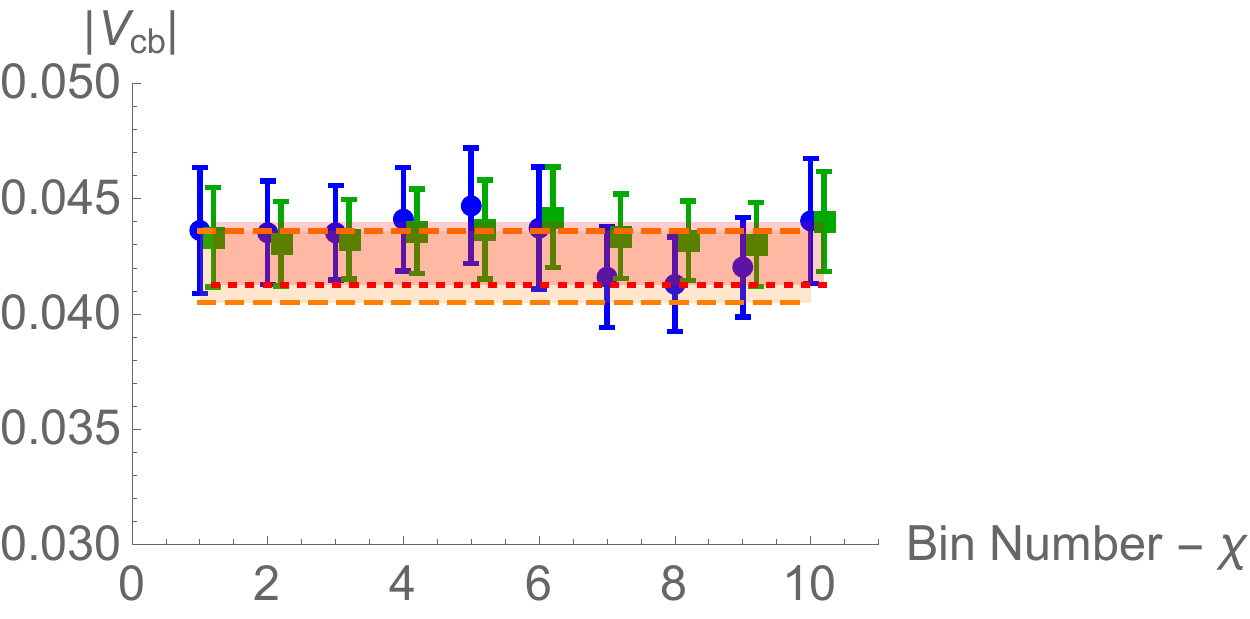}
\label{fig:VcbfinalvalueBIS_D}}
\caption{\textit{The final estimates of $\vert V_{cb} \vert$ adopting the alternative strategy explained above, which would result from the use of the blinded FNAL/MILC input~\cite{Aviles-Casco:2019zop}. The colour code of the points and of the bands is the same as in Fig.\,\ref{VcbfinalvalueDstar}.}
\hspace*{\fill} \small}
\label{VcbfinalvalueBIS}
\end{center}
\end{figure}

The validity of this test suggests the following strategy to get rid of the systematic effects in the original correlation matrix. We compute the correlations $\rho_{ij}\vert_{ratio}$ of the $\left[ (d\Gamma/dx) / \Gamma \right]_{exp}$ bootstrap events and then we derive a \emph{new} covariance matrix of the experimental data given by
\begin{equation}
\label{newcorrmatrix}
C_{ij}\vert_{exp,NEW} = \rho_{ij}\vert_{ratio} \times \sigma_{i,exp}  \sigma_{j,exp},
\end{equation}
where $\sigma_{exp}$ are the uncertainties associated to the experimental differential decay widths. At this point, we repeat the whole procedure for the extraction of $\vert V_{cb} \vert$ starting from new bootstraps for the experimental data, extracted through the matrix $C_{ij}\vert_{exp,NEW}$. Fig.\,\ref{VcbfinalvalueBIS} shows the distributions of $\vert V_{cb} \vert$ for each bin together with the values of $\vert V_{cb} \vert$ for each kinematical variable and for each experiment. No underestimate of $\vert V_{cb} \vert$ is observed in this case. Equations\,(\ref{eq28Carr})-(\ref{eq28Carrb}) allow us to combine our results in a final estimate of $\vert V_{cb} \vert$, which reads
\begin{equation}
\label{VcbestLASTBD*milcBIS}
\vert V_{cb} \vert \times 10^{3} = 41.4 \pm 1.5.
\end{equation}
With respect to Eq.\,(\ref{VcbestLASTBD*milcold_full}), the mean value is higher and the error is substantially decreased. Our interpretation is that probably the original correlation matrix of the data of the first Belle measurement \cite{Abdesselam:2017kjf} was affected by calibration errors, which are reduced by redefining the correlation matrix as in Eq. (\ref{newcorrmatrix}). 

\subsubsection{JLQCD input}

\begin{figure}[htb!]
\begin{center}
{\includegraphics[scale=0.55]{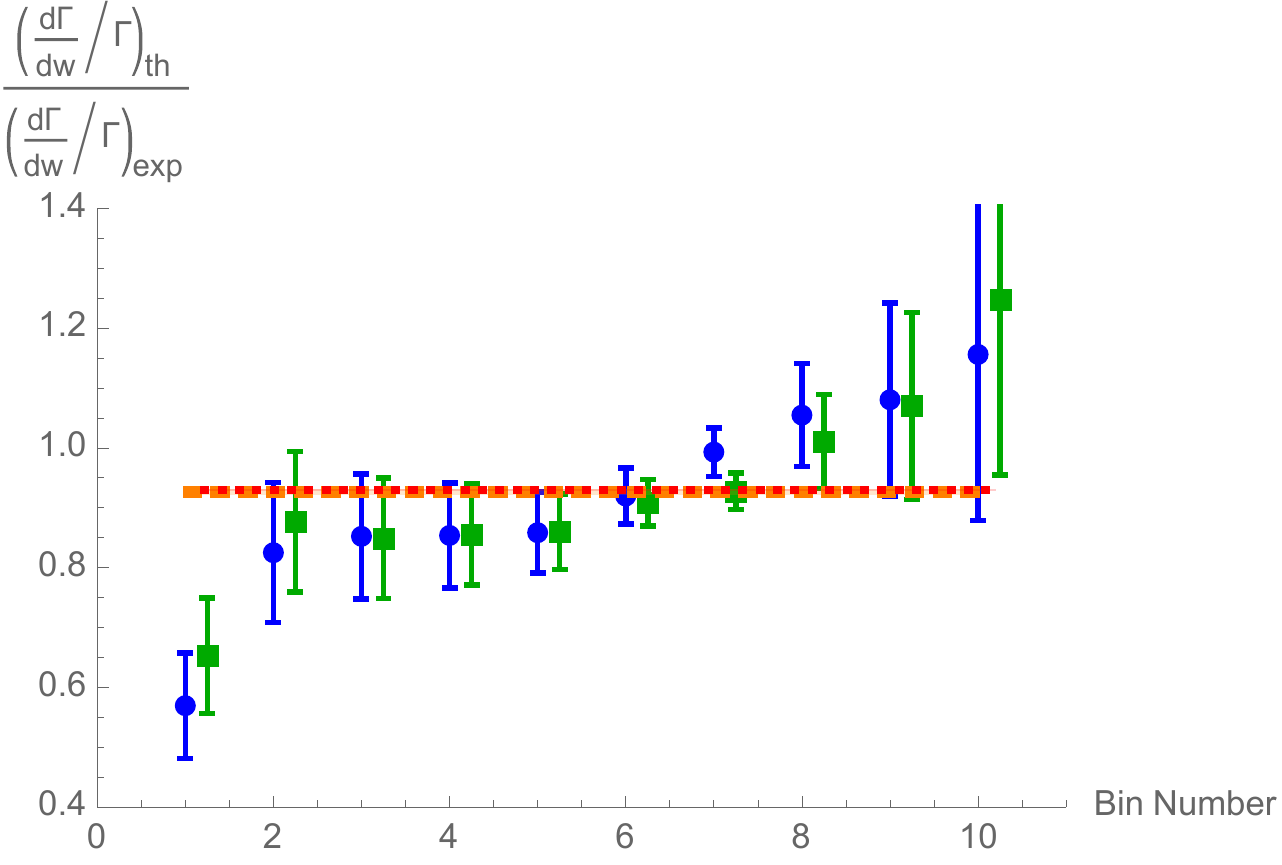}
\label{fig:dGammaratiofinalvalueBIS_A}}
{\includegraphics[scale=0.55]{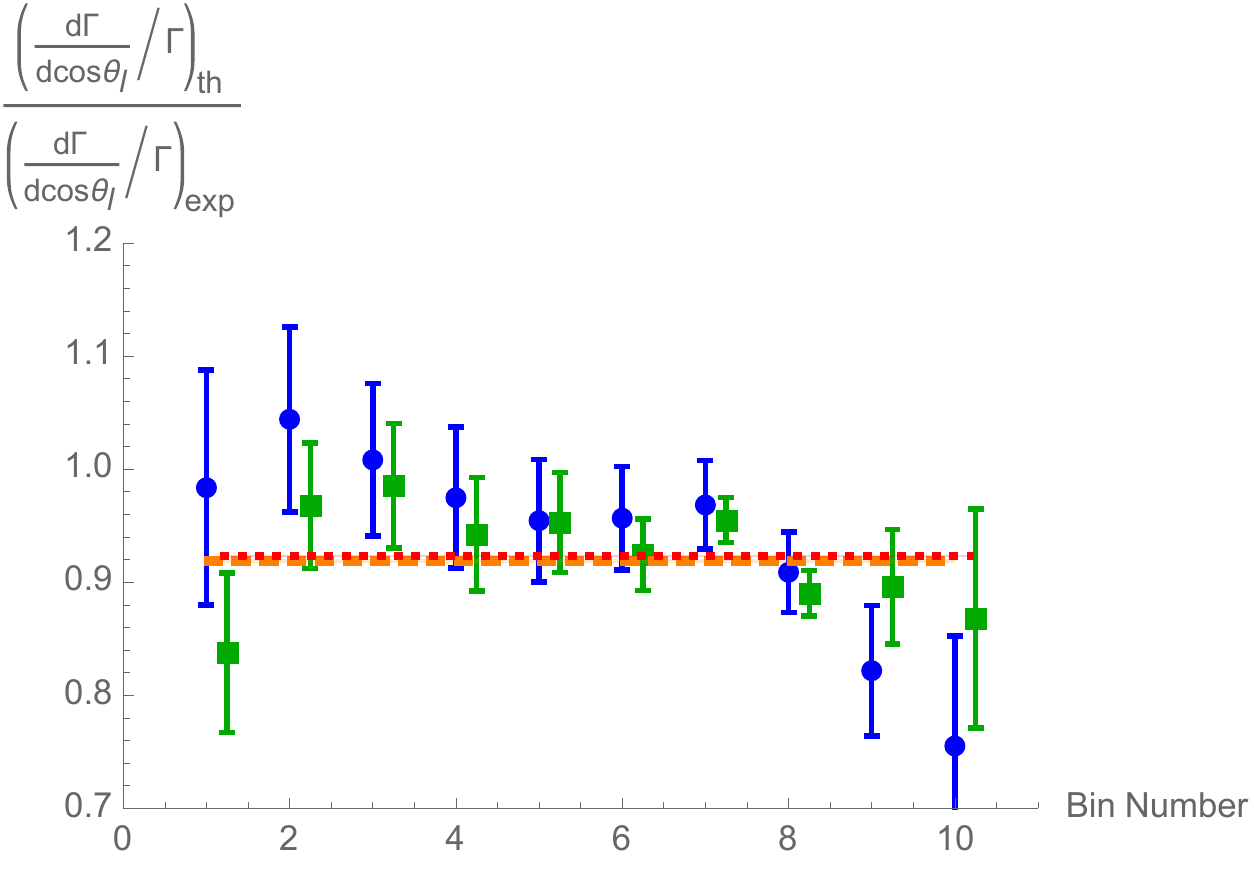}
\label{fig:dGammaratiofinalvalueBIS_B}}\\ 
{\includegraphics[scale=0.55]{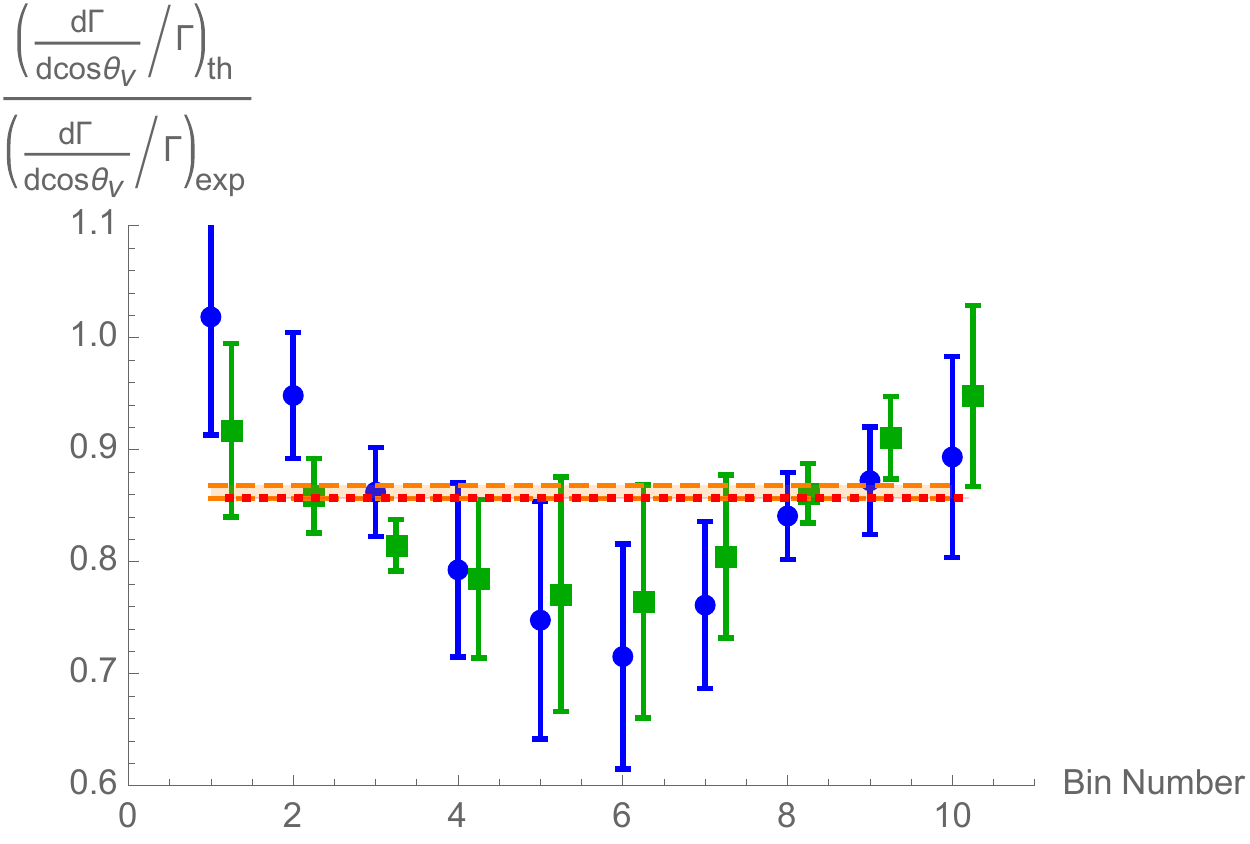}
\label{fig:dGammaratiofinalvalueBIS_C}}
{\includegraphics[scale=0.55]{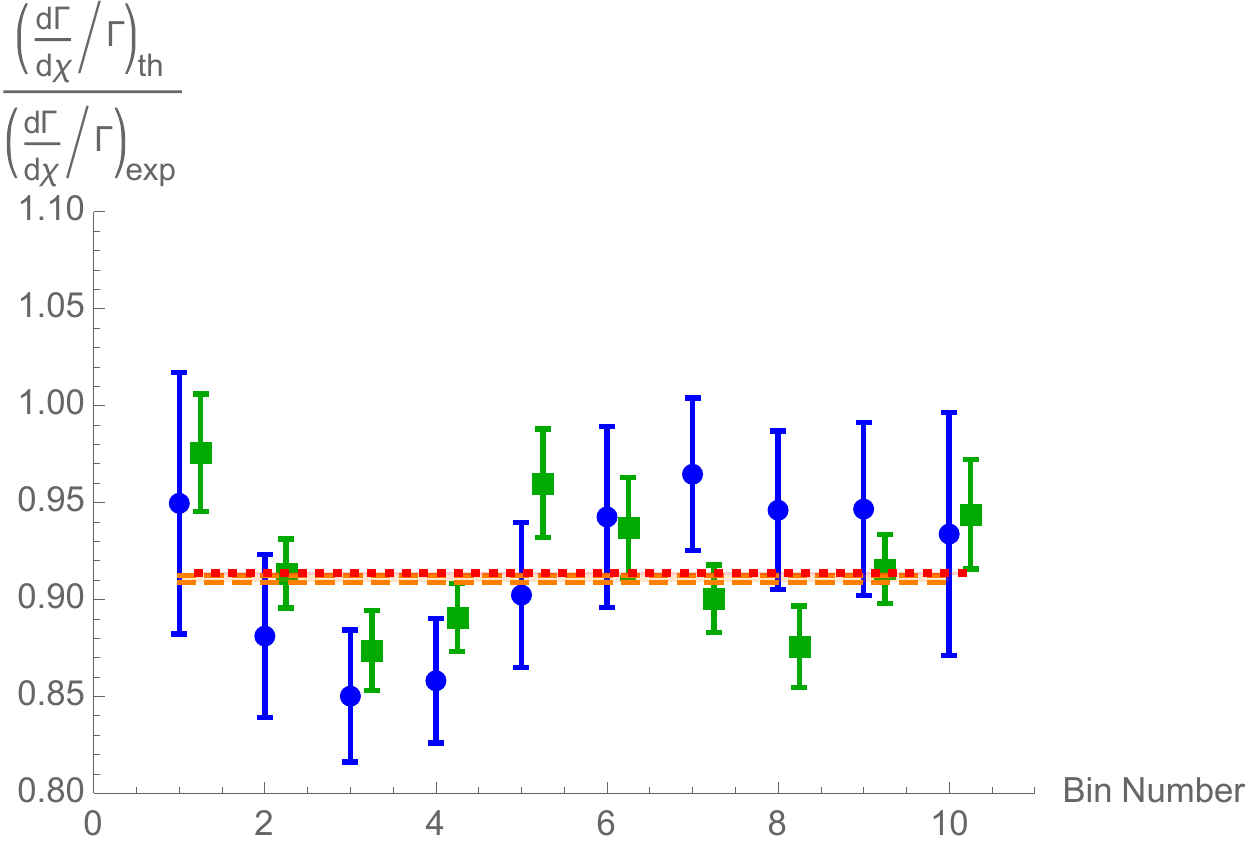}
\label{fig:dGammaratiofinalvalueBIS_D}}
\caption{\textit{The final estimates of the quantities (\ref{dGammadxGammaEXPR}) for all the experimental bins and for each kinematical variable, resulting from JLQCD input \cite{Kaneko:2019vkx}. The colour code of the points and of the bands is the same of Fig.\,\ref{VcbfinalvalueDstar}.}
\hspace*{\fill} \small}
\label{dGammaratiofinalvalueBIS}
\end{center}
\end{figure}

Let us now examine the JLQCD data. In this case, the problems discussed for the FNAL/MILC input data occur again only for the "blue" experimental measurements \cite{Abdesselam:2017kjf}, but not for the "green" ones \cite{Waheed:2018djm}. In the latter case, the standard procedure gives 
\begin{equation}
\label{VcbestLASTBD*jlqcdmiddle}
\vert V_{cb} \vert \times 10^{3} = 40.3 \pm 1.4,
\end{equation}
while in the former case 
\begin{equation}
\label{VcbestLASTBD*jlqcdmiddle2}
\vert V_{cb} \vert \times 10^{3} = 38.9 \pm 3.0.
\end{equation}
The explanation of the large uncertainty in Eq.(\ref{VcbestLASTBD*jlqcdmiddle2}) is the same one of the FNAL/MILC case. If we then consider \cite{Abdesselam:2017kjf, Waheed:2018djm} together, the combined result reads
\begin{equation}
\label{VcbestLASTBD*jlqcdold}
\vert V_{cb} \vert \times 10^{3} = 39.6 \pm 2.4.
\end{equation}
Hence, we have implemented the improved strategy based on (\ref{dGammadxGammaEXPR}). In Fig.\,\ref{dGammaratiofinalvalueBIS} we show the distributions of the quantities (\ref{dGammadxGammaEXPR}) for the JLQCD case, having determined the mean values and the corresponding uncertainties through the bootstraps of the experimental data and of the FFs as before. Also in this case the final mean values do not lie systematically below the experimental points.

\begin{figure}[htb!]
\begin{center}
{\includegraphics[scale=0.55]{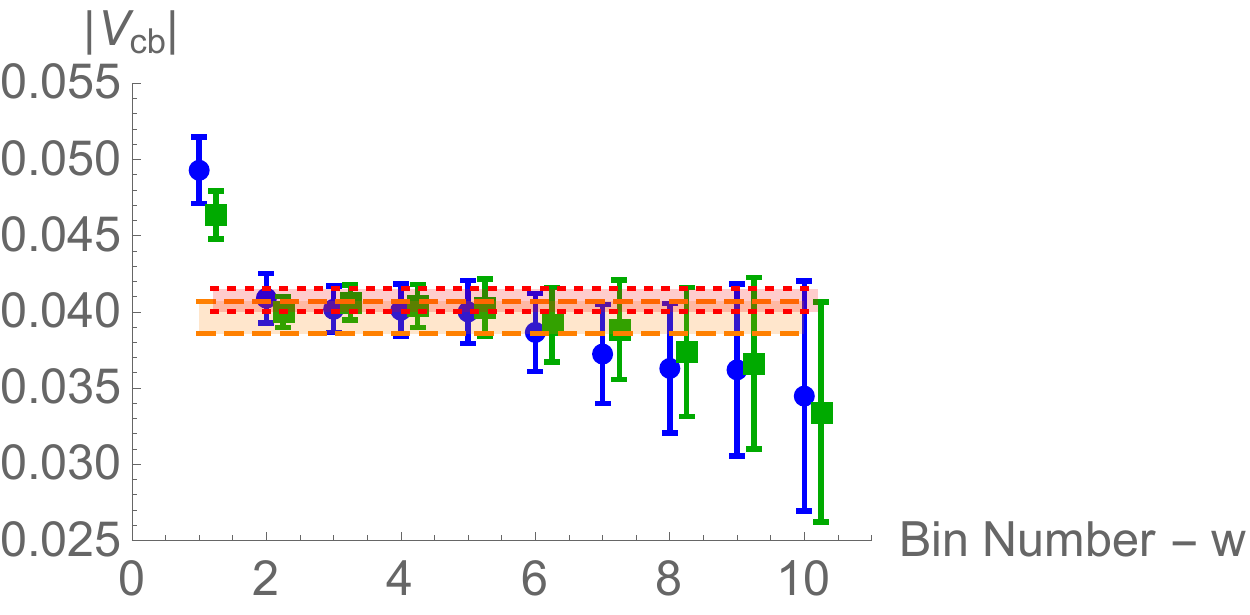}
\label{fig:Vcbfinalvalue2_A}}
{\includegraphics[scale=0.55]{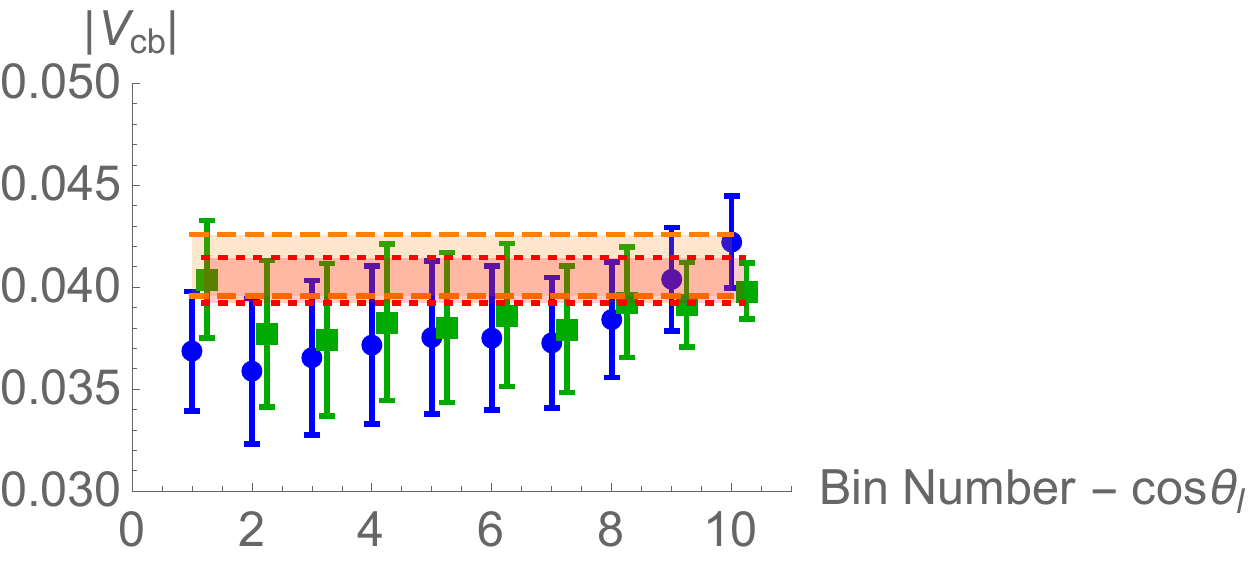}
\label{fig:Vcbfinalvalue2_B}}\\ 
{\includegraphics[scale=0.55]{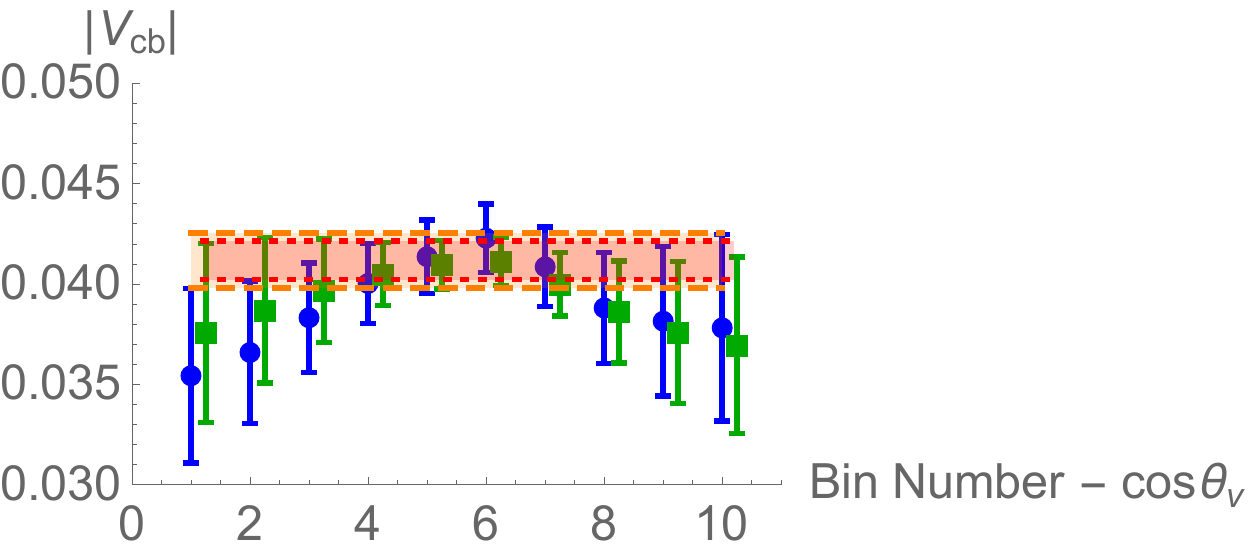}
\label{fig:Vcbfinalvalue2_C}}
{\includegraphics[scale=0.55]{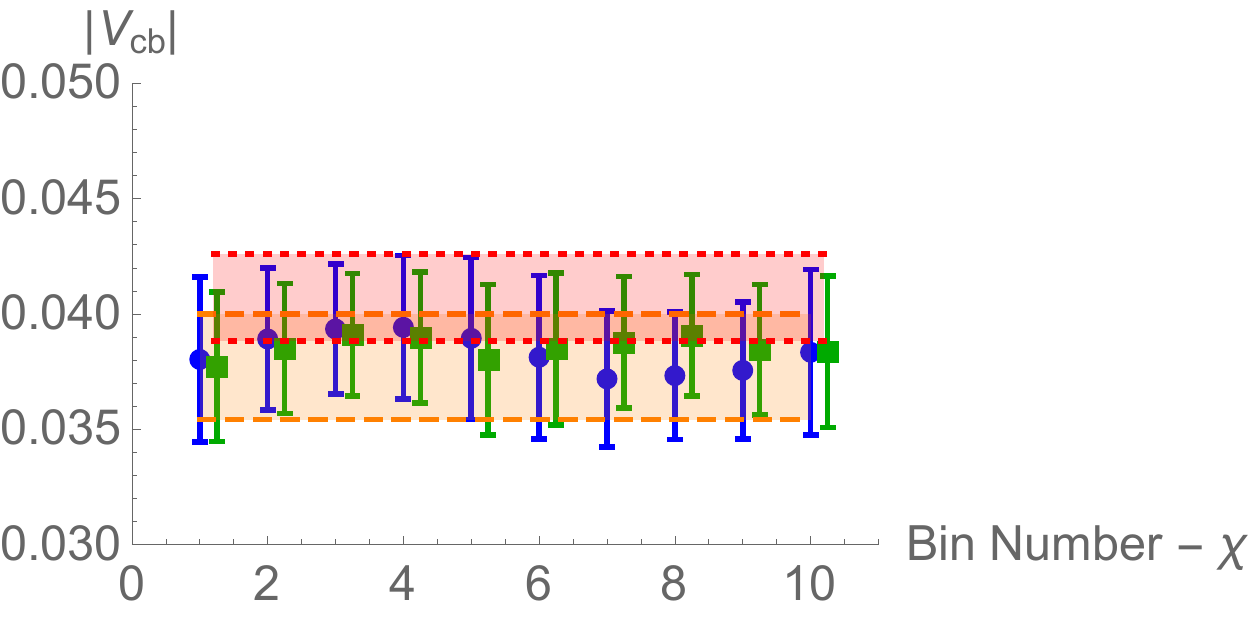}
\label{fig:Vcbfinalvalue2_D}}
\caption{\textit{The final estimates of $\vert V_{cb} \vert$ for the JLQCD data inputs\,\cite{Kaneko:2019vkx}, adopting the alternative strategy explained before. The colour code of the points and of the bands is the same of Fig.\,\ref{VcbfinalvalueDstar}.}
\hspace*{\fill} \small}
\label{Vcbfinalvalue2}
\end{center}
\end{figure}

Fig.\,\ref{Vcbfinalvalue2} shows the distributions of $\vert V_{cb} \vert$ for each bin and the separate mean values for $\vert V_{cb} \vert$. Equations\,(\ref{eq28Carr})-(\ref{eq28Carrb}) allow us to combine our results in a second estimate of $\vert V_{cb} \vert$, which reads
\begin{equation}
\label{VcbestLASTBD*jlqcdBIS}
\vert V_{cb} \vert \times 10^{3} = 40.4 \pm 1.8.
\end{equation}

\subsubsection{FNAL/MILC+JLQCD input}

If we combine FNAL/MILC and JLQCD data, we obtain a third estimate of $\vert V_{cb} \vert$ starting from the FF bands in Fig.\,\ref{FFMM222}. The analysis, developed as in the separate two cases, gives the result
\begin{equation}
\label{VcbestLASTBD*milcjlqcdBIS}
\vert V_{cb} \vert \times 10^{3} = 40.6 \pm 1.6.
\end{equation}

\subsubsection{Other determinations of $\vert V_{cb} \vert$ in the literature}

Let us now discuss how the inclusive and the other exclusive estimates of $\vert V_{cb} \vert$ compare to each other. The most recent inclusive determination of $\vert V_{cb} \vert$ reads \cite{Gambino:2016jkc, Aoki:2019cca}
\begin{equation}
\label{Vcbestincl}
\vert V_{cb} \vert_{incl} \times 10^{3} = 42.00 \pm 0.65.
\end{equation}
For the other exclusive determinations (from $B \to D^* \ell \nu$ decays \emph{only}), we mention:
\begin{eqnarray*}
\vert V_{cb} \vert_{excl} \times 10^{3} &=& 39.6^{+1.1}_{-1.0} \qquad \quad ~ \text{\cite{Gambino:2019sif}} ~ , ~ \\
\vert V_{cb} \vert_{excl} \times 10^{3} &=& 39.56^{+1.04}_{-1.06} \qquad \, ~ \text{\cite{Jaiswal:2020wer} }~ , ~ \\
\vert V_{cb} \vert_{excl} \times 10^{3} &=& 38.30 \pm 0.8 \qquad \text{\cite{Aoki:2019cca}} ~ , ~ 
\end{eqnarray*}
where the authors had implemented BGL-like analyses of the same experimental data. In these papers, however, only the LQCD computation $h_{A_1}(1)=0.906(13)$ \cite{Bailey:2014tva} was included. All these determinations are compatible with each other, although there is a non-negligible tension with the inclusive determination at the $\sim 2\sigma$ level.

\subsubsection{Summary of this Section}

In conclusion, let us highlight the main features of our procedure to extract $\vert V_{cb}\vert$ from exclusive experiments. First of all, we choose to keep distinct the lattice and the experimental data. In other words, only LQCD computations are used in order to derive the allowed unitarity bands of the FFs as functions of $z$ thanks to the DM method. Then, the experimental measurements are considered only for determining $\vert V_{cb} \vert$. Secondly, as stressed already in the previous Section, in our study the pseudoscalar FF $P_1(z)$ plays a central role also in the determination of $\vert V_{cb}\vert$. In fact, the KC (\ref{KC2}) allows us to constrain the $\mathcal{F}_1(z)$ band obtaining a better precision in the region not explored by lattice computations, $i.e.$ at large values of $z$.

Since the preliminary FNAL/MILC lattice data contain a small, but unknown blinding factor~\cite{Aviles-Casco:2019zop} common to all FFs, our final estimate of $\vert V_{cb} \vert$ is given by Eq.~(\ref{VcbestLASTBD*jlqcdBIS}) obtained using only the preliminary unblinded JLQCD data\,\cite{Kaneko:2019vkx}. It is compatible with other exclusive determinations obtained in the literature\,\cite{Gambino:2019sif,Jaiswal:2020wer,Aoki:2019cca}. Moreover, we obtain consistency also with the inclusive determination~(\ref{Vcbestincl}), though, we remind, we still make use of preliminary lattice results for the FFs.

Since the values of $\vert V_{cb} \vert$ extracted from the differential rate in $w$ play a fundamental role in the alternative strategy explained above, we have implemented the following exercise to compare FNAL/MILC, JLQCD and FNAL/MILC+JLQCD data. Starting from Eqs.\,(\ref{finaldiff333BDst})-(\ref{helampl}), the differential distribution is given by
\begin{equation}
\label{modquadrHi}
\frac{1}{\kappa}\,\frac{d\Gamma}{dw}(w) = (m_B^2+m_D^2-2m_Bm_Dw)\times(\vert H_0(w) \vert^2 +\vert H_+(w) \vert^2 +\vert H_-(w) \vert^2),
\end{equation}
where 
\begin{equation*}
\kappa = \frac{G_F^2 \vert V_{cb} \vert^2 \eta_{EW}^2  m_{D^*}^2 \sqrt{w^2-1}}{48 \pi^3 m_B}.
\end{equation*}

\begin{figure}[htb!]
\begin{center}
\includegraphics[width=0.90\textwidth]{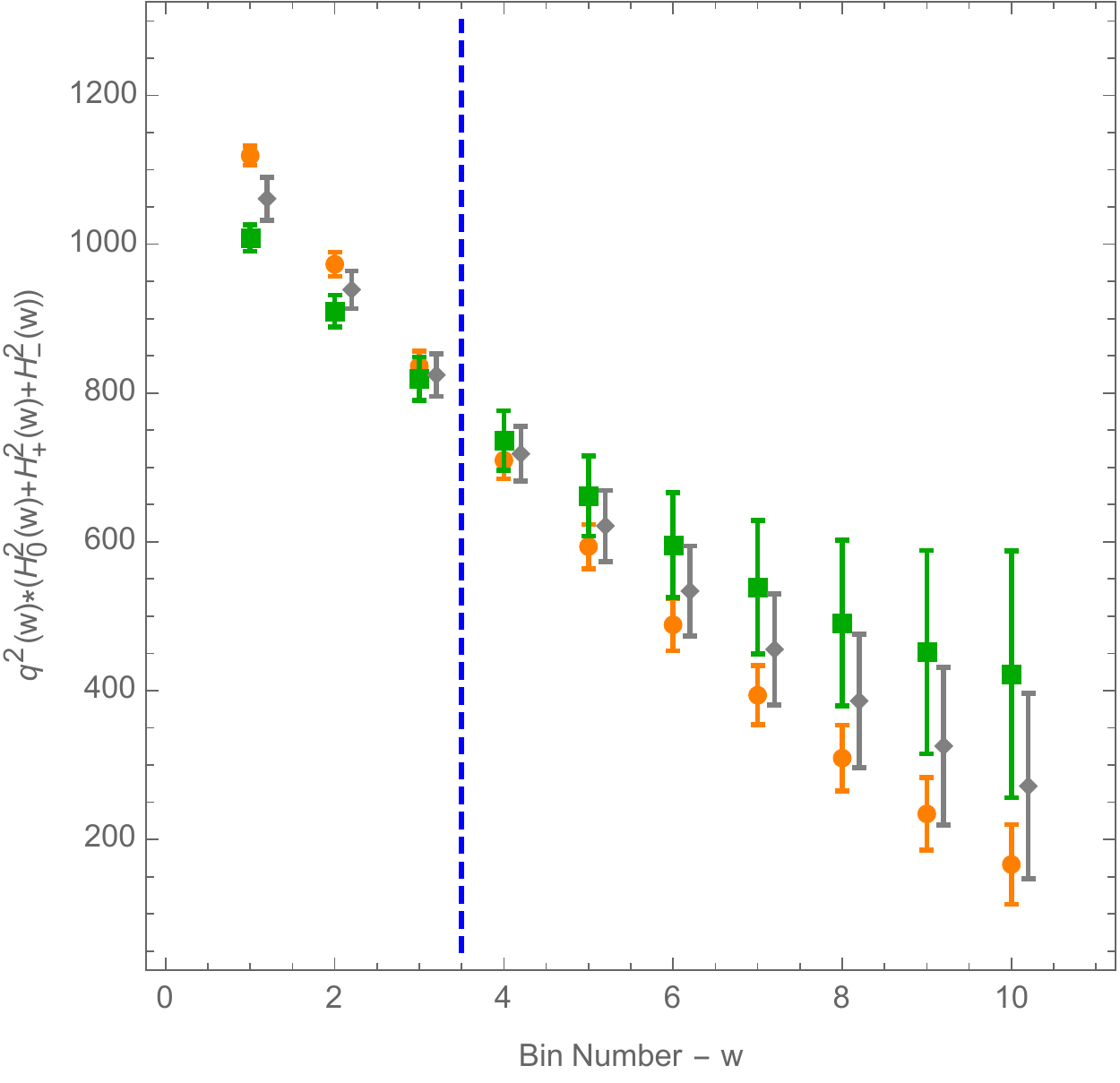}
\caption{\textit{The quantity given by the r.h.s.~of Eq.\,(\ref{modquadrHi}) determined by the DM method adopting FNAL/MILC (orange points), JLQCD (green squares) and FNAL/MILC+JLQCD (gray diamonds) input data. The location of latter ones has been slightly shifted to the right for a better reading. The region at the right of the dashed blue line is not accessible by LQCD computations, thus in this case the points result uniquely from the unitarity and the kinematical constraints on the FFs obtained within the DM method. The strongest tension between FNAL/MILC and JLQCD is present to the left of the dashed blue line, where direct LQCD computations of the FFs are available.}\hspace*{\fill}}
\label{modquadrHifig}
\end{center}
\end{figure}

We have thus decided to plot the quantity on the r.h.s.~of Eq.\,(\ref{modquadrHi}) in order to highlight the differences between the FNAL/MILC, the JLQCD and the FNAL/MILC+JLQCD input data. The results are shown in Fig.\,\ref{modquadrHifig}. The largest differences between the orange points (FNAL/MILC) and the green squares (JLQCD) are present at small recoil, where both the sets of data are more precise since they come from direct computations on the lattice. 

\subsection{New estimate of $R(D^*)$ and of the polarization observables}

In Table \ref{tab:pheno} we report all the numerical results of the phenomenological application of the DM method for the $B \to D^*$ transition. We show separately the effects of the FNAL/MILC, the JLQCD and the FNAL/MILC+JLQCD input lattice data, together with the experimental measurements of each observable of interest, as we will explain in what follows. 

\begin{figure}[htb!]
\begin{center}
{\includegraphics[scale=0.85]{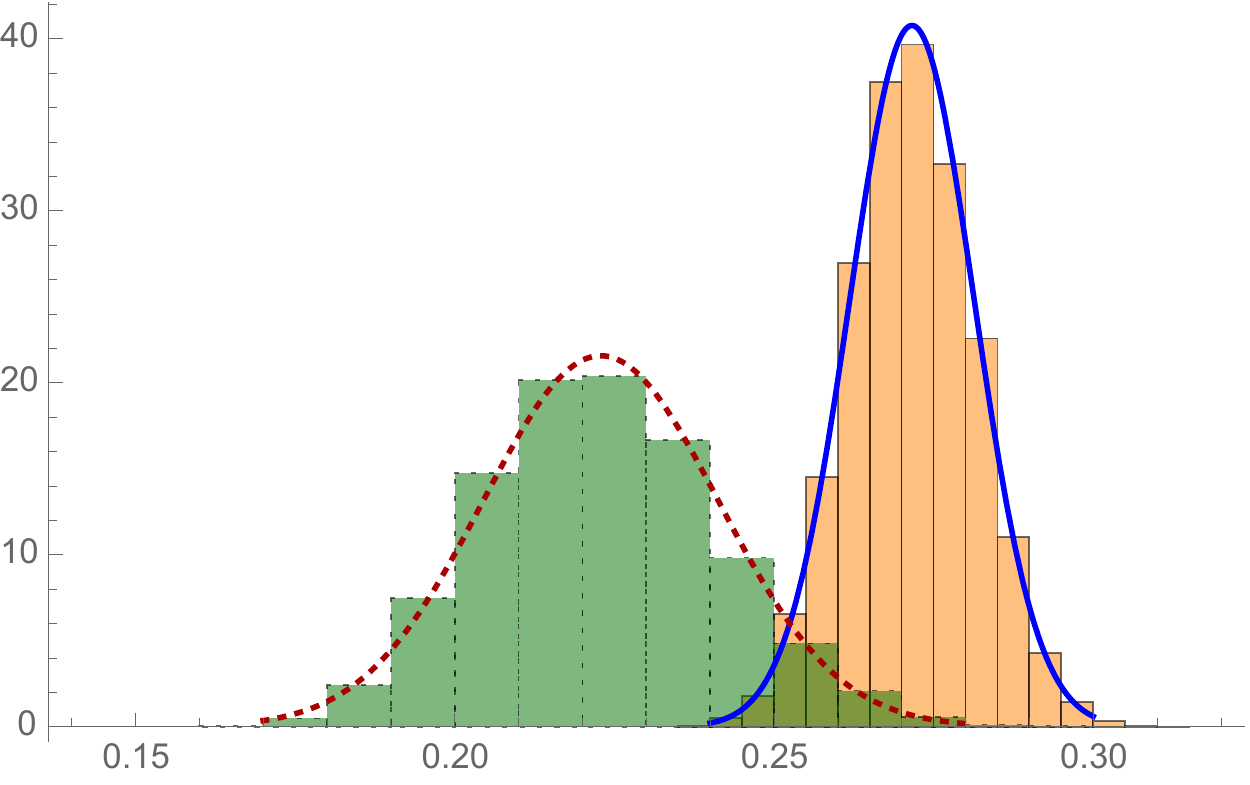}
\label{fig:RDstar_A}}\\
{\includegraphics[scale=0.85]{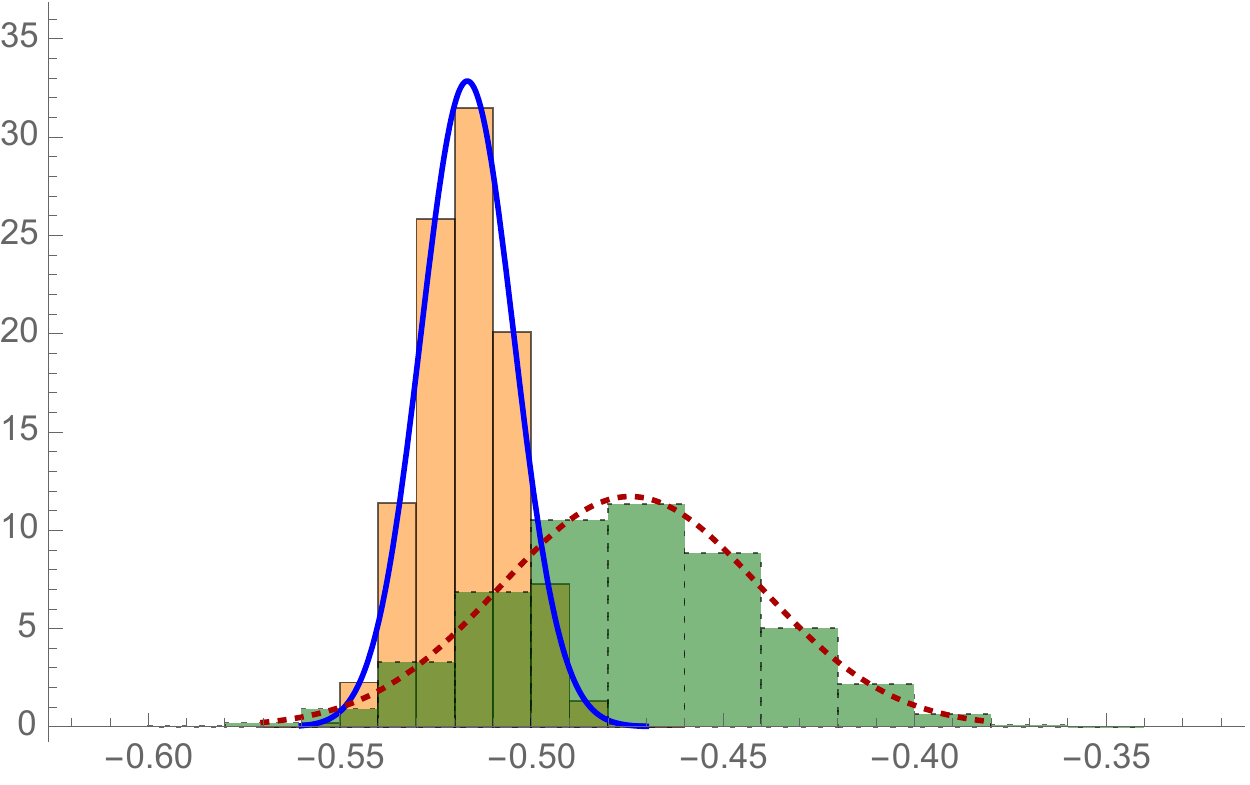}
\label{fig:RDstar_B}}\\ 
{\includegraphics[scale=0.85]{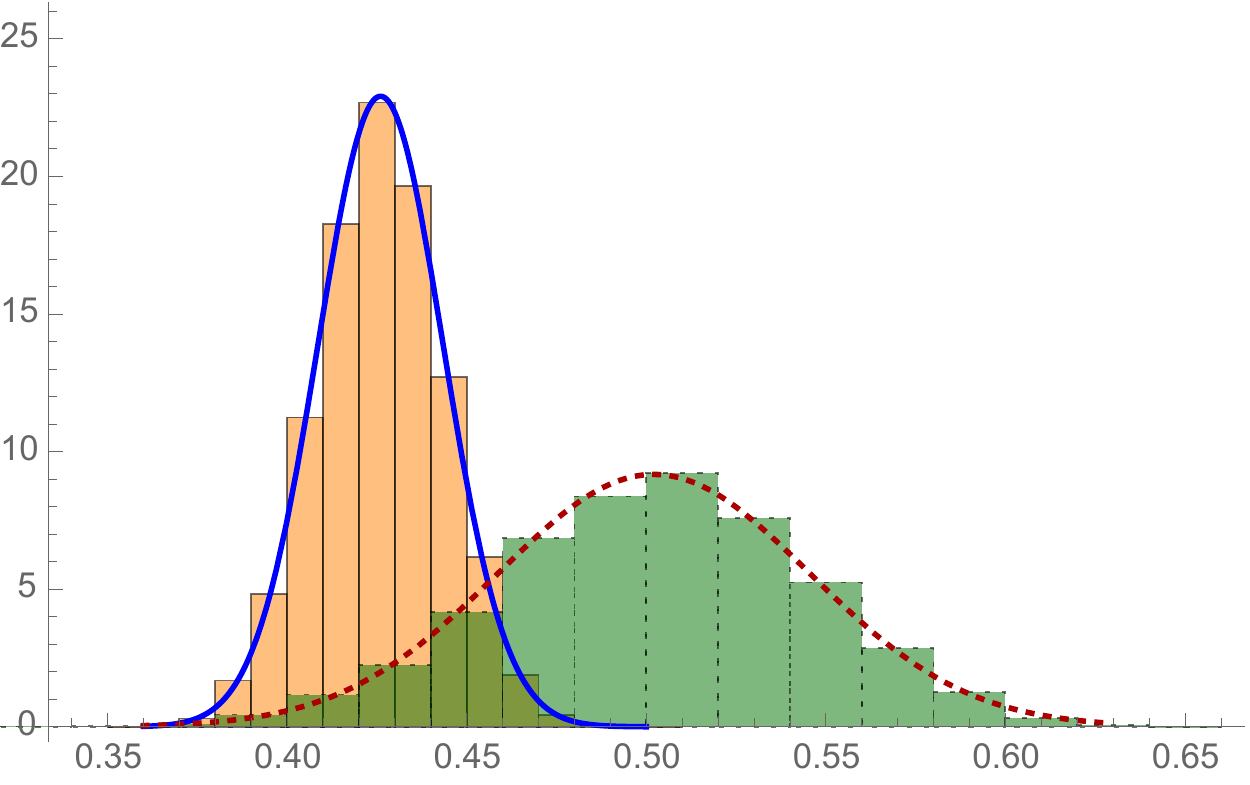}
\label{fig:RDstar_C}}
\caption{\textit{The bins represent the PDFs of the bootstrap events respectively for $R(D^*)$ (top), $P_{\tau}$ (middle) and $F_L$ (bottom). The curves are the best Gaussian fits of the bootstraps themselves. Colour code: the solid orange (dotted green) bins and the solid blue (dotted red) curve result from the analysis with the FNAL/MILC (JLQCD) input data.}
\hspace*{\fill} \small}
\label{RDstar}
\end{center}
\end{figure}

The ratio $R(D^*)$ is a powerful test of the Lepton Flavour Universality (LFU), one of the pillars of the SM. Its definition is 
\begin{equation}
R(D^*) \equiv \frac{\Gamma(B \to D^* \tau \nu_{\tau})}{\Gamma(B \to D^* \ell \nu)},
\end{equation}
where $\ell$ is a light lepton, namely an electron or a muon. For more explicit formul\ae\, in terms of the various FFs, see \cite{Ivanov:2016qtw, Bigi_2017}. In order to obtain new estimates of $R(D^*)$, we proceed as follows. We compute $N_{boot}$ values of $R(D^*)$ by using the bootstrap events of the FFs previously extracted. We then fit the histogram of these events with a normal distribution, in order to obtain a final expectation value and a final uncertainty, see Fig.\,\ref{RDstar}a for a graphical representation of this procedure for both the FNAL/MILC and the JLQCD inputs. Our results are summarized in the first row of Table \ref{tab:pheno}, together with the most recent average of the measurements of $R(D^*)$ \cite{HFLAV}  
\begin{equation}
\label{Rdsexp}
R(D^*) = 0.295 \pm 0.011 \pm 0.008,
\end{equation}
where the first error is statistical and the second one systematic. The FNAL/MILC and the JLQCD results are in tension to each other at the $\sim2.3\sigma$ level. Moreover, the difference between theory and experiment is large only in the JLQCD and in the FNAL/MILC+JLQCD cases, respectively with a $\sim 3.1\sigma$ and a $\sim 2.0\sigma$ tension. The FNAL/MILC estimate, instead, is compatible with the experimental data. We have also compared our values of $R(D^*)$ with the average of the theoretical estimates computed by HFLAV \cite{HFLAV}, which is based on \cite{Jaiswal_2017, Bernlochner:2017jka, Bigi_2017} and reads
\begin{equation}
\label{RdsHFLAV}
R(D^*) = 0.258 \pm 0.005.
\end{equation}
Since our bands of the FFs are not constrained by experimental data in the high-$z$ regime, the uncertainties of our results are larger than the one reported by HFLAV in (\ref{RdsHFLAV}). While the FNAL/MILC and the FNAL/MILC+JLQCD estimates are compatible with HFLAV, the JLQCD one presents a $\sim 1.8\sigma$ tension with it.

\begin{table}[htb!]
\renewcommand{\arraystretch}{1.1}
\begin{center}
{\small
\begin{tabular}{|c||c||c||c||c|}
\hline
& FNAL/MILC \, \cite{Aviles-Casco:2019zop} & JLQCD \, \cite{Kaneko:2019vkx} & FNAL/MILC+JLQCD & Experiments\\
\hline
\hline
$R(D^*)$ & $0.272 \pm 0.010$ & $0.224 \pm 0.018$ & $0.249 \pm 0.019$ & $0.295 \pm 0.011 \pm 0.008$ \, \cite{HFLAV} \\
$P_{\tau}(D^*)$ & $-0.52 \pm 0.01$ & $-0.47 \pm 0.03$ & $-0.50 \pm 0.03$ & $-0.38 \pm 0.51^{+0.21}_{-0.16}$ \, \cite{Hirose:2016wfn}\\
$F_L(D^*)$ & $0.43 \pm 0.02$ & $0.50 \pm 0.04$ & $ 0.46 \pm 0.04$ & $0.60 \pm 0.08 \pm 0.04$ \, \cite{Abdesselam:2019wbt}\\
\hline
\end{tabular}
}
\caption{\textit{Summary of the extrapolated values of the anomaly $R(D^*)$, the $\tau$-polarization $P_{\tau}$ and the $D^*$ longitudinal polarization $F_L$ adopting respectively the FNAL/MILC, the JLQCD and the FNAL/MILC+JLQCD lattice data as inputs for the matrix method.}
\hspace*{\fill} \small}
\label{tab:pheno}
\end{center}
\end{table}

Since the specific blinding factor adopted by FNAL/MILC in Ref.~\cite{Aviles-Casco:2019zop} has a negligible impact on the evaluation of the ratio $R(D^*)$, we quote as our final estimate the weighted average of the results obtained using either FNAL/MILC or JLQCD lattice data, namely
\be
    R(D^*) = 0.261 \pm 0.020 ~ , ~
    \label{eq:RDstar_final}
\ee 
where a scale factor of $\simeq 2.3$ has been applied to get the final uncertainty.

There are other interesting observables that can be computed: the $\tau$-polarization $P_{\tau}$ and the $D^*$ longitudinal polarization $F_L$ . All the necessary formul\ae\, in terms of the various FFs can be read directly from \cite{Ivanov:2016qtw, Bigi_2017, Bhattacharya:2018kig}. For these two quantities, we follow the same procedure described for the $R(D^*)$ case. In Figs\,\ref{RDstar}b-\ref{RDstar}c we show the distributions of the events and the relative Gaussian fits for both the FNAL/MILC and the JLQCD inputs. Our results are summarized in the second and in the third rows of Table \ref{tab:pheno} together with the Belle measurements \cite{Hirose:2016wfn, Abdesselam:2019wbt}: 
\begin{eqnarray}
\label{Ptauexp}
P_{\tau}(D^*)\vert_{\exp} &=& -0.38 \pm 0.51^{+0.21}_{-0.16},\\
\label{FLexp}
F_L(D^*)\vert_{\exp} &=& 0.60 \pm 0.08 \pm 0.04.
\end{eqnarray}
Other theoretical predictions can be found in \cite{Bhattacharya:2018kig, Gambino:2019sif, Jaiswal:2020wer}. For what concerns $P_{\tau}$, we see that all the theoretical results are compatible with the measurement (\ref{Ptauexp}). For $F_L$ the JLQCD and the FNAL/MILC+JLQCD cases are in agreement  with the experiments, while we have a tension of $\sim1.9\sigma$ for the FNAL/MILC one.

\section{SEMILEPTONIC $B \to D$ DECAYS}

%%%%%%%%%%%%%%% BD %%%%%%%%%%%%%%%%%%%

In this Section, we study semileptonic $B \to D$ decays. Our goal is to extract $\vert V_{cb} \vert$ and the ratio $R(D)$ through the DM method. The inputs are the FNAL/MILC data~\cite{Bailey_2015}, summarized in the Table \ref{tab:LQCDMILC}. There exists another lattice computation performed by the HPQCD Collaboration~\cite{Na_2015}. The HPQCD values of the FFs are compatible with the FNAL/MILC results within larger uncertainties and, for this reason, they will not be considered in the following. As for the susceptibilities, we use the results of our non-perturbative lattice computation, see the last column of Table \ref{tab:chi}.

\begin{table}[htb!]
\renewcommand{\arraystretch}{1.1}
\begin{center}
{\small
\begin{tabular}{ c|c|cccccc }
& FNAL/MILC &&& Correlation & Matrix &&\\
\hline
$f_+(1)$ & 1.1994(095) & 1.\,\,\,\,\, & 0.9674 & 0.8812 & 0.8290\,\,\,\,\, & 0.8533\,\,\,\,\, & 0.8032\\
$f_+(1.08)$ & 1.0941(104) & 0.9674\,\,\,\,\, & 1. & 0.9532 & 0.8241\,\,\,\,\, & 0.8992\,\,\,\,\, & 0.8856\\
$f_+(1.16)$ & 1.0047(123) & 0.8812\,\,\,\,\, & 0.9523 & 1. & 0.7892\,\,\,\,\, & 0.8900\,\,\,\,\, & 0.9530\\
$f_0(1)$ & 0.9026(072) & 0.8290\,\,\,\,\, & 0.8241 & 0.7892 & 1.\,\,\,\,\, & 0.9650\,\,\,\,\, & 0.8682\\
$f_0(1.08)$ & 0.8609(077) & 0.8533\,\,\,\,\, & 0.8992 & 0.8900 & 0.9650\,\,\,\,\, & 1.\,\,\,\,\, & 0.9519\\
$f_0(1.16)$ & 0.8254(094) & 0.8032\,\,\,\,\, & 0.8856 & 0.9530 & 0.8682\,\,\,\,\, & 0.9519\,\,\,\,\, & 1.\\
\end{tabular}
}
\caption{\textit{Values of the LQCD computations of the FFs $f_{+,0}(w)$ and their correlations as reported by FNAL/MILC Collaboration in \cite{Bailey_2015}.}
\hspace*{\fill} \small}
\label{tab:LQCDMILC}
\end{center}
\end{table}

\subsection{Theoretical expression of the differential decay width}

The hadronic matrix element reads
\begin{equation}
\label{FINALmatrelem}
 \Braket{D(p_{D})| V^{\mu} |B(p_{B})}=f^{+}(q^2)\left(p_{B}^{\mu}+p_{D}^{\mu} - \frac{m_{B}^2-m_{D}^2}{q^2}q^{\mu}\right) +f^{0}(q^2) \frac{m_{B}^2-m_{D}^2}{q^2} q^{\mu},
\end{equation}
where
\begin{equation}
\label{fpfmf0}
f^{0}(q^2) = \frac{q^2}{m_{B}^2-m_{D}^2}f^-(q^2) + f^+(q^2).
\end{equation}
and $q^{\mu} = p_B^{\mu} - p_D^{\mu}$. The two FFs in Eq.\,(\ref{FINALmatrelem}) are constrained by the kinematical relation
\begin{equation}
\label{KC}
f^{0}(0) = f^{+}(0).
\end{equation} 

A direct computation gives the final expression of the differential decay width
\begin{equation}
\begin{aligned}
\label{finaldiff333}
&\frac{d\Gamma}{dq^2}=\frac{G_F^2 \vert V_{cb} \vert^2 \eta_{EW}^2}{24\pi^3} \left(1-\frac{m_{\ell}^2}{q^2}\right)^2\\
&\hskip 0.65truecm\times \left[\vert \vec{p}_{D}\vert^3 \left(1+\frac{m_{\ell}^2}{2q^2}\right) \vert f^+(q^2) \vert^2 + m_{B}^2 \vert \vec{p}_{D} \vert \left( 1-\frac{m_{D}^2}{m_{B}^2}\right)^2 \frac{3m_{\ell}^2}{8q^2} \vert f^0(q^2) \vert^2\right],
\end{aligned}
\end{equation}
where $G_F$ is the Fermi constant, $\vec{p}_{D}$ the 3-momentum of the $D$ meson, $i.e.$
\begin{equation}
\label{3momD}
\vert \vec{p}_{D} \vert = \left[ \left( \frac{m_{B}^2+m_{D}^2-q^2}{2m_{B}}  \right)^2  -m_{D}^2 \right]^{1/2},
\end{equation}
$\eta_{EW} = 1+\alpha\ln(M_Z/m_B)/\pi \simeq 1.0066$ the leading electromagnetic correction and $m_{\ell}$ the mass of the produced lepton. Since the Belle collaboration measured the differential decay width (\ref{finaldiff333}) for the production of an electron or a muon, one can neglect the mass of the lepton. Thus, the expression (\ref{finaldiff333}) simplifies to
\begin{equation}
\label{finaldiff2}
\frac{d\Gamma}{dq^2}\simeq\frac{G_F^2 \vert V_{cb} \vert^2\eta_{EW}^2}{24\pi^3} \vert \vec{p}_{D} \vert^3 \vert f^+(q^2) \vert^2.
\end{equation}

\subsection{Application of the DM method to the description of the FFs}

In order to write down the matrices (\ref{eq:Delta}) for $f_{+,0}(z)$, we need the following kinematical functions
\begin{eqnarray}
\label{eq:fac_chinese}
\phi_{+}(z) & = & \frac{8r^2}{m_B} \sqrt{\frac{16}{3 \pi}} \, 
                            \frac{(1+z)^2 \sqrt{1-z}}{\left[(1+r)(1-z)+2\sqrt{r}(1+z)\right]^5}\ , \\
\label{eq:fac_chinese2}
\phi_{0}(z) & = & r (1 - r^2) \sqrt{\frac{16}{ \pi}} \,
                           \frac{(1-z^2) \sqrt{1-z}}{\left[(1+r)(1-z)+2\sqrt{r}(1+z)\right]^4}\,  
\end{eqnarray} 
with $r \equiv m_D / m_B$.
Each of them is then modified according to Eq.\,(\ref{chinesePOLES}) using the poles in Table III of \cite{Bigi_2017}. The susceptibilities are those in the last column of Table \ref{tab:chi}.

\begin{figure}[htb!]
\begin{center}
\includegraphics[width=0.95\textwidth]{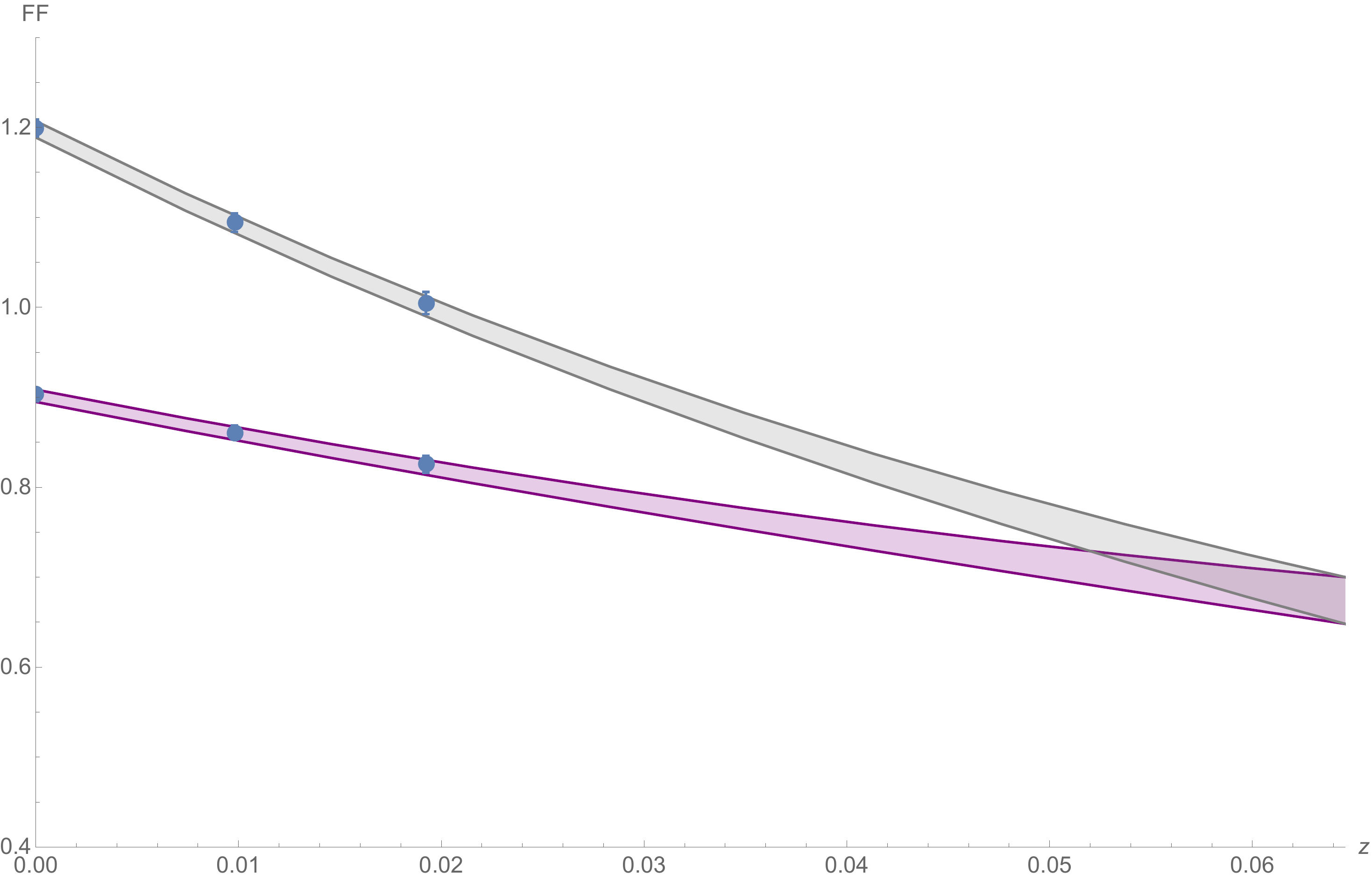}
\caption{\textit{The bands of the FFs entering $B \to D \ell \nu$ decays computed through the DM method. The colour code is (lower) violet band for $f_{0}(z)$, (higher) gray one for $f_{+}(z)$. The blue points are the FNAL/MILC data \cite{Bailey_2015}, summarized in Table \ref{tab:LQCDMILC}.}
\hspace*{\fill} \small}
\label{FFMMbd}
\end{center}
\end{figure} 

In Fig.\,\ref{FFMMbd} we show the bands of the FFs resulting from the DM method. In order to obtain these bands we used the sceptical approach: with the extraction of 100 values of $r$ (common to both FFs), we are able to recover all the generated bootstraps, which then contribute to the final bands of the FFs\footnote{The distribution  of the values of the sceptical parameter r turns out to be peaked around $r = 1$.}. The extrapolation at $z_{max}$ is crucial in order to analyse experimental data. The matrix description gives the result
\begin{equation}
\label{fplusBDzmax}
f(z_{max})=0.674 \pm 0.026,
\end{equation}
which is compatible with the LCSR estimate \cite{Gubernari:2018wyi}
\begin{equation*}
f(z_{max})_{\rm{LCSR}}=0.65 \pm 0.08.
\end{equation*}
We stress that the value (\ref{fplusBDzmax}) comes from a non-perturbative and model-independent approach. These two properties are important differences in comparison with the popular parametrizations present in the literature, like for example the BGL \cite{Boyd:1995sq, Boyd:1995cf, Boyd:1997kz} and the CLN \cite{Caprini:1995wq,Caprini:1997mu} ones. We use the results of this Section to obtain updated values of both $\vert V_{cb} \vert$ and $R(D)$.

\subsection{New estimate of $\vert V_{cb} \vert$}

In order to obtain an updated value of $\vert V_{cb} \vert$, we put together  our description of the lattice FFs in the whole kinematical range and the experiments. Let us briefly describe the experimental state-of-the-art. The most recent measurement of the differential decay width $d\Gamma/dw$ has been performed at Belle \cite{Glattauer_2016}. In Table II of this work, the authors report the results of the measurements, dividing the kinematical region into 10 bins in the recoil variable $w$. The correlation matrix of the systematic errors is also given. 

We follow the procedure \cite{Riggio:2017zwh}, that has been used for the extraction of $\vert V_{cd} \vert$ and $\vert V_{cs} \vert$ in the case of the semileptonic $D \to \pi \ell \nu$ and $D \to K \ell \nu$ decays. First of all, we re-express Eq.\,(\ref{finaldiff2}) as
\begin{equation}
\label{VcbFINAL}
\vert V_{cb} \vert = \sqrt{\frac{d\Gamma}{dq^2}\vert_{exp} \times \frac{24 \pi^3}{G_F^2 \eta_{EW}^2 \vert \vec{p}_{D} \vert^3 \vert f^+(q^2) \vert^2_{th}}}.
\end{equation}
Then, we generate $N_{boot}$ bootstraps of the experimental differential decay width for every bin in $w$ through a multivariate Gaussian distribution and similarly we extract $N_{boot}$ bootstrap events for the FFs $f_{+,0}(w)$ for each of the bins $w_i$ ($i=1,\cdots,10$). The mean value and the covariance matrix of the distribution can be directly computed through our DM method. Finally, we compute $\vert V_{cb} \vert$ for each recoil bin through the expression (\ref{VcbFINAL}).

\begin{figure}[htb!]
\begin{center}
\includegraphics[width=0.95\textwidth]{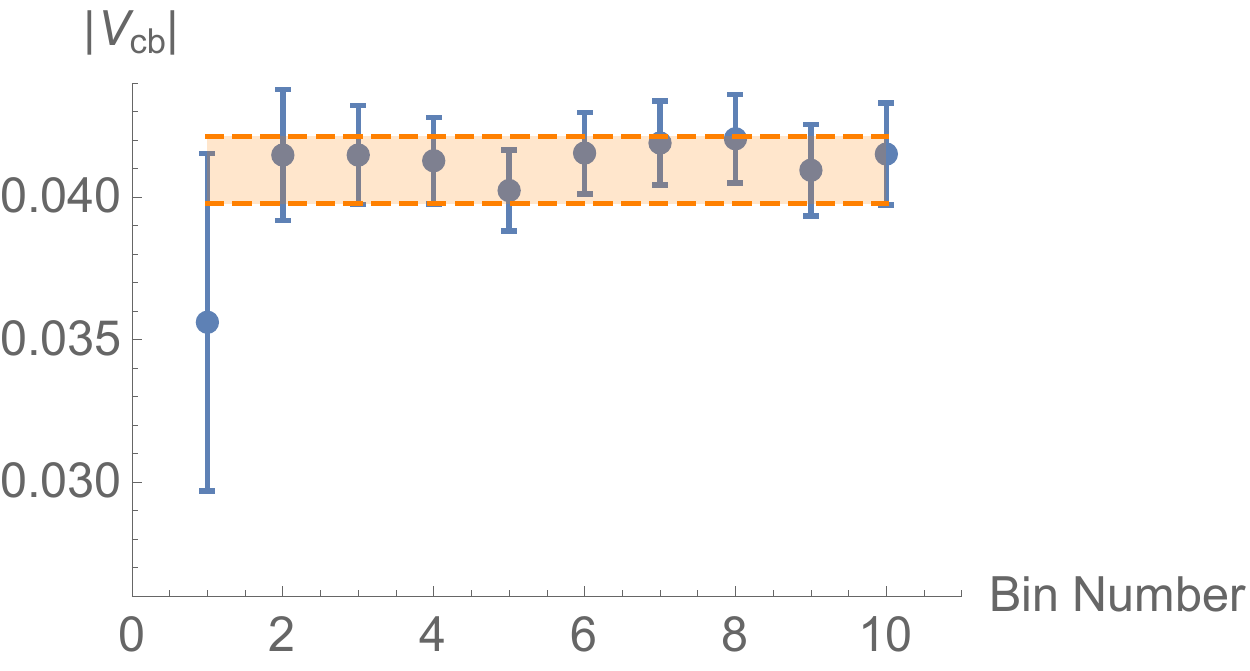}
\caption{\textit{Values of the CKM matrix element $\vert V_{cb} \vert$ resulting from the combination of the sceptical matrix description of the FFs and the experimental data by Belle. The orange band represents the result of the weighted mean described in the text.}
\hspace*{\fill} \small}
\label{Vcbfinalvalue}
\end{center}
\end{figure}

Let us now fix a particular bin. In order to extract a  mean value and uncertainty for $\vert V_{cb} \vert$, we fit the histogram of these events with a normal distribution and save the values of the corresponding marginalized parameters. We then combine the resulting 10 values of the CKM matrix element, one for each $w$ bin. The application of Eq.\,(\ref{muVcbfinal}) gives us the following result for $\vert V_{cb} \vert$
\begin{equation}
\label{VcbestLAST}
\vert V_{cb} \vert \times 10^{3} = 41.0 \pm 1.2.
\end{equation}
In Fig.\,\ref{Vcbfinalvalue} we show the 10 values of the CKM matrix element, one for each $w$ bin, and the final band corresponding to Eq.\,(\ref{VcbestLAST}).

Let us now compare our value of $\vert V_{cb} \vert$ with the inclusive and the other exclusive estimates. Our estimate (\ref{VcbestLAST}) results to be compatible with the inclusive one (\ref{Vcbestincl}) at less than 1$\sigma$. For what concerns instead the other exclusive determinations ($B \to D \ell \nu$ \emph{only}), some of the results present in literature are 
\begin{eqnarray*}
\vert V_{cb} \vert_{excl} \times 10^{3} &=& 40.49 \pm 0.97 \qquad \text{\cite{Bigi_2016}}~ , ~\\
\vert V_{cb} \vert_{excl} \times 10^{3} &=& 41.0 \pm 1.1 \qquad \quad \text{\cite{Jaiswal_2017} }~ , ~\\
\vert V_{cb} \vert_{excl} \times 10^{3} &=& 40.1 \pm 1.0 \qquad \quad \text{\cite{Aoki:2019cca}} ~ . ~\\
\end{eqnarray*}
All these determinations are compatible with our result in Eq.\,(\ref{VcbestLAST}).

In conclusion, while in the analyses of Refs.~\cite{Bigi_2016, Jaiswal_2017, Aoki:2019cca} the lattice and the experimental data are fitted \emph{all together} in order to constrain the shape of the FFs, in this work the two sources of information are always kept separate. To be more precise, the LQCD computations are used in order to derive the allowed bands of the FFs, while the experimental measurements are considered only for the final determination of $\vert V_{cb} \vert$, avoiding in this way any possible bias of the experimental distribution on the theoretical predictions and hence on the extracted value of $\vert V_{cb}\vert$. This difference justifies the larger uncertainty of our estimate of $\vert V_{cb} \vert$ with respect to the other calculations.

\subsection{New estimate of $R(D)$}

The ratio $R(D)$ is defined as 
\begin{equation}
R(D) \equiv \frac{\Gamma(B \to D \tau \nu_{\tau})}{\Gamma(B \to D \ell \nu)}.
\end{equation}
Since in the $B \to D$ case we have only two FFs to deal with, we can write the rather compact expression 
\begin{equation}
\small{
R(D) = \frac{\int_{m_\tau^2}^{(m_B - m_D)^2} dq^2 \vert \vec{p}_D \vert^3 f_+^2(q^2) \left( 1-\frac{m_{\tau}^2}{q^2} \right)^2 \left[ 1+\frac{m_{\tau}^2}{2q^2} + \frac{m_B^2}{\vert \vec{p_D} \vert^2} \left( 1-\frac{m_{D}^2}{m_B^2} \right)^2 \frac{3 m_{\tau}^2}{8q^2} \frac{f_0^2(q^2)}{f_+^2(q^2)} \right]}{\int_0^{(m_B - m_D)^2} dq^2  \vert \vec{p}_D \vert^3 f_+^2(q^2)},
}
\end{equation}
where $m_{\tau}$ is the mass of the $\tau$ lepton and we have considered the electron and the muon as massless.

\begin{figure}[htb!]
\begin{center}
\includegraphics[width=0.95\textwidth]{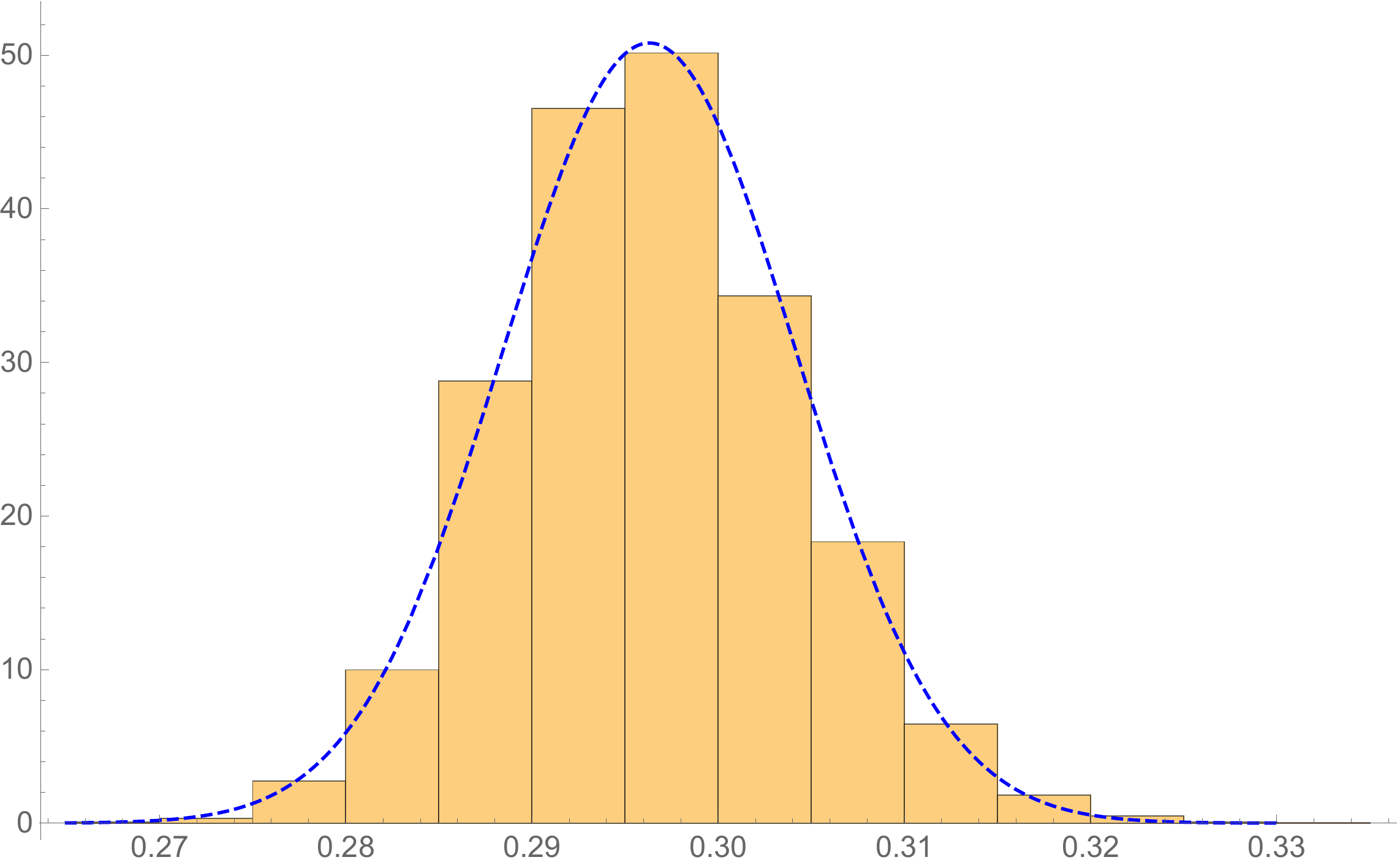}
\caption{\textit{The orange bins represent the PDF of the bootstrap events for $R(D)$. The dashed blue curve is the best Gaussian fit of the bootstraps themselves.}
\hspace*{\fill} \small}
\label{RDplotplot}
\end{center}
\end{figure}

In order to obtain a new estimate of $R(D)$ through the matrix description of the FFs, we proceed as for $R(D^*)$. Using the bootstrap events for the FFs extracted as explained before, we compute $N_{boot}$ values of the ratio $R(D)$, and fit the histogram of these events with a normal distribution, in order to obtain the expectation value and the uncertainty, as in Fig.\,\ref{RDplotplot}. Our prediction is
\begin{equation*}
R(D) = 0.296 \pm 0.008.
\end{equation*}

We compare our estimate of $R(D)$ with the average of the theoretical estimates computed by HFLAV \cite{HFLAV}, based on \cite{Bigi_2016, Jaiswal_2017, Bernlochner:2017jka}
\begin{equation*}
R(D) = 0.299 \pm 0.003.
\end{equation*}
The two results are compatible to each other. Note that our uncertainty is larger than the one by HFLAV, since in \cite{Bigi_2016, Jaiswal_2017, Bernlochner:2017jka} the authors used experimental data to constrain the FFs. We then compare our estimate with the most recent average of the measurements \cite{HFLAV}  
\begin{equation*}
R(D) = 0.340 \pm 0.027 \pm 0.013,
\end{equation*}
where the first error statistical is and the second one systematic. Hence, we have a $\sim 1.4\sigma$ tension between  the theoretical and the experimental determinations of this quantity.

\section{Conclusion}

In this paper we have re-analysed the lattice and the experimental data concerning $B \to D^{(*)} \ell \nu$ decays. We have shed a new light onto the two phenomenological problems that at present affect these processes, $i.e.$ the $\vert V_{cb} \vert$ puzzle and the anomalies $R(D^{(*)})$. The most original contribution of our analysis is the new approach to the description of the FFs, namely the application of non-perturbative and model-independent matrix method described in \cite{DiCarlo:2021dzg} and applied in the present analysis to semileptonic $B \to D^{(*)}$ decays. For our numerical study, we used the non-perturbative values of the susceptibilities, computed on the lattice, as discussed in all details in \cite{Martinelli:2021frl}. 

The lattice data used for the $B \to D^* \ell \nu$ decays are the preliminary unblinded data by the JLQCD Collaboration~\cite{Kaneko:2019vkx} and the preliminary blinded ones by the FNAL/MILC Collaboration~\cite{Aviles-Casco:2019zop}. However, our interest for these decays is mainly methodological, namely we want to apply a new method to study the $B \to D^{(*)} \ell \nu$ transitions, having in mind that the whole analysis has to be repeated once the final calculations for the FFs and their correlations will be available. Our approach has highlighted a problem related to the experimental correlation matrix of the data of Ref.\,\cite{Abdesselam:2017kjf}, which we discussed in detail in Section III D, while confirming previous results for the $B \to D \ell \nu$ transitions. 

Our final results for $\vert V_{cb} \vert$ can be summarized as  
\begin{itemize}
\item for the $B \to D$ case using the final FNAL/MILC\,\cite{Bailey_2015} lattice data
\begin{equation*}
\vert V_{cb} \vert \times 10^{3} = 41.0 \pm 1.2
\end{equation*}
\item for the $B \to D^*$ case using the preliminary JLQCD\,\cite{Kaneko:2019vkx} lattice data
\begin{equation*}
\vert V_{cb} \vert \times 10^{3} = 40.4 \pm 1.8 ~ . ~
\end{equation*}
\end{itemize}
They are compatible with each other and lower than, but still consistent with the inclusive determination $\vert V_{cb} \vert_{incl} = (42.00 \pm 0.65) \cdot 10^{-3}$\,\cite{Gambino:2016jkc, Aoki:2019cca} at the 1$\sigma$ level. 
In the $B \to D$ case the uncertainty of our result is comparable with those obtained in literature using experimental data to constrain the shape of the FFs (see Refs.\,\cite{Bigi_2016,Jaiswal_2017,Aoki:2019cca}), while for $B \to D^*$ it is greater, but nevertheless still remarkably good (see Refs.\,\cite{Gambino:2019sif,Jaiswal:2020wer,Aoki:2019cca}).
Furthermore, using the final FNAL/MILC\,\cite{Bailey_2015} lattice results for the $B \to D$ case and the preliminary JLQCD\,\cite{Kaneko:2019vkx} and FNAL/MILC\,\cite{Aviles-Casco:2019zop} lattice data for the $B \to D^*$ case we have obtained the following pure theoretical estimates of the ratios $R(D^{(*)})$:
\begin{equation*}
R(D) = 0.296 \pm 0.008,\,\,\,\,\,R(D^*)= 0.261 \pm 0.020 ~ , ~
\end{equation*}
which differ by $\sim 1.4\sigma$ from the latest experimental determinations\,\cite{HFLAV}.

It is important to state that the dispersion matrix method can be applied to whatever semileptonic process. In particular, it will be interesting to take into consideration the exclusive semileptonic decays of $B_s$ mesons, namely $B_s^0 \to D^-_s \mu^+ \nu_\mu$ and $B_s^0 \to D^{*\, -} _s \mu^+ \nu_\mu$ \cite{Aaij:2020hsi,Aaij:2020xjy}, which have been recently measured at LHCb and give larger values for $\vert V_{cb} \vert$. Finally, the same approach can be extended also to the baryons, $i.e.$  $\Lambda_b \to \Lambda_c \ell \nu$, and to the $b \to u$ transitions, first of all $B \to \pi \ell \nu$, with the aim of determining $\vert V_{ub} \vert$.

\begin{acknowledgments}
We are grateful to Giulio d'Agostini, Luca Silvestrini, Ayan Paul and Mauro Valli for very useful discussions. We acknowledge PRACE for awarding us access to Marconi at CINECA, Italy under the grant the PRACE project PRA067.  We also acknowledge use of CPU time provided by CINECA under the specific initiative INFN-LQCD123.
G.M. and S.S. thank MIUR  (Italy)  for  partial  support  under  the  contract  PRIN 2015. 
S.S are supported by the Italian Ministry of Research (MIUR) under grant PRIN 20172LNEEZ.
\end{acknowledgments}

\appendix
\section{A way to combine LQCD computations and to determine the final estimate of $\vert V_{cb} \vert$ for $B \to D^*$ decays}

Let us focus only on the $B \to D^* \ell \nu$ decays. In order to combine the FNAL/MILC and the JLQCD lattice values of the FFs and to obtain a final result for $\vert V_{cb} \vert$, given the various estimates for each of the $\{w,  \cos \theta_l, \cos \theta_v, \chi\}$ variables, we have used the following strategy \cite{Carrasco:2014cwa}. Let us assume that we have $N$ determinations of a physical quantity $x$, each of them with mean value $x_k$ and uncertainty $\sigma_k$ ($k=1,\cdots,N$). Our goal is to combine them in a final estimate. By assigning the same weight to the $N$ values, we have that
\begin{eqnarray}
\label{eq28Carr}
\mu_{x} &=& \frac{1}{N} \sum_{k=1}^N x_k,\\
\sigma^2_{x} &=& \frac{1}{N} \sum_{k=1}^N \sigma_k^2 + \frac{1}{N} \sum_{k=1}^N(x_k-\mu_{x})^2.
\label{eq28Carrb}
\end{eqnarray}
As explained in the central text, Eq.\,(\ref{eq28Carrb}) gives a large value of the uncertainty whenever the $x_k$ are very different to each other because of the second term in its r.h.s.

\section{Correlation matrix of LQCD data for semileptonic $B \to D^*$ decays}

In Table \ref{tab:LQCDMILCJLQCDcorr} we present the correlation matrix that we have used for our study of the semileptonic $B \to D^*$ transitions. Our assumption is to consider a high correlation between the values of the same FF computed at the three different recoils while we assume zero correlation between the values of the different FFs.

\begin{table}[htb!]
\renewcommand{\arraystretch}{1.1}
\begin{center}
{\small
\begin{tabular}{ c|cccccccccccc}
&&&&&Correlation &matrix &&&\\
\hline
$h_V(w_1)$& 1.\,\,\,\,\,\,\,\,\,\, & 0.9\,\,\,\,\,\,\,\,\,\,& 0.8\,\,\,\,\,\,\,\,\,\,\,\, & 0. & 0. & 0.\,\,\,\,\,\,\,\,\,\, & 0.\,\,\,\,\,\,\,\,\,\,& 0.\,\,\,\,\,\,\,\,\,\, & 0.\,\,\,\,\,\,\,\,\,\, & 0.\,\,\,\,\,\,\,\,\,\, & 0.\,\,\,\,\,\,\,\,\,\, & 0. \\
$h_V(w_2)$& 0.9\,\,\,\,\,\,\,\,\,\, & 1.\,\,\,\,\,\,\,\,\,\,& 0.9\,\,\,\,\,\,\,\,\,\,\,\, & 0. & 0. & 0.\,\,\,\,\,\,\,\,\,\, & 0.\,\,\,\,\,\,\,\,\,\,& 0.\,\,\,\,\,\,\,\,\,\, & 0.\,\,\,\,\,\,\,\,\,\, & 0.\,\,\,\,\,\,\,\,\,\, & 0.\,\,\,\,\,\,\,\,\,\, & 0. \\
$h_V(w_3)$&  0.8\,\,\,\,\,\,\,\,\,\, & 0.9\,\,\,\,\,\,\,\,\,\,& 1.\,\,\,\,\,\,\,\,\,\,\,\, & 0. & 0. & 0.\,\,\,\,\,\,\,\,\,\, & 0.\,\,\,\,\,\,\,\,\,\,& 0.\,\,\,\,\,\,\,\,\,\, & 0.\,\,\,\,\,\,\,\,\,\, & 0.\,\,\,\,\,\,\,\,\,\, & 0.\,\,\,\,\,\,\,\,\,\, & 0. \\
$h_{A_1}(w_1)$&  0.\,\,\,\,\,\,\,\,\,\, & 0.\,\,\,\,\,\,\,\,\,\,& 0.\,\,\,\,\,\,\,\,\,\,\,\, & 1. & 0.9 & 0.8\,\,\,\,\,\,\,\,\,\, & 0.\,\,\,\,\,\,\,\,\,\,& 0.\,\,\,\,\,\,\,\,\,\, & 0.\,\,\,\,\,\,\,\,\,\, & 0.\,\,\,\,\,\,\,\,\,\, & 0.\,\,\,\,\,\,\,\,\,\, & 0. \\
$h_{A_1}(w_2)$&  0.\,\,\,\,\,\,\,\,\,\, & 0.\,\,\,\,\,\,\,\,\,\,& 0.\,\,\,\,\,\,\,\,\,\,\,\, & 0.9 & 1. & 0.9\,\,\,\,\,\,\,\,\,\, & 0.\,\,\,\,\,\,\,\,\,\,& 0.\,\,\,\,\,\,\,\,\,\, & 0.\,\,\,\,\,\,\,\,\,\, & 0.\,\,\,\,\,\,\,\,\,\, & 0.\,\,\,\,\,\,\,\,\,\, & 0.  \\
$h_{A_1}(w_3)$&  0.\,\,\,\,\,\,\,\,\,\, & 0.\,\,\,\,\,\,\,\,\,\,& 0.\,\,\,\,\,\,\,\,\,\,\,\, & 0.8 & 0.9 & 1.\,\,\,\,\,\,\,\,\,\, & 0.\,\,\,\,\,\,\,\,\,\,& 0.\,\,\,\,\,\,\,\,\,\, & 0.\,\,\,\,\,\,\,\,\,\, & 0.\,\,\,\,\,\,\,\,\,\, & 0.\,\,\,\,\,\,\,\,\,\, & 0.  \\
$h_{A_2}(w_1)$&  0.\,\,\,\,\,\,\,\,\,\, & 0.\,\,\,\,\,\,\,\,\,\,& 0.\,\,\,\,\,\,\,\,\,\,\,\, & 0. & 0. & 0.\,\,\,\,\,\,\,\,\,\, & 1.\,\,\,\,\,\,\,\,\,\,& 0.9\,\,\,\,\,\,\,\,\,\, & 0.8\,\,\,\,\,\,\,\,\,\, & 0.\,\,\,\,\,\,\,\,\,\, & 0.\,\,\,\,\,\,\,\,\,\, & 0. \\
$h_{A_2}(w_2)$&  0.\,\,\,\,\,\,\,\,\,\, & 0.\,\,\,\,\,\,\,\,\,\,& 0.\,\,\,\,\,\,\,\,\,\,\,\, & 0. & 0. & 0.\,\,\,\,\,\,\,\,\,\, & 0.9\,\,\,\,\,\,\,\,\,\,& 1.\,\,\,\,\,\,\,\,\,\, & 0.9\,\,\,\,\,\,\,\,\,\, & 0.\,\,\,\,\,\,\,\,\,\, & 0.\,\,\,\,\,\,\,\,\,\, & 0. \\
$h_{A_2}(w_3)$&  0.\,\,\,\,\,\,\,\,\,\, & 0.\,\,\,\,\,\,\,\,\,\,& 0.\,\,\,\,\,\,\,\,\,\,\,\, & 0. & 0. & 0.\,\,\,\,\,\,\,\,\,\, & 0.8\,\,\,\,\,\,\,\,\,\,& 0.9\,\,\,\,\,\,\,\,\,\, & 1.\,\,\,\,\,\,\,\,\,\, & 0.\,\,\,\,\,\,\,\,\,\, & 0.\,\,\,\,\,\,\,\,\,\, & 0.  \\
$h_{A_3}(w_1)$&  0.\,\,\,\,\,\,\,\,\,\, & 0.\,\,\,\,\,\,\,\,\,\,& 0.\,\,\,\,\,\,\,\,\,\,\,\, & 0. & 0. & 0.\,\,\,\,\,\,\,\,\,\, & 0.\,\,\,\,\,\,\,\,\,\,& 0.\,\,\,\,\,\,\,\,\,\, & 0.\,\,\,\,\,\,\,\,\,\, & 1.\,\,\,\,\,\,\,\,\,\, & 0.9\,\,\,\,\,\,\,\,\,\, & 0.8 \\
$h_{A_3}(w_2)$&  0.\,\,\,\,\,\,\,\,\,\, & 0.\,\,\,\,\,\,\,\,\,\,& 0.\,\,\,\,\,\,\,\,\,\,\,\, & 0. & 0. & 0.\,\,\,\,\,\,\,\,\,\, & 0.\,\,\,\,\,\,\,\,\,\,& 0.\,\,\,\,\,\,\,\,\,\, & 0.\,\,\,\,\,\,\,\,\,\, & 0.9\,\,\,\,\,\,\,\,\,\, & 1.\,\,\,\,\,\,\,\,\,\, & 0.9  \\
$h_{A_3}(w_3)$&  0.\,\,\,\,\,\,\,\,\,\, & 0.\,\,\,\,\,\,\,\,\,\,& 0.\,\,\,\,\,\,\,\,\,\,\,\, & 0. & 0. & 0.\,\,\,\,\,\,\,\,\,\, & 0.\,\,\,\,\,\,\,\,\,\,& 0.\,\,\,\,\,\,\,\,\,\, & 0.\,\,\,\,\,\,\,\,\,\, & 0.8\,\,\,\,\,\,\,\,\,\, & 0.9\,\,\,\,\,\,\,\,\,\, & 1. \\
\end{tabular}
}
\caption{\textit{Correlation matrix adopted for FNAL/MILC and JLQCD lattice data in our study.}
\hspace*{\fill} \small}
\label{tab:LQCDMILCJLQCDcorr}
\end{center}
\end{table}

\bibliography{BDstar}

\end{document}